\documentclass[twocolumn]{aastex62}

\usepackage{graphicx}
\usepackage{txfonts}
\usepackage{multirow}
\usepackage{array}

\begin{document}

\title{Plasma Diagnostics From Active Region and Quiet Sun Spectra Observed by \textit{Hinode}/EIS: \\ Quantifying the Departures from  a Maxwellian Distribution}

\correspondingauthor{J. L\"{o}rin\v{c}\'{i}k}
\email{juraj.lorincik@asu.cas.cz}

\author[0000-0002-9690-8456]{Juraj L\"{o}rin\v{c}\'{i}k}
\affil{Institute of Astronomy, Charles University, V Hole\v{s}ovi\v{c}k\'{a}ch 2, CZ-18000 Prague 8, Czech Republic}
\affil{Astronomical Institute of the Czech Academy of Sciences, Fri\v{c}ova 298, 251 65 Ond\v{r}ejov, Czech Republic}

\author[0000-0003-1308-7427]{Jaroslav Dud\'{i}k}
\affil{Astronomical Institute of the Czech Academy of Sciences, Fri\v{c}ova 298, 251 65 Ond\v{r}ejov, Czech Republic}

\author[0000-0002-4125-0204]{Giulio del Zanna}
\affil{DAMTP, Centre for Mathematical Sciences, Wilberforce road, Cambridge, CB3 OWA, UK}

\author[0000-0003-2629-6201]{Elena Dzif\v{c}\'{a}kov\'{a}}
\affil{Astronomical Institute of the Czech Academy of Sciences, Fri\v{c}ova 298, 251 65 Ond\v{r}ejov, Czech Republic}

\author[0000-0002-6418-7914]{Helen E. Mason}
\affil{DAMTP, Centre for Mathematical Sciences, Wilberforce road, Cambridge, CB3 OWA, UK}

\begin{abstract}

We perform plasma diagnostics, including that of the non-Maxwellian $\kappa$-distributions, in several structures observed in the solar corona by the Extreme-Ultraviolet Imaging Spectrometer (EIS) onboard the \textit{Hinode} spacecraft. To prevent uncertainties due to the in-flight calibration of EIS, we selected spectral atlases observed shortly after the launch of the mission. One spectral atlas contains an observation of an active region, while the other is an off-limb quiet Sun region. To minimize the uncertainties of the diagnostics, we rely only on strong lines and we average the signal over a spatial area within selected structures. Multiple plasma parameters are diagnosed, such as the electron density, differential emission measure, and the non-Maxwellian parameter $\kappa$. To do that, we use a simple, well-converging iterative scheme based on refining the initial density estimates via the DEM and $\kappa$. We find that while the quiet Sun spectra are consistent with a Maxwellian distribution, the coronal loops and moss observed within active region are strongly non-Maxwellian with $\kappa$ $\lessapprox$ 3. These results were checked by calculating synthetic ratios using DEMs obtained as a function of $\kappa$. Ratios predicted using the DEMs assuming $\kappa$-distributions converged to the ratios observed in the quiet Sun and coronal loops. To our knowledge, this work presents a strong evidence of a presence of different electron distributions between two physically distinct parts of the solar corona.

\end{abstract}

\keywords{Sun: UV radiation--Sun: corona--techniques: imaging spectroscopy--radiation mechanisms: non-thermal}

\section{Introduction} \label{sec_introduction}

Dynamic phenomena which take place in the solar atmosphere can violate plasma equilibrium, leading to non-Maxwellian (non-thermal) distributions of particles. Populations of non-Maxwellian particles are expected to be present in different parts and structures of the solar atmosphere, where acceleration mechanisms such as turbulence, shocks and magnetic reconnection are likely to occur \citep[see][and references therein]{Dudik17}. The most prominent among these are probably the solar flares, releasing enormous amounts of non-thermal electrons accelerated to speeds reaching fractions of $c$ \citep[see e.g.,][]{fletcher11,bian14, oka18}. A typical example of the non-thermal emission is observed e.g. in the form of power laws in X-ray spectra of instruments such as \textit{RHESSI} or recently \textit{NuSTAR}. It usually originates in bremsstrahlung and microflares \citep[e.g.,][]{christe08,hannah08,wright17}, which energy outputs are lower compared to those of solar flares \citep[e.g.,][]{Lin84}. Non-thermal electrons are also predicted to accompany nanoflares \citep{bakke18, Che18} nowadays commonly discussed in terms of coronal heating \citep[see e.g.,][]{klimchuk2006,reep13,viall17,Priest18}.

An example of a non-Maxwellian distribution of particle energies or velocities, is the $\kappa$- (kappa) distribution. This is defined as \citep[e.g.,][]{owocki83,livadiotis17}:
\begin{equation}
f_\kappa(E) = \left(\frac{m}{2\pi k_{\text{B}}T}\right)^{3/2} \frac{A_\kappa}{\left(1 + \frac{E}{(\kappa -3/2)k_{\text{B}}T}\right)^{\kappa + 1}},
\label{Eq:kappa}
\end{equation}
where $A_\kappa$ is a normalization constant $\Gamma(\kappa + 1)/[\Gamma(\kappa - 1/2)(\kappa - 3/2)^{3/2}]$, $k_{\text{B}}$ is the Boltzmann's constant, and $m$ is the electron mass. The distribution has two parameters: temperature $T \in (0, \infty)$ for which the physical meaning is the same as that of the kinetic temperature $T$ in the Maxwellian distribution \citep{livadiotis09} and $\kappa \in (3/2, \infty)$, which describes the system's departure from the Maxwellian. $\kappa \rightarrow \infty$ corresponds to the Maxwellian distribution, while $\kappa \rightarrow 3/2$ describes its furthest departure. The $\kappa$-distributions are characterized by a nearly-Maxwellian core and a suprathermal tail. The fraction of
particles corresponding to this tail can, e.g. in the case of the $\kappa$\,=\,2 distribution, contain more than 80\% of the total energy of electrons in the system \citep[][]{oka13}.

The $\kappa$-distributions, or distributions with enhanced high-energy tail, are expected to occur due to a range of processes, such as acceleration due to electric fields \citep[e.g.,][]{Burge12,Gordovskyy13,Gordovskyy14,Ripperda17,Threlfall18}, or turbulence \citep{Hasegawa85,Laming07,Che14,bian14}, wave-particle interactions \citep{Vocks08,Vocks16}, density or temperature gradients \citep{Roussel-Dupre80a,Shoub83,Ljepojevic88}. More generally, they occur wherever the Knudsen number is larger than about 0.01 \citet{Scudder19}, a condition commonly expected in solar and stellar coronae \citep{Scudder13}.

The $\kappa$-distributions have an influence on the ionization equlibrium \citep{dzif92,Wannawichian03,dzifdudik13}, excitation rates \citep[e.g.,][]{dzif06,dzif08}, and affect other processes and quantities \citep[e.g.,][and references therein]{Marsch06,Lazar16,MeyerV17,deAvillez15,deAvillez18,livadiotis17}. Therefore, the $\kappa$-distributions alter relative intensities of spectral lines in an optically thin plasma. Details of the spectral synthesis using the original excitation cross-sections for iron ions can be found in \citet{dudik14}. Approximations of the excitation cross-sections for all astrophysically relevant ions, based on modifications of the rates as available in the CHIANTI v7.1 database \citep{Dere97, Landi13}, are contained in the KAPPA package \citep{dzif15}.

{ The first discovery of a $\kappa$-distributions was obtained from in-situ measurements of electron velocities in the Earth's magnetosphere \citep{Olbert68, Vasyliunas68} and later on in the solar wind \citep[e.g.,][]{maksimovic97, Nieves08, LeChat10,Martinovic16}}. Ever since, distributions of particles with suprathermal tails have been detected in various kinds of space plasmas \citep[see e.g., the review of][]{pierrard10}. A question thus arises whether the non-Maxwellians in solar wind indeed originate in the solar corona. The presence of $\kappa$-distributions of ions is manifested in broad profiles of emission lines formed e.g. in flare conditions \citep{jeffrey16,jeffrey17,polito18}, which authors fitted with $\kappa$ as low as 2. Comparable values of $\kappa$ were used by \citep{Dudik17a} to fit lines of \ion{Si}{4}, \ion{O}{4}, and \ion{S}{4} observed by the \textit{Interface Region Imaging Spectrograph} (IRIS) in the centre of the studied active region. Furthermore, \citet{dzif17} found that high-energy tails indeed significantly affect these transition region lines. Electron $\kappa$-distributions can be investigated in two ways. First, by the direct fitting of high-energy tails of HXR spectra \citet{kasparova09,oka13,oka15,oka18} and using ratios of line intensities. 

A method for diagnostics of $\kappa$ utilizing ratios of line intensities was developed by \citet{dzifkuli10}. It is based on comparing observed and theoretical line intensity ratios, one sensitive to $T$ and the other to $\kappa$, plotted on both axes of the ratio-ratio diagrams constructed for known densities. Since we use this method in this manuscript, it is further described in Section \ref{sec_kappadiag}.

Theoretical combinations of line ratios sensitive to $\kappa$ observed by the Extreme-ultraviolet and Imaging Spectrometer \citep[EIS;][]{culhane07} onboard the \textit{Hinode} satellite \citep{kosugi07} were investigated by \citet{dzifkuli10} and \citet{dudik14,dudik19}. In all cases, one line in a ratio sensitive to $\kappa$ is observed in the short-wavelength and the other in the long-wavelength channel. Measurements of the parameter $\kappa$ using the ratio-ratio diagrams in the solar corona were performed by \citet{mackovjak13} and \citet{Dudik15} using data observed by EIS. \citet{mackovjak13} attempted to diagnose the electron distributions using lines of oxygen and sulphur, but were unable to precisely measure the parameter $\kappa$ because some lines were weak, or unresolved blends were present, both issues leading to large uncertainties. \citet{Dudik15} used multiple ratios of Fe line intensities to find extremely non-Maxwellian distributions with $\kappa \leq$ 2 in a transient coronal loop. The authors also accounted for the multithermal effects in the observed plasma. The observed ratios were found to be reproduced best by synthetic ratios calculated for the \mbox{$\kappa$\,=\,2} distribution. Finally, \citet{dzif18} applied the ratio-ratio method to flare spectra observed by the Extreme-Ultraviolet Variabiity Experiment \citet[EVE][]{Woods12} onboard the \textit{Solar Dynamics Observatory}. The authors showed that plasma is strongly non-Maxwellian with $\kappa\,\leq$\,2 during the early and impulsive phases of the flare and thermalizes in the graudal phase.  

As we mentioned before, the ratio-ratio technique for diagnostics of $\kappa$ works for known electron densities. Therefore, the electron densities need to be determined before the diagnostics of $\kappa$ can be applied. \citet{mackovjak13} and \citet{Dudik15} did so by using measurements of density-sensitive line intensity ratios. These ratios are however also slightly sensitive to both $T$ and $\kappa$, and as a result, the authors were only able to constrain the range of possible densities, which for some structures were as large as 0.8 dex in log($N_\mathrm{e}$\,[cm$^{-3}$]). Since the measurements of $\kappa$ are dependent on the electron density, this increases the uncertainties in the diagnostics.

Furthermore, \citet{Dudik15} discussed the EIS calibration and its degradation as another possible source of uncertainties. Since the launch of \textit{Hinode} in 2006, there have been several studies quantifying the changes in the in-flight calibration, in particular the decrease of sensitivity of the long-wavelength channel of EIS compared to the short-wavelength one \citep[e.g.,][]{mariska13,delzanna13calib, warren14calib}. Two in-flight calibration routines were developed to revise the effective areas as well as account for the sensitivity decay with time \citep{delzanna13calib,warren14calib}. As either of these routines have strong effects on several diagnostic ratios, the uncertainties in the previous results are considerable. 

In this work, we present the measurements of plasma from datasets obtained soon after the launch of \textit{Hinode} taken near in time, to reduce problems with the degradation of sensitivity. Furthermore, the diagnostics of $\kappa$ is here coupled with accurate measurements of the electron density. This manuscript is organized as follows. In Section \ref{sec_data} we describe the observations and the data reduction. Section \ref{sec_methods} outlines the diagnostic methods used for measurements of plasma parameters. We employ an iterative method (Section \ref{sec_iterations}) that significantly decreases the uncertainties in the measurements of the electron density (\ref{sec_nediag}), coupled with diagnostics of DEM (\ref{sec_demdiag}), as well as temperature and $\kappa$ (\ref{sec_kappadiag}). The results are presented in Section \ref{sec_results}. In Section \ref{sec_discussion}, a discussion of the results is provided. Finally, our findings are summarized in Section \ref{sec_conclusions}. 

\begin{figure*}[!t]
  \centering    
    \includegraphics[width=6.50cm, clip,   viewport= 30 20 425 380]{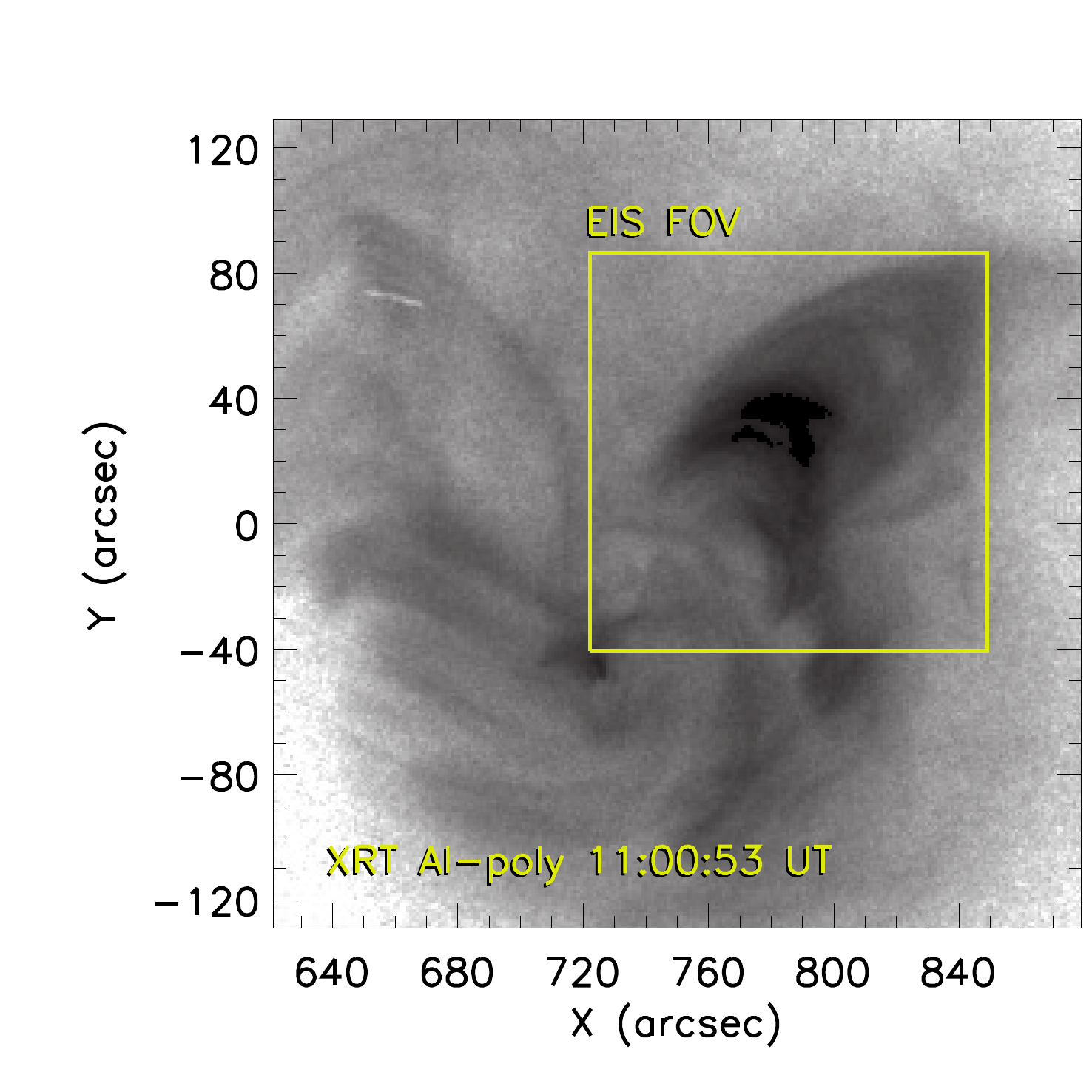}
    \includegraphics[width=5.265cm, clip,   viewport= 105 20 425 380]{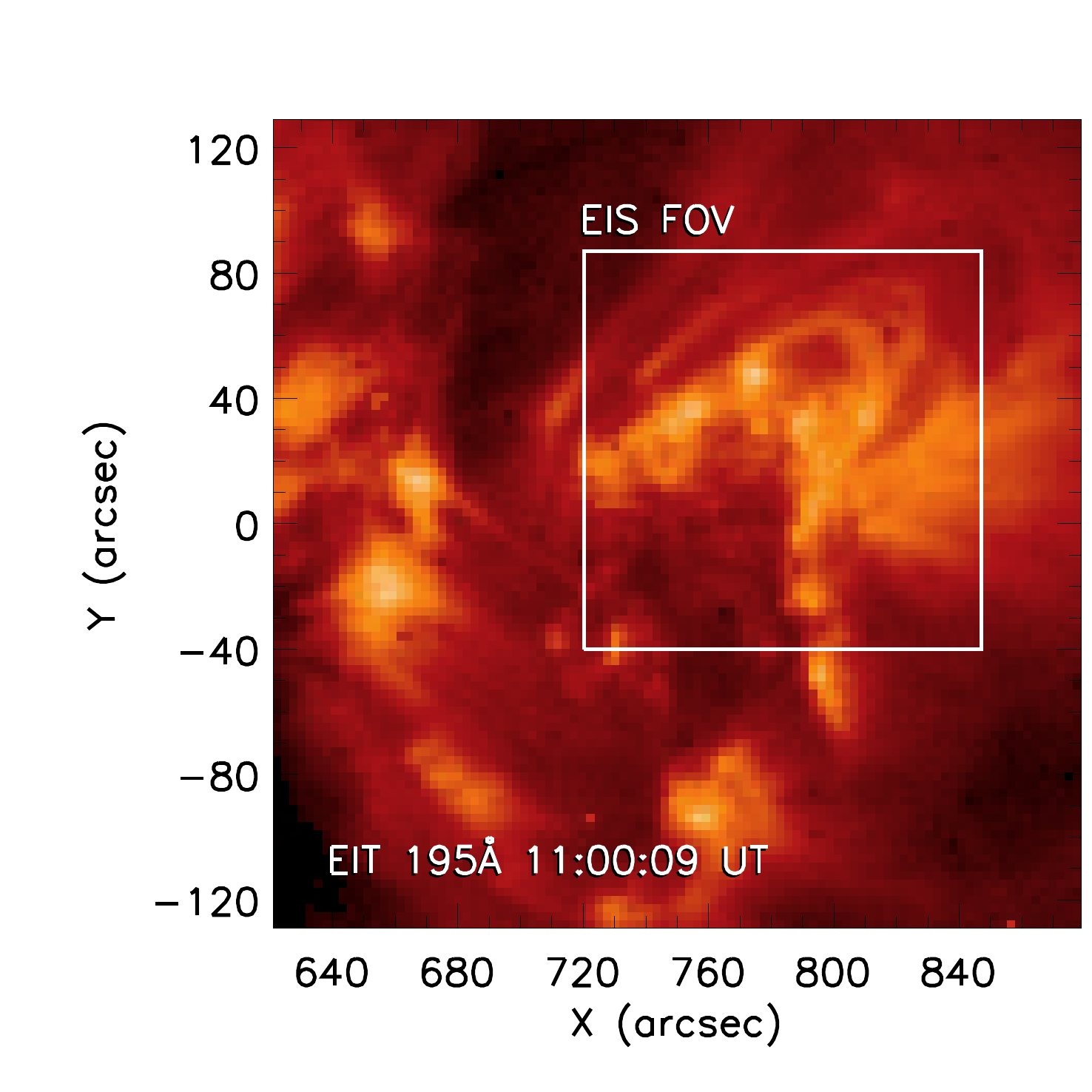}
    \includegraphics[width=5.265cm, clip,   viewport= 105 20 425 380]{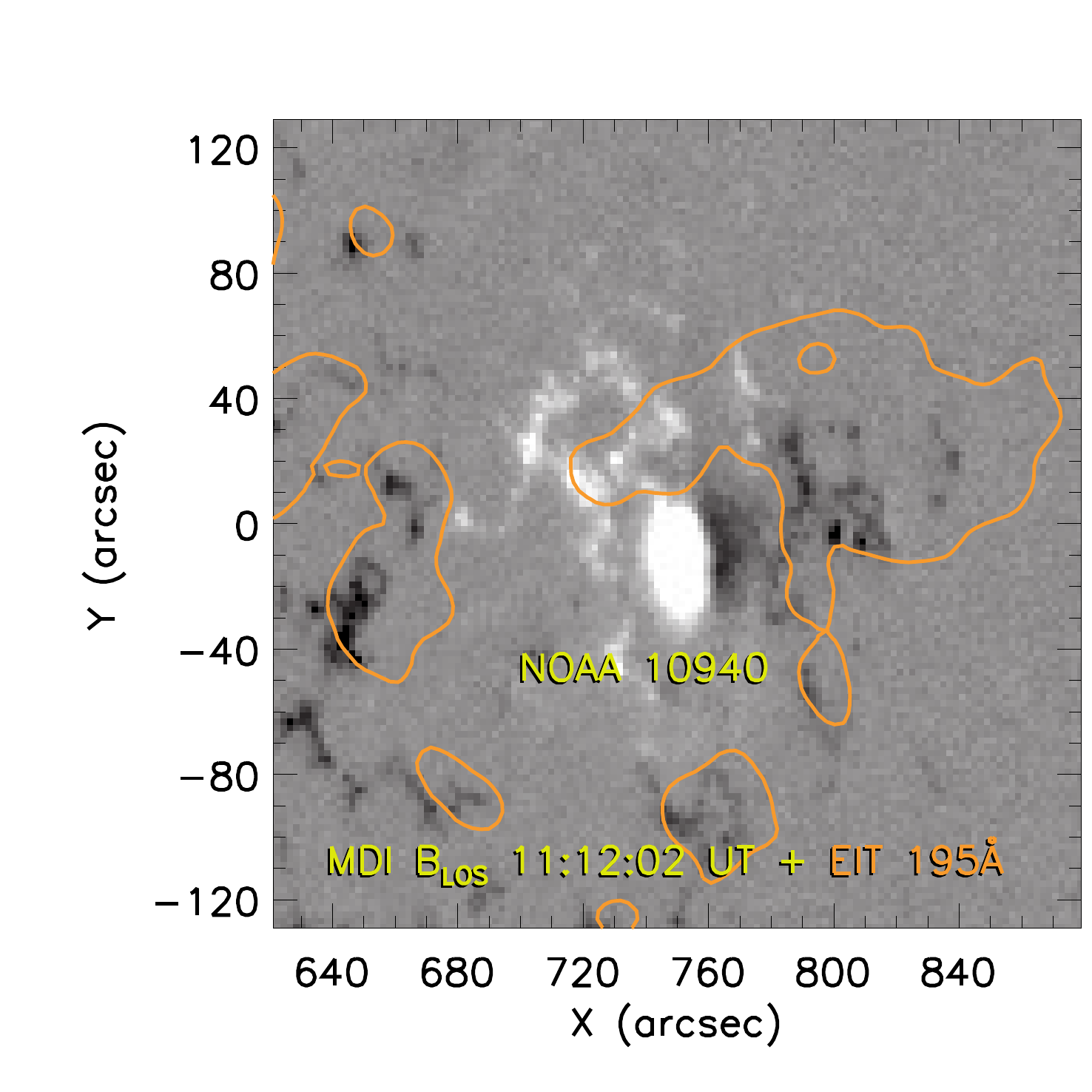}

  \caption{Context observations of the NOAA 10940 active region observed in the Al-poly filter channel of XRT ({\textit{Left}) and 195\AA~filter channel of EIT (\textit{middle}). FOV of EIS is indicated using yellow and white frames. \textit{Right} panel shows MDI magnetogram overlayed with contours corresponding to 5 DN s$^{-1}$ pix$^{-1}$ in the 195\AA~channel of EIT. \label{figure_context}}}
\end{figure*}
\begin{figure*}[!t]
  \centering
    \includegraphics[width=6.62cm, clip,  viewport= 15 40 340 330]{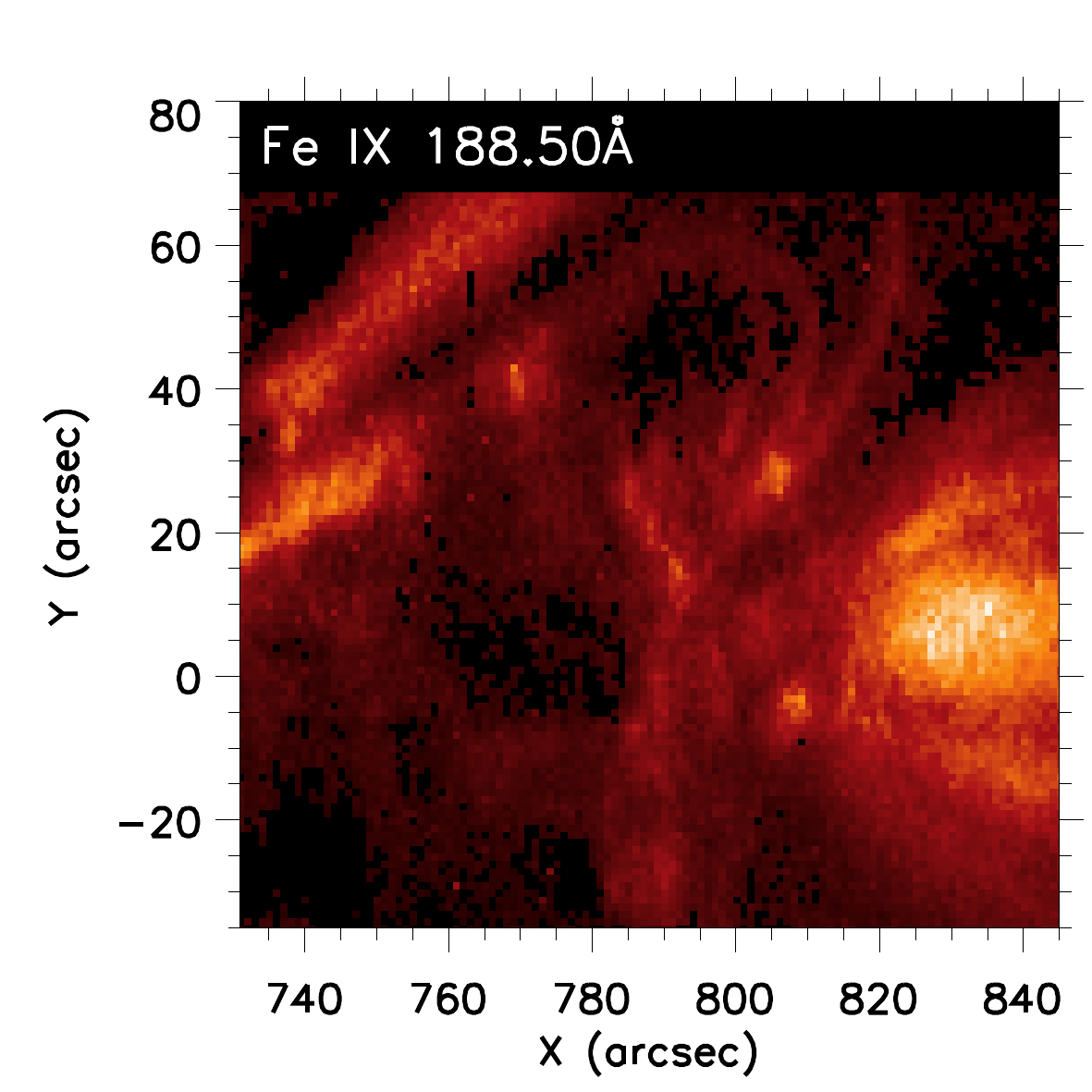}
    \includegraphics[width=5.6cm,  clip,  viewport= 65 40 340 330]{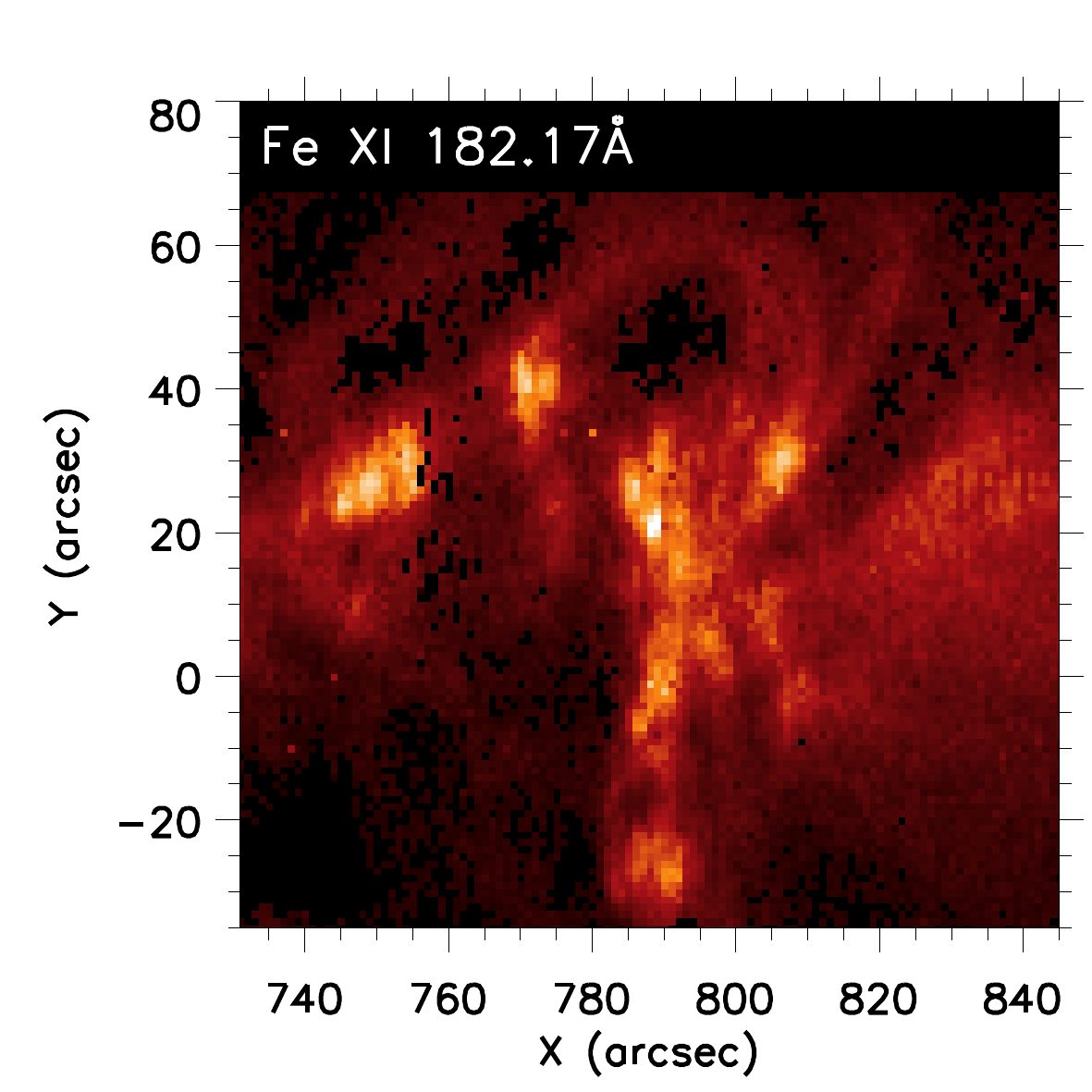}
    \includegraphics[width=5.6cm,  clip,  viewport= 65 40 340 330]{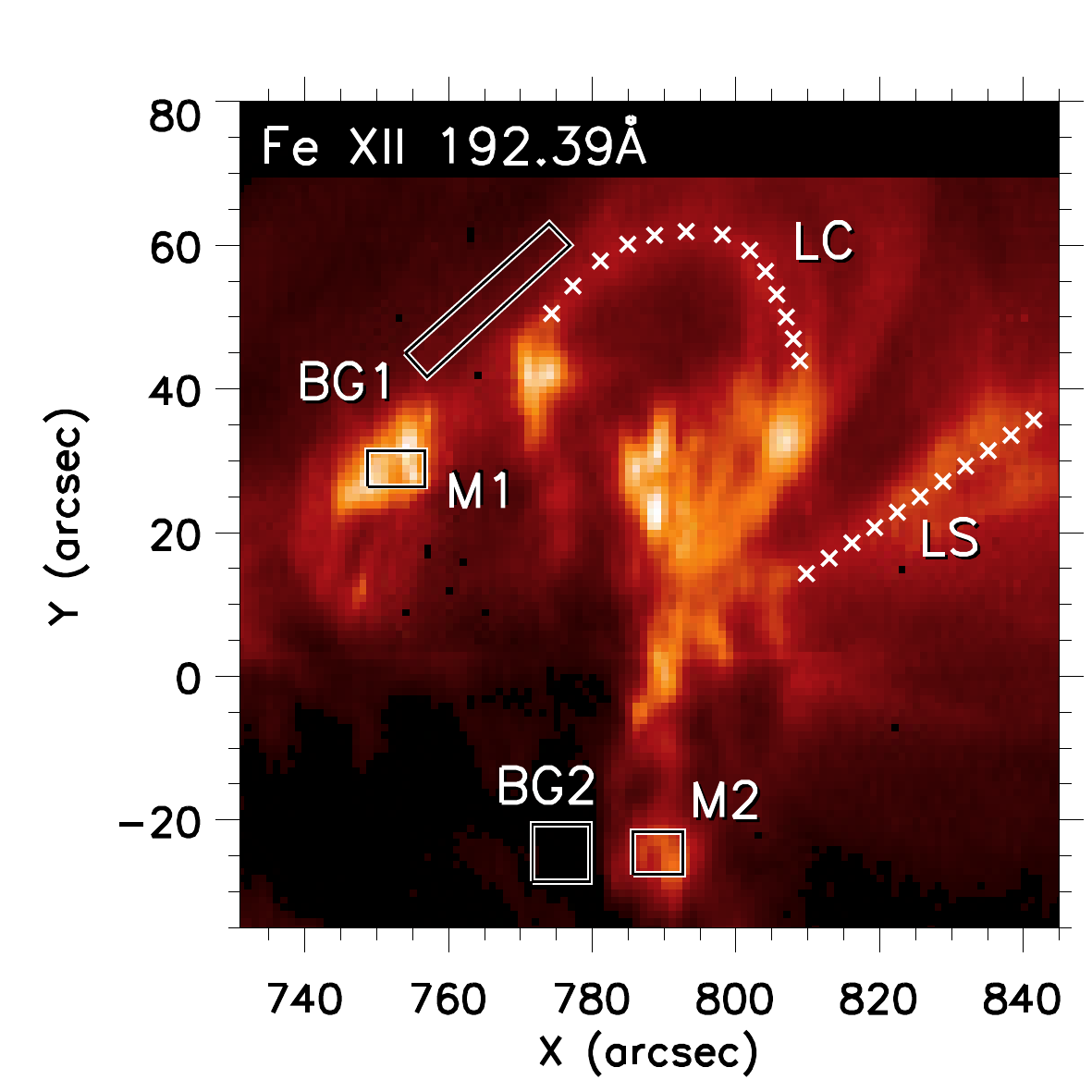}
    \\
    \includegraphics[width=6.62cm, clip,  viewport= 15 00 340 317]{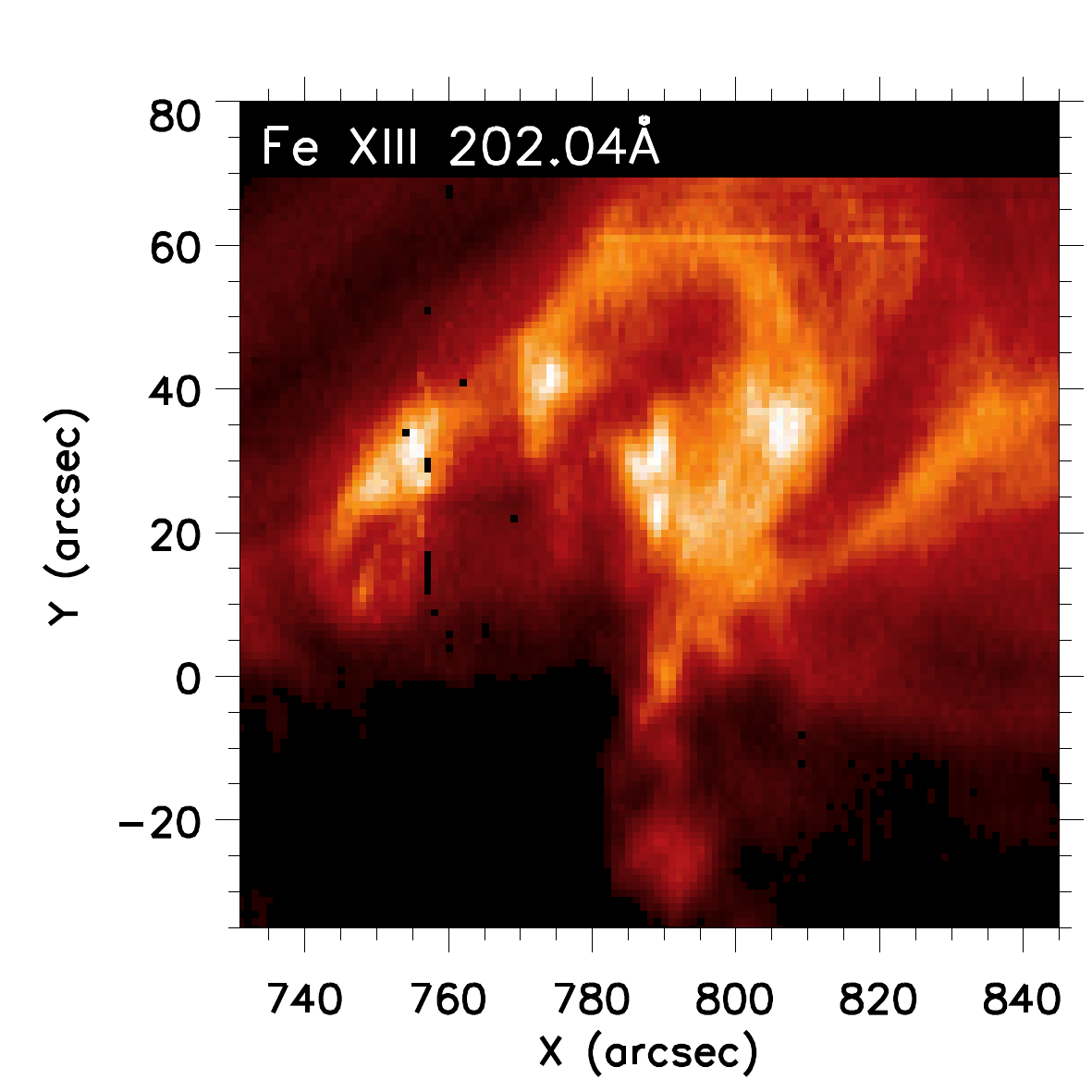}
    \includegraphics[width=5.6cm,  clip,  viewport= 65 0 340 317]{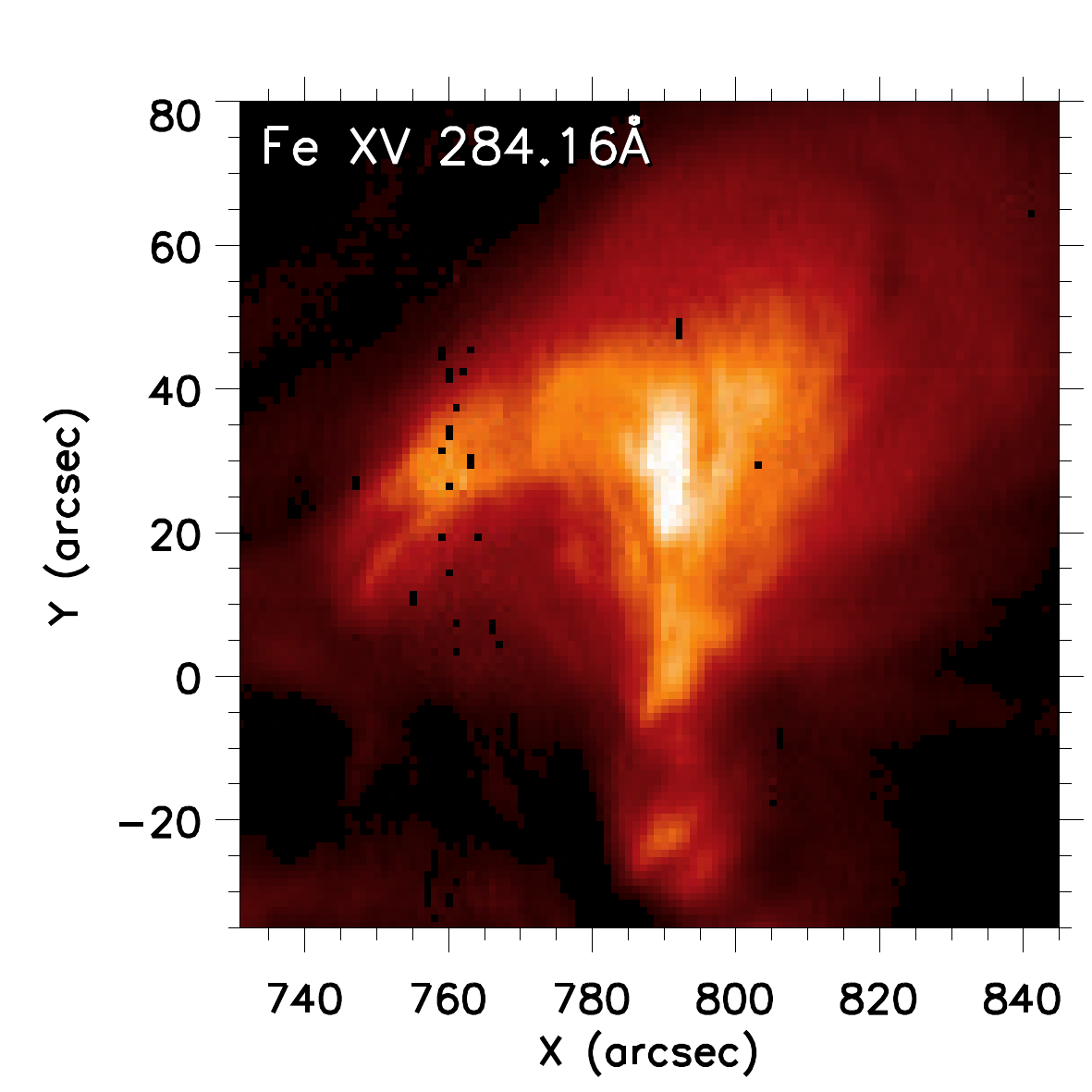}
    \includegraphics[width=5.6cm,  clip,  viewport= 65 0 340 317]{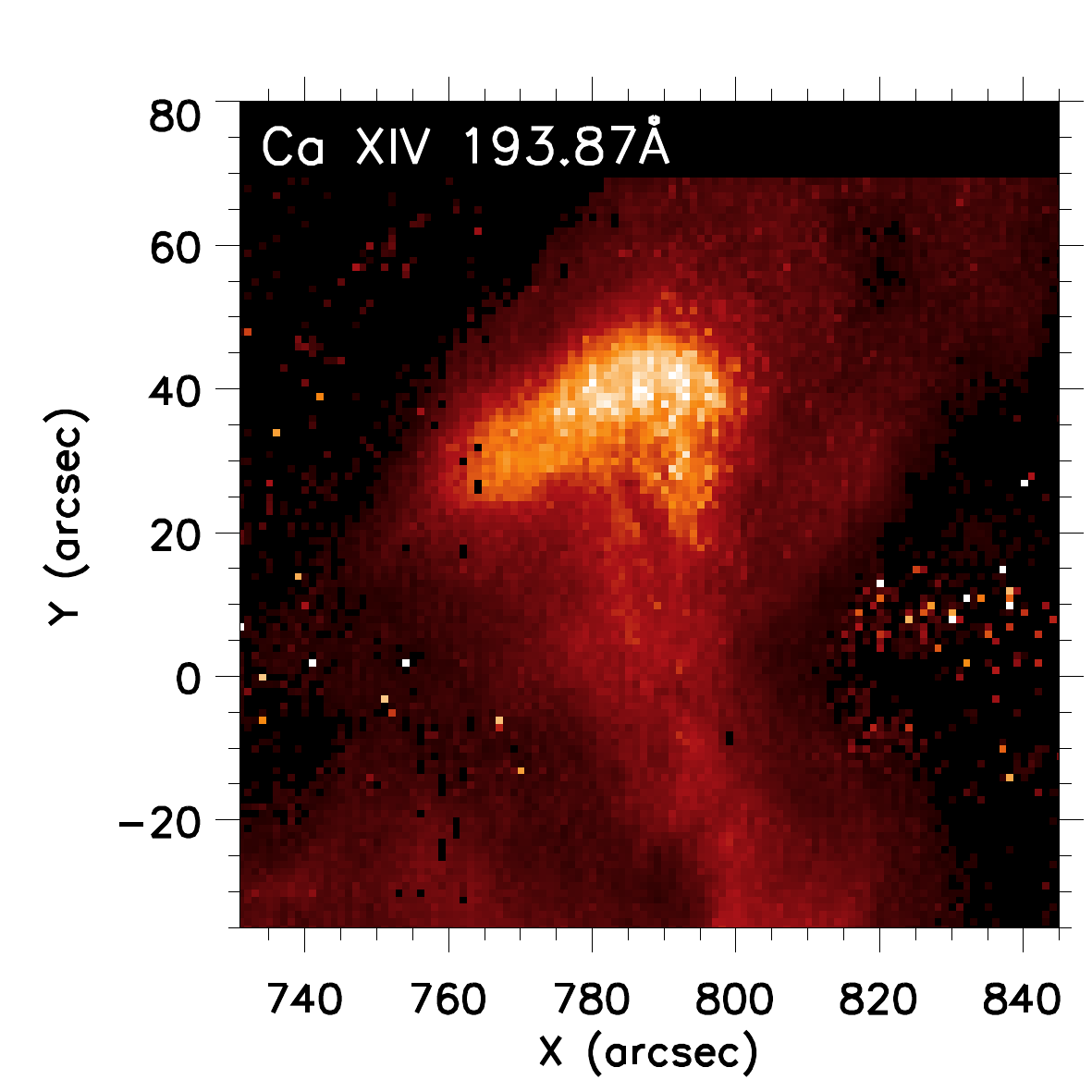}
  \caption{Context observations of the NOAA 10940 active region
    observed on 2007 February 5 in different spectral lines observed
    by EIS. Acronyms shown in the \ion{fe}{12} 192.39\AA~line mark
    structures in which we have averaged intensities and used for
    later diagnostics of the plasma. \label{figure_overview_AR}}
\end{figure*}
\section{\textit{Hinode}/EIS Observations \\of the Active Region and Quiet Sun} \label{sec_data}

To analyze plasma properties in different regions within the solar corona, we used data containing observations of an active region and quiet Sun observed by \textit{Hinode}/EIS. To carry out this study we use the spectral atlases, i.e. observations in which the whole spectral range of the instrument is observed. 

We have selected observations of the active region NOAA 10940 (hereinafter, AR) observed on 2007 February 5. This active region was rastered from 10:52:12 UT in 30\,s exposures using a 1$\arcsec$ slit. The spectral atlas containing observations of an off-limb quiet Sun (hereinafter, QS) was rastered on 2007 March 11. The observations were carried out using 1$\arcsec$ slit, 90\,s exposures, and started at 02:32:12 UT.

\begin{figure*}[!t]
  \centering
   
    \includegraphics[width=6.62cm, clip,  viewport= 15 40 340 330]{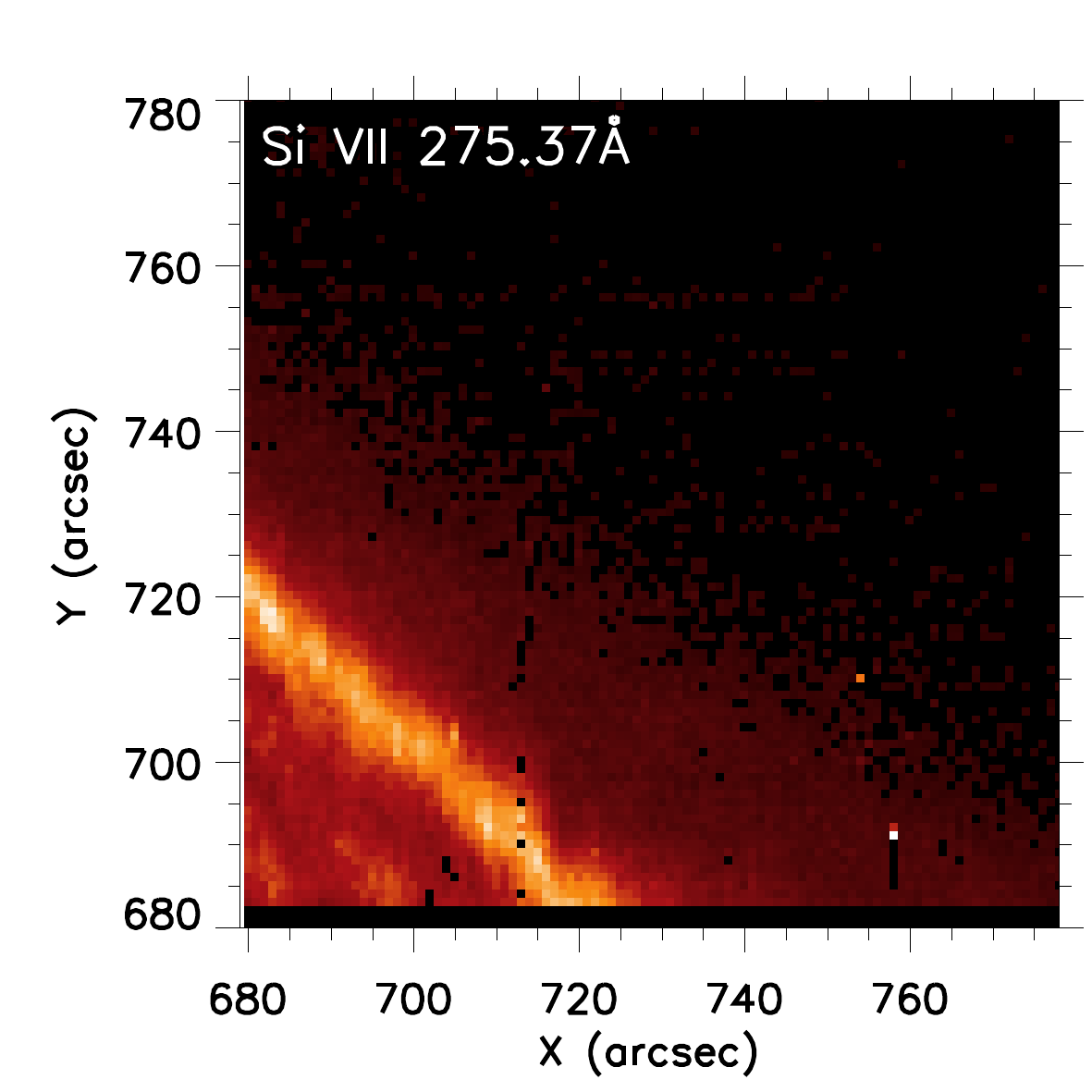}
    \includegraphics[width=5.6cm,  clip,  viewport= 65 40 340 330]{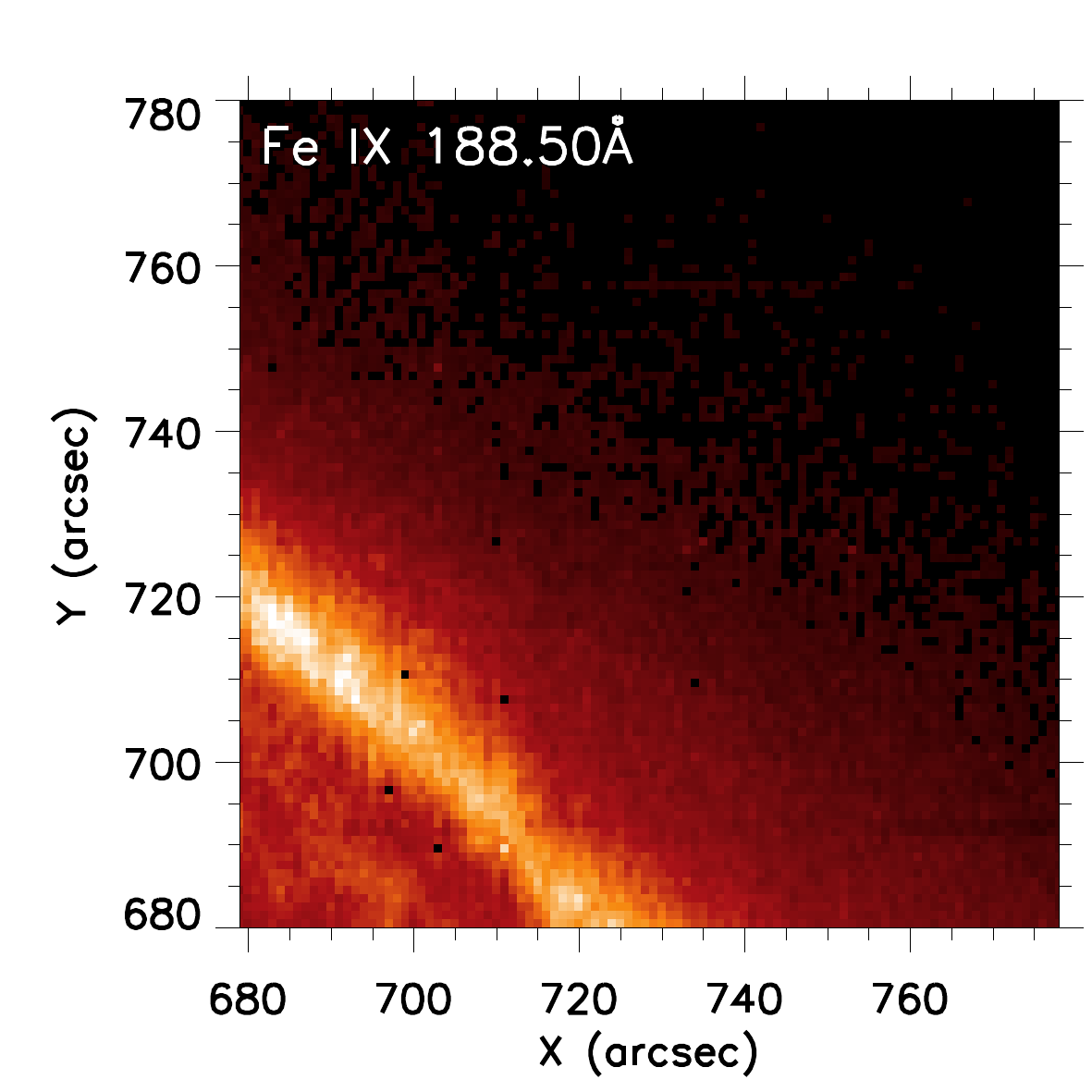}
    \includegraphics[width=5.6cm,  clip,  viewport= 65 40 340 330]{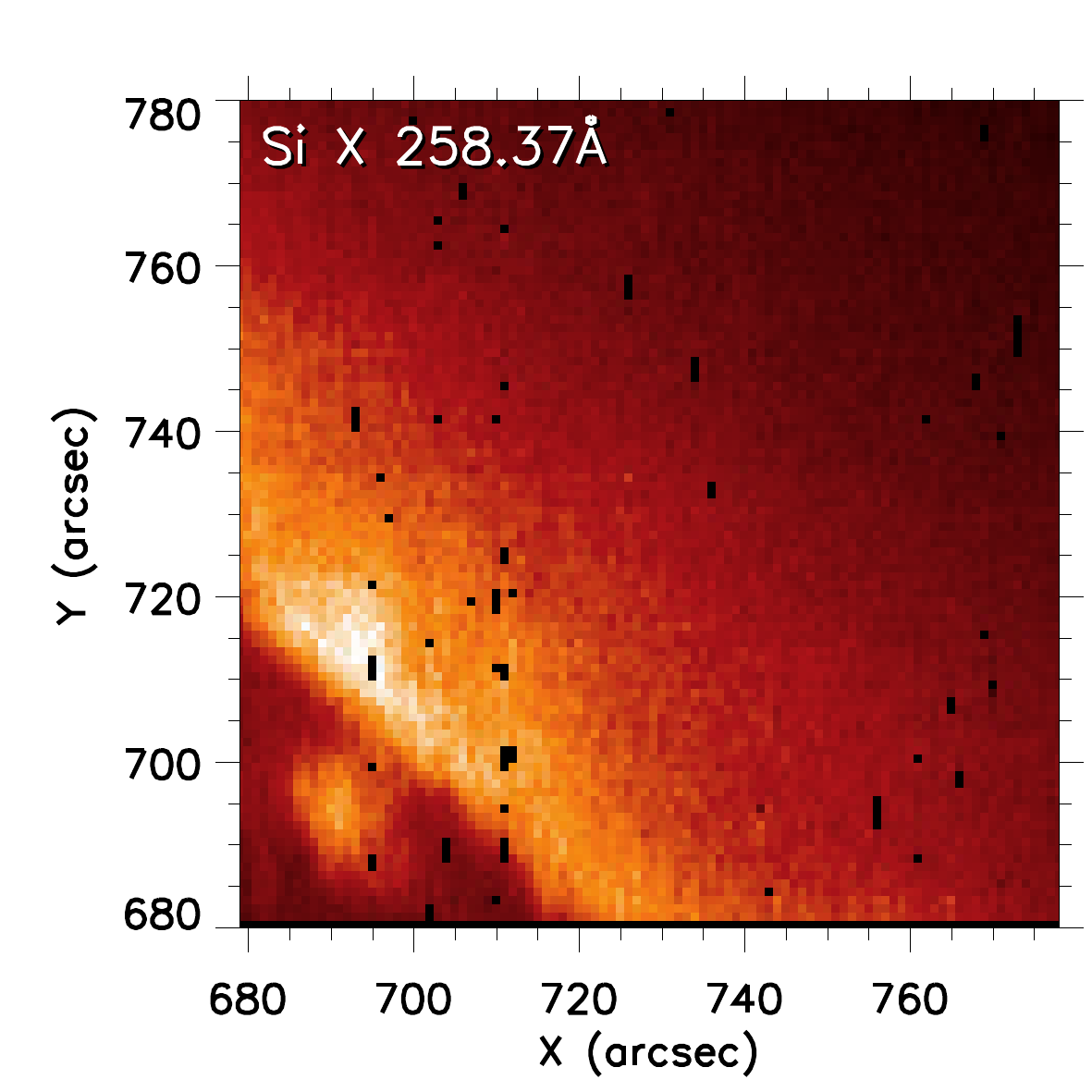}
    \\
    \includegraphics[width=6.62cm, clip,  viewport= 15 0 340 320]{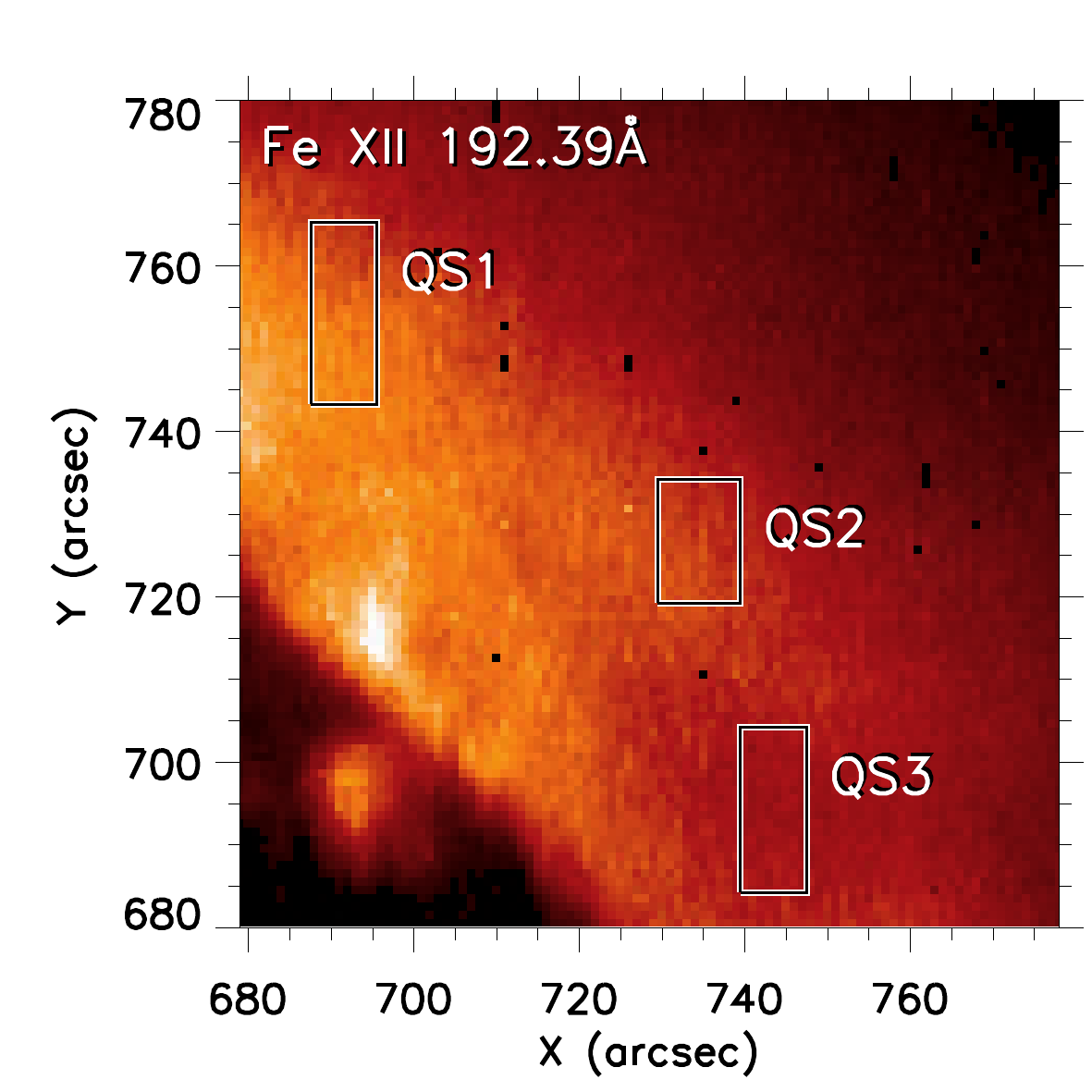}
    \includegraphics[width=5.6cm,  clip,  viewport= 65 0 340 320]{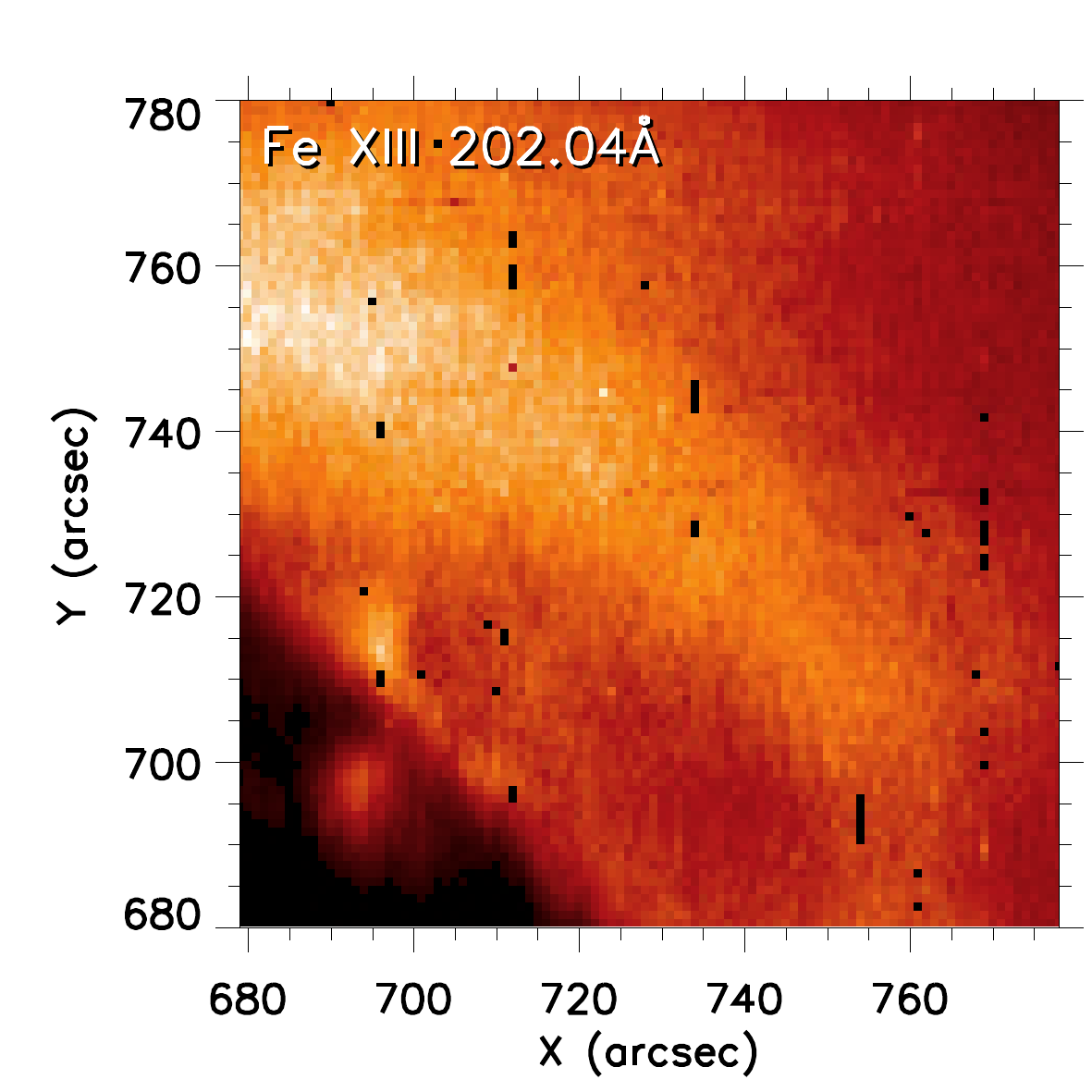}
    \includegraphics[width=5.6cm,  clip,  viewport= 65 0 340 320]{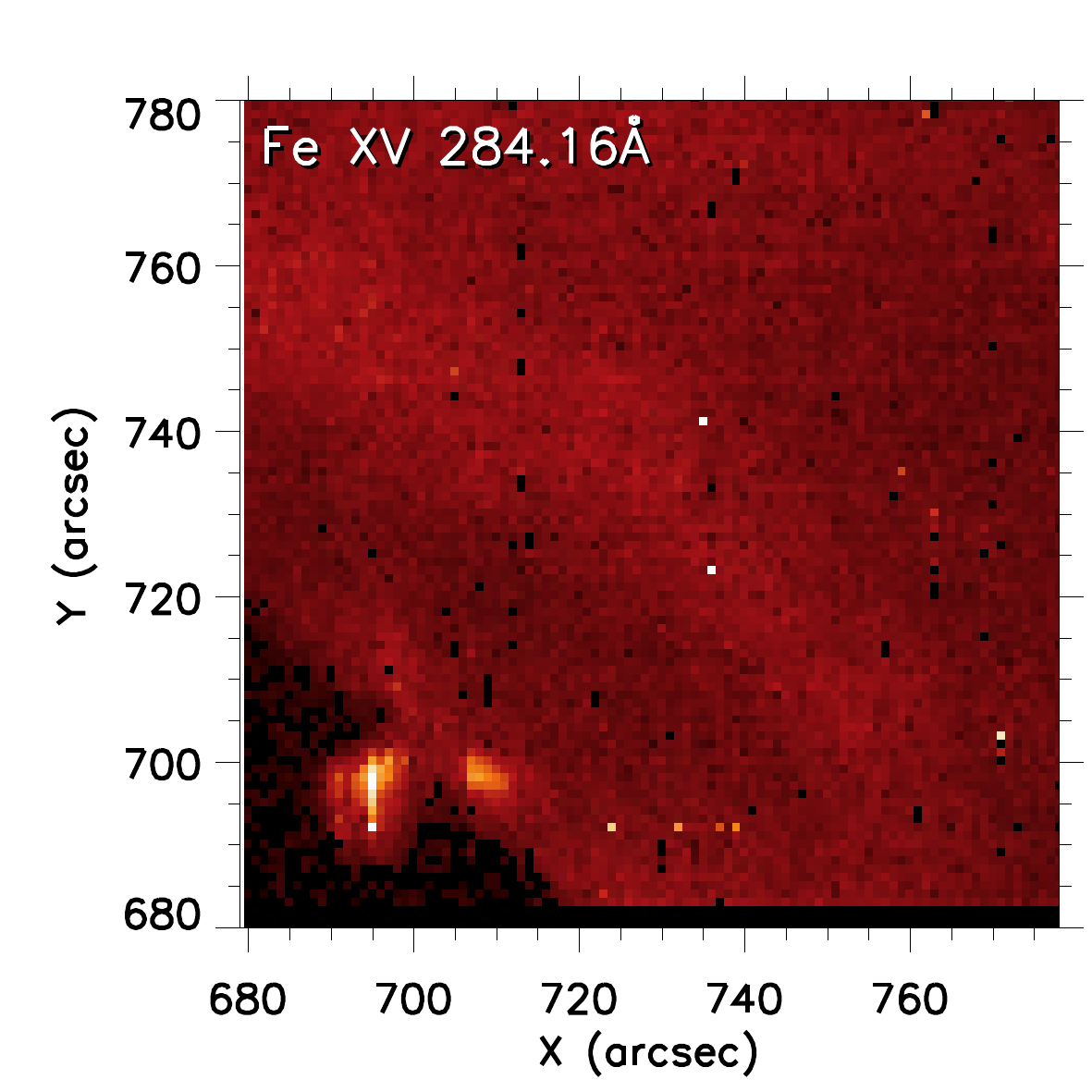}
  \caption{Context observations of the quiet Sun observed on 2007
    March in different spectral lines observed by EIS. Boxes QS1--QS3
    mark areas in which we averaged intensities which we used for later diagnostics of plasma.  \label{figure_overview_QS}}
\end{figure*}

The observations of the AR are shown in Figures \ref{figure_context} and \ref{figure_overview_AR}, while the QS observations are shown in Figure \ref{figure_overview_QS}. To have an overview of the temperature structure of the regions observed, Figures \ref{figure_context} and \ref{figure_overview_AR} show EIS spectral lines of ions formed at different temperatures. To obtain these images, the selected spectral lines were fitted in the whole EIS FOV using the automatic fitting routine \texttt{auto\_fit}. Where applicable, multi-Gaussian fits were used to fit the observed spectra.

Note that an extensive discussion on the temperature structure of the observed plasma is left to Sections \ref{sec_res_demdiag} and \ref{sec_res_kappadiag}. 
\subsection{Data reduction}

Both spectral atlases were processed in the same manner. Data were first converted into level-1 using the \texttt{eis\_prep} routine. The correction for spectrum rotation was then applied, with $Y$- offsets found using the standard \texttt{eis\_ccd\_offset} procedure. We then had to shift the data in the long-wavelength channel by 2 $\arcsec$ in $X$, because of their relative shift with respect to the short-wavelength channel data. We also found that the raster steps in solar $X$ are not equal to the slit width of 1$\arcsec$ \citep[see][]{delzanna11b}. However, since we perform diagnostics from EIS measurements only, these small inconsistencies, below the spatial resolution of EIS, were neglected. 

Both AR and QS datasets contained less than $1\%$ of missing pixels, which we excluded from the statistics. Most of them were visible at wavelengths of about 193\,\AA~and were located at certain positions along the slit, at approximately $Y$\,$\approx$\,690\arcsec in the QS, and $Y$\,$\approx$\,0$\arcsec$ in the AR data.

Since both data sets were observed relatively shortly after the launch of the instrument, we used the ground calibration \citep{culhane07} for the absolute calibration of intensities. The data were not corrected further for degradation. Any such degradation in the early 2007 would be only a few per cent \citep[see Figure 9 in][]{delzanna13calib}.


%
\subsection{AR Observations}

\subsubsection{Context observations}

Since EIS observed only a portion of the AR 10940, we first examined
the context observations provided by other instruments taken at the
same time as the EIS observations. The X-ray Telescope
\citep[XRT,][]{golub07} was imaging the AR  mainly in the Al-poly
channel. Data were processed using the standard routines
\texttt{xrt\_prep}, \texttt{xrt\_jitter}, and then manually co-aligned
with EIS using lines formed at temperatures corresponding to the AR
core. The XRT image, shown in the left panel of Figure
\ref{figure_context}, reveals the bright core of the active region, an arcade of the active region loops, as well as fainter loops rooted on the eastern and southern side of the FOV. The EIS FOV is indicated by the yellow box. 

We also examined imaging data produced by the instruments on board the
\textit{Solar and Heliospheric Observatory (SOHO)}. The Extreme
ultraviolet Imaging Telescope \citep[EIT;][]{delab95} observed this
active region in the 195\AA~filter channel at a cadence of 12
minutes. EIT data were then manually co-aligned with the \ion{Fe}{12}
195.12\AA~intensities observed by EIS and are shown in the middle
panel of Figure \ref{figure_context}. Upon inspection of the EIT data,
we did not find any significant changes in the morphology of the
active region during the period of the EIS rastering. In particular,
no major brightenings occurred within the AR, which is also supported
by the flat light curve of the \textit{GOES} X-ray flux during the
period studied (not shown).

The magnetic structure of the active region was examined using the $B_{\text{LOS}}$ data measured by the Michelson Doppler Imager \citep[MDI;][]{scherrer95} onboard \textit{SOHO}. The right panel of Figure \ref{figure_context} shows level 1.8 MDI data saturated to $\pm$ 1000 G, coaligned with EIT. To compare the morphology of the active region with the distribution of the underlying magnetic field, we overplotted the $B_{\text{LOS}}$ data with 195\,\AA~filter channel contours (orange) corresponding to 5 DN s$^{-1}$ pix$^{-1}$ produced using data smoothed by a 5 $\times$ 5 boxcar. 

\subsubsection{Observations of the Active region} \label{sec_obs_ar}

Figure \ref{figure_overview_AR} shows that in the lines of \ion{Fe}{11}--\ion{Fe}{13}, which are primarily used for diagnostics, as well as in other lines, the observed active region is structured. Relatively short active region loops are present together with long loops overlying the active region, as well as coronal moss located at the footpoints of the hot core emission. 

In the \ion{Fe}{9} 188.50\,\AA~and \ion{Fe}{11} 182.17\,\AA~(log $T_{\text{max}}\text{ [K]}=6.15$) line images, we see short and relatively-faint active region loops located at about [800$\arcsec$, 50$\arcsec$]. At $\approx$[830$\arcsec$, 0$\arcsec$], lower portions of long and bright coronal loops overlying the active region are visible. Conjugate footpoints of these loops are distributed in several patches, one of which can be seen at $\approx$[740$\arcsec$, 40$\arcsec$], while the other ones are outside of the EIS FOV. The coronal moss is located at the footpoints of hot core loops, for example, in a small bright region at $\approx$[740$\arcsec$, 20$\arcsec$] and in an elongated region located at $X$\,$\approx$ 790$\arcsec$. 

The structure of the active region is similar in the \ion{Fe}{12} 192.39\,\AA~and \ion{Fe}{13} 202.04\,\AA~line images. The short active region loops seen in \ion{Fe}{9} and \ion{Fe}{11} belong to a bright arcade, a portion of which is beyond the EIS FOV. One of these loops, hereinafter refer as to the 'curved loop' ('LC'), is in the \ion{Fe}{12} 192.39\,\AA~image highlighted with crosses. We selected this loop for further diagnostics. Note that this loop can be split into several strands in cooler lines. 

Furthermore, there is a fainter, broad bundle of long fan loops rooted on the western side of the active region. We selected one relatively bright loop for further diagnostics, and in the remainder of this paper, we will refer to it as to the 'straight loop' ('LS', see Figure \ref{figure_overview_AR}). 

Observations in hotter spectral lines, such as the \ion{Fe}{15} 284.16\AA~line (log($T_\mathrm{max}$\,[K])\,=\,6.35 for Maxwellian conditions) reveal typical thermal structure of the active region, with hot emission concentrated in the core. In the 'hottest line' shown in Figure \ref{figure_overview_AR}, the \ion{Ca}{14} 193.87\,\AA~line (log($T_\mathrm{max}$\,[K])\,=\,6.55), the core of the active region is similar to the XRT morphology (Figure \ref{figure_context}, left).

We also selected two regions of coronal moss. These regions are indicated with boxes 'M1' and 'M2' in the \ion{Fe}{12} 192.39\,\AA~image. In the \ion{Fe}{12} image, we also show the areas selected for background subtraction. The intensities in these areas were averaged and later subtracted from the intensities averaged in the respective structures. We emphasize that the subtraction of coronal background is crucial for any diagnostics, as the background intensity can add up to several tens of percent of the intensity to the observed coronal structure \citep[e.g.,][]{delzanna03}. When choosing the backgrounds, we were trying to select areas which a) contain as many pixels as possible in order to minimize the uncertainties of the observed intensities and b) be spatially close to the respective structure.  

Background for LC and M1 was selected in the upper-left part of the EIS FOV (inclined box 'BG1'), in a narrow dark region close to both LC and M1. The background BG2 for the moss M2 was chosen close by, in a region devoid of any emission in almost all of the spectral lines used in this work. Unfortunately, no background matching our criteria was found in the vicinity of LS. This is due to the fan loops spanning a large area. Although there is a relatively dark region located to the north of LS in \ion{Fe}{12}; cooler loops overlying the active regions can be seen in \ion{Fe}{9}, as well as hotter emission in \ion{Fe}{15}. For these reasons, we use the background BG1 for LS, as no other appropriate choice can be made. 

In the remainder of this paper we will use background-subtracted intensities only. To the structures which intensities have been averaged and background intensities subtracted we will further refer as to 'M1s', 'M2s', 'LCs', and 'LSs'. 

\subsection{Observations of the quiet Sun} \label{sec_obs_qs}

The observed off-limb Quiet Sun area together with a portion of the solar disk are shown in Figure \ref{figure_overview_QS}. Limb brightening is seen in cool lines such as \ion{Si}{7} 257.37\,\AA~and \ion{Fe}{9} 188.5\,\AA~(log$T_\mathrm{max}$ [K])\,$\approx$\,5.8 for Maxwellian conditions). There is also a faint, arc-like, off-limb structure spanning the EIS FOV. It is most evident in the \ion{Fe}{13} 202.04\,\AA~, and traces of it can also be seen in \ion{Fe}{12} and \ion{Fe}{15}. For purposes of plasma diagnostics, we selected three boxes QS1--QS3, which are shown in the bottom-left panel of Figure \ref{figure_overview_QS}. Since the QS contains diffuse emission, no background subtraction was performed. Note that there is also emission originating from the disk, in the form of a bright point present at coordinates of about [700$\arcsec$, 700$\arcsec$], seen in lines of ions with log($T_\mathrm{max}$ [K])\,$\geq$\,6.15. 

Finally, we note that the same observation was used by \citet{delzanna12} when producing an atlas of coronal lines. Intensities averaged in area which corresponds to our box QS1 can be found in Table A.1. therein. 
\section{Diagnostic Method}  \label{sec_methods}

The properties of the observed optically thin coronal plasma are diagnosed using standard techniques based on comparisons of the observed line intensities with synthetic ones. Diagnostics of the electron density $N_\mathrm{e}$ (Section \ref{sec_nediag}), temperature $T$, and the $\kappa$ parameter (Section \ref{sec_kappadiag}) are based on the line ratio technique, while the multithermal nature of the plasma is quantified using the differential emission measure (see Section \ref{sec_demdiag}).

The intensities of the spectral lines of diagnostic interest were obtained via line fitting, which included the subtraction of the neighboring continuum. Details of the fitting procedure are given in Appendix \ref{sec_appendix} and the intensities of lines used in this work are listed in Table \ref{tab_fit_int_ar}. The method for calculation of synthetic spectra is described in the following section.

\subsection{Synthetic spectra}
\label{Sect:3.1}

\subsubsection{Synthetic line intensities}
\label{Sect:3.1.1}

The synthetic spectra are calculated here in the optically thin and coronal approximations. In optically thin conditions, the line intensity $I_{ji}$ arising from a transition $j \to i$ between energy levels $j > i$ is given by the integral of the emissivity $\varepsilon_{ji}$ along the line of sight $l$ \citep[cf.,][]{Mason94,Phillips08}:
\begin{equation}
	I_{ji} 	  = \int \varepsilon_{ji} (T,N_\mathrm{e},\kappa) \mathrm{d}l = \int A_X G_{X,ji}(T,N_\mathrm{e},\kappa) N_\mathrm{e} N_\mathrm{H} \mathrm{d}l\,,
	\label{Eq:line_intensity}
\end{equation}
where the $\varepsilon_{ji}$ is given by the product of the relative abundance $A_X$ of the element $X$, the factor $N_\mathrm{e} N_\mathrm{H}$\,$\approx$\,0.83$N^2_\mathrm{e}$, and the contribution function $G_{X,ji}$
\begin{equation}
	G_{X,ji}(T,N_\mathrm{e},\kappa) = \frac{hc}{\lambda_{ji}} \frac{A_{ji}}{N_\mathrm{e}} \frac{N(X_j^{+k})}{N(X^{+k})} \frac{N(X^{+k})}{N(X)}\,.
	\label{Eq:G(T)}
\end{equation}
There, the $\lambda_{ji}$ represents the wavelength of the emission line, $hc/\lambda_{ji}$ is the photon energy, and $A_{ji}$ is the Einstein coefficient for the spontaneous emission. The fractions $N(X_j^{+k}) / N(X^{+k})$ and $N(X^{+k})/N(X)$ represent the fractions of the ion $X^{+k}$ with the electron on the upper excited level $j$ and the relative ion abundance of the ion $X^{+k}$, respectively. In the coronal approximation, these fractions can be calculated separately, as the ionization and recombination processes occur dominantly from and to the ground level. This means that the ionization and recombination processes do not influence the relative level populations $N(X_j^{+k}) / N(X^{+k})$ of the ion $X^{+k}$.

In the optically thin solar corona, the observed emission along the line of sight can originate at many different plasma temperatures. In such a case, the Equation (\ref{Eq:line_intensity}) for line intensity is customarily rewritten to
\begin{equation}
	I_{ji} 	  = \int A_X G_{X,ji}(T,N_\mathrm{e},\kappa) \mathrm{DEM}_{\kappa}(T) \mathrm{d}T\,,
	\label{Eq:line_intensity_DEM}
\end{equation}
where the quantity DEM$_{\kappa}(T)$\,=\,$N_\mathrm{e} N_\mathrm{H} \mathrm{d}l / \mathrm{d}T$ is the differential emission measure.

This definition assumes that there is a single-valued function, i.e. there is a particular distribution of plasma along the line of sight for which a DEM can be defined. We note that the assumption of optically thin plasma may not be valid everywhere in the solar corona. In particular, the well-known, bright \ion{Fe}{12} 195.12\,\AA~self-blend can be partially optically thick in active region conditions \citep{DelZanna19}. Throughout this work, we use the \ion{Fe}{12} 192.39\,\AA~line instead. This line originates from the same 3s$^2$\,3p$^2$\,3d $^4 P$ system \citep[see Table B.4 in ][]{dudik14}, meaning that its intensity with respect to the 195.12\,\AA~selfblend is almost independent of plasma conditions, namely $N_\mathrm{e}$ and $\kappa$.

\subsubsection{Atomic data}
\label{Sect:3.1.2}

The atomic data used for spectral synthesis described in Section \ref{Sect:3.1.1} are from the latest version 9 of the  CHIANTI database \citep{Dere97,Dere19}. The ionization equilibrium for the non-Maxwellian $\kappa$-distributions is obtained using the method of \citet{dzifdudik13}.

For the iron ions of importance for the diagnostics of $\kappa$ in this work, we directly use the excitation cross-sections from \citet[][\ion{Fe}{8}]{DelZanna14a}, \citet[][\ion{Fe}{9}]{DelZanna14b}, \citet[][\ion{Fe}{10}]{DelZanna12e}, \citet[][\ion{Fe}{11}]{DelZanna13b}, except levels 37, 39, and 41, for which the data of \citet{delzanna10} are used; \citet[][\ion{Fe}{12}]{DelZanna12a}, \citet[][\ion{Fe}{13}]{DelZanna12b}, \citet[][]{Liang10a} and \citet{Landi12} for \ion{Fe}{14}, \citet[][\ion{Fe}{15}]{Berrington05}, \citet[][\ion{Fe}{16}]{Liang09}, and \citet[][\ion{Fe}{17}]{Liang10b}. These ions are the most important ones for the DEM diagnostics (Section \ref{sec_demdiag}). Finally, for the \ion{Si}{10}, which is used for density diagnostics, we use the cross-sections calculated by \citet{Liang12}. To obtain the corresponding excitation and de-excitation rates using these cross-sections, the cross-sections are integrated directly over the $\kappa$-distributions \citep{dudik14}.

\subsection{Iterative diagnostic procedure}  \label{sec_iterations}

The line intensities are functions of three individual parameters, the electron density $N_\mathrm{e}$, temperature $T$, and $\kappa$, the diagnostics of which is our objective. Of these, the $T$ and $\kappa$ are parameters of the distribution (Equation \ref{Eq:kappa}), and must be diagnosed simultaneously. The line ratios sensitive to $\kappa$ are also a function of electron density \citep{dzifkuli10,mackovjak13,dudik14,Dudik15,dudik19,dzif18}. Therefore, the electron density $N_\mathrm{e}$ is diagnosed prior to diagnostics of $T$ and $\kappa$. However, the theoretical curves for the diagnostics of the $N_\mathrm{e}$ from density-sensitive line intensity ratios also depend slightly on $T$ and $\kappa$ \citep[see, e.g., Figures 4--7 in][]{dudik14}. At the same time, for the diagnostics of $T$ and $\kappa$, precise measurements of the electron density irrespective of $T$ and $\kappa$ are needed. In the work of \citet{Dudik15}, only constraints on the electron density were derived. To improve upon this situation, we use a simple {iterative-like approach, where the diagnosed quantities are refined in multiple steps}.

{In accordance with approach of \citet{mackovjak13} and \citet{Dudik15}, we first obtain, in each structure, constraints on the electron densities}, providing us with ranges of possible densities. This is done using the density-sensitive line intensity ratios, without any assumptions on the temperature structure of the emitting plasma, or the value of $\kappa$ {\citep[see][]{dzifkuli10}}. Details on deriving these density ranges are provided in Section \ref{sec_nediag}.
These ranges can be large, up to 0.8 dex \citep[e.g.,][]{Dudik15}, and can be narrowed only if the thermal structure of the emitting region is accounted for. We achieve this in conjunction with the DEM$_\kappa(T)$ as follows. For the range of possible densities, a grid of DEM$_\kappa(T)$ are reconstructed for all values of $\kappa$ and $N_\mathrm{e}$. 
Then, the density-sensitive ratio curves are weighted using the DEM$_\kappa(T)$ obtained, thus removing the spread due to the unknown $T$. These DEM-weighted density-sensitive ratios are plotted for two extreme values of $\kappa$\,=\,2 and a Maxwellian (Figure \ref{figure_dem_densdiag}). This leads to an initial estimate on density, $N_\mathrm{e,0}$. 

The $N_\mathrm{e,0}$ is dependent on $\kappa$ only slightly, with the DEM-weighted ratios for $\kappa$\,=\,2 yielding densities about 0.1--0.2 dex lower than the corresponding Maxwellian curves. This behavior of the density-sensitive coronal lines ratios with $\kappa$ is well-known \citep[e.g.,][]{dzifkuli10,dudik14,Dudik15,dzif18} and occurs only for low $\kappa$\,$\to$\,2 values. The resulting uncertainty in density is small enough to permit an initial estimate of the $\kappa$ value using the ratio-ratio technique. Since the DEM-weighted density-sensitive ratios do not change appreciably for $\kappa$\,$\gtrsim$\,3, we can restrict the initial estimate of $\kappa$ to two extreme ranges, $\kappa$\,$\in$\,(3, $\infty$) and $\kappa$\,$\lesssim$\,2.

The next iteration consists of repeating the diagnostics of density. If the initial estimate yielded $\kappa$\,$\gtrsim$\,3, the Maxwellian DEM-weighted density-sensitive ratios are used to obtain $N_\mathrm{e}$. Conversely, if the initial estimate of $\kappa$ yielded $\kappa$\,$\lesssim$\,2, the DEM-weighted density-sensitive ratios for $\kappa$\,=\,2 are used. The resulting densities are then used again to plot the ratio-ratio diagram and obtain the next iteration of $\kappa$.

In principle, this procedure could be repeated until converging values of $N_\mathrm{e}$, DEM$_\kappa(T)$, and $\kappa$ are found. In practice, \textit{(i)} the small difference of the DEM-weighted density-sensitive ratios for the Maxwellian and $\kappa$\,=\,2, together with \textit{(ii)} the insensitivity of DEM to $N_\mathrm{e}$, \textit{(iii)} calculation of the spectra for only the integer values of $\kappa$ being used \citep{dzif15}, and \textit{(iv)} a rather large photon-noise uncertainty of the measured ratios sensitive to $\kappa$ mean that two iterations, as described above, are sufficient for this diagnostics to converge.

\subsubsection{Diagnostics of electron density}
\label{sec_nediag}

The electron density $N_\mathrm{e}$ is diagnosed by comparing the observed and theoretical ratios of line intensities. This method is well-known and has been utilized in multiple studies, using ions of Fe \citep[e.g.,][]{young09, watanabe09, dudik14, Dudik15, polito17, mulay17a}, or other elements \citep[e.g.,][]{mackovjak13, mulay17b}. EIS observes numerous strong spectral lines which can be used for measurements of density. Lines typically used are of \ion{Fe}{11} \citep{delzanna10}, \ion{Fe}{12} lines such as the 186.89\AA~and 195.12\AA~lines \citep{delzanna12}, or lines of \ion{Fe}{13}, such as the 196.53\AA~, 202.04\AA~, or 203.8\AA~line \citep[e.g.,][]{young07, young09, watanabe09, delzanna11}. 

Because the densities measured from different ratios might differ \citep[e.g.,][]{Dudik15}, here we combine results from four different line ratios of three different ions, being the \ion{Si}{10} 258.37\,\AA\,/\,261.06\,\AA, \ion{Fe}{12} 186.89\AA\,/\,192.39\AA, \ion{Fe}{13} 196.53\,\AA\,/\,202.04\,\AA, and \ion{Fe}{13} 203.83\,\AA\,/\,202.04\,\AA~line ratios. Note that the sensitivity of the \ion{Si}{10} ratio is weak for log($N_\mathrm{e}$\,[cm$^{-3}$]) $ > 9.5$. On the other hand, this ratio permits measurements of density below log($N_\mathrm{e}$\,[cm$^{-3}$]) $< 8$ which is important for constraining density in the quiet Sun. In the \ion{Fe}{12} ratio, we opted to use the 186.89\,\AA~line with the 192.39\,\AA~line instead of the 195.12\,\AA~one. The \ion{Fe}{12} 186.89\,\AA\,/\,192.39\,\AA~ratio is sufficiently density sensitive in the range of log($N_\mathrm{e}$\,[cm$^{-3}$]) $\approx$ 7.5--11. 

\begin{figure*}[h]
  \centering
    
    \includegraphics[width=0.41\textwidth, clip,  viewport= 0 0 505 370]{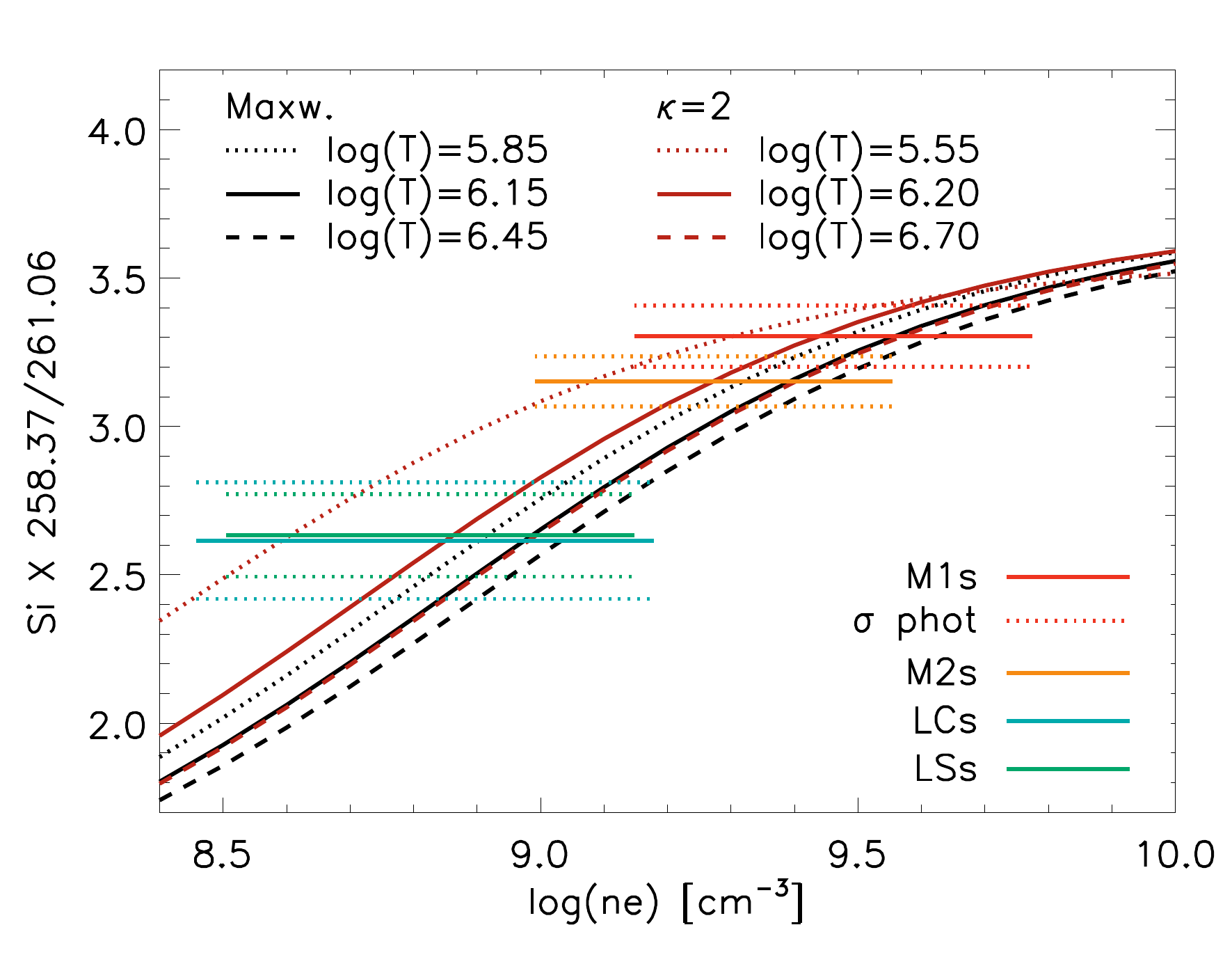}
    \includegraphics[width=0.41\textwidth, clip,  viewport= 0 0 505 370]{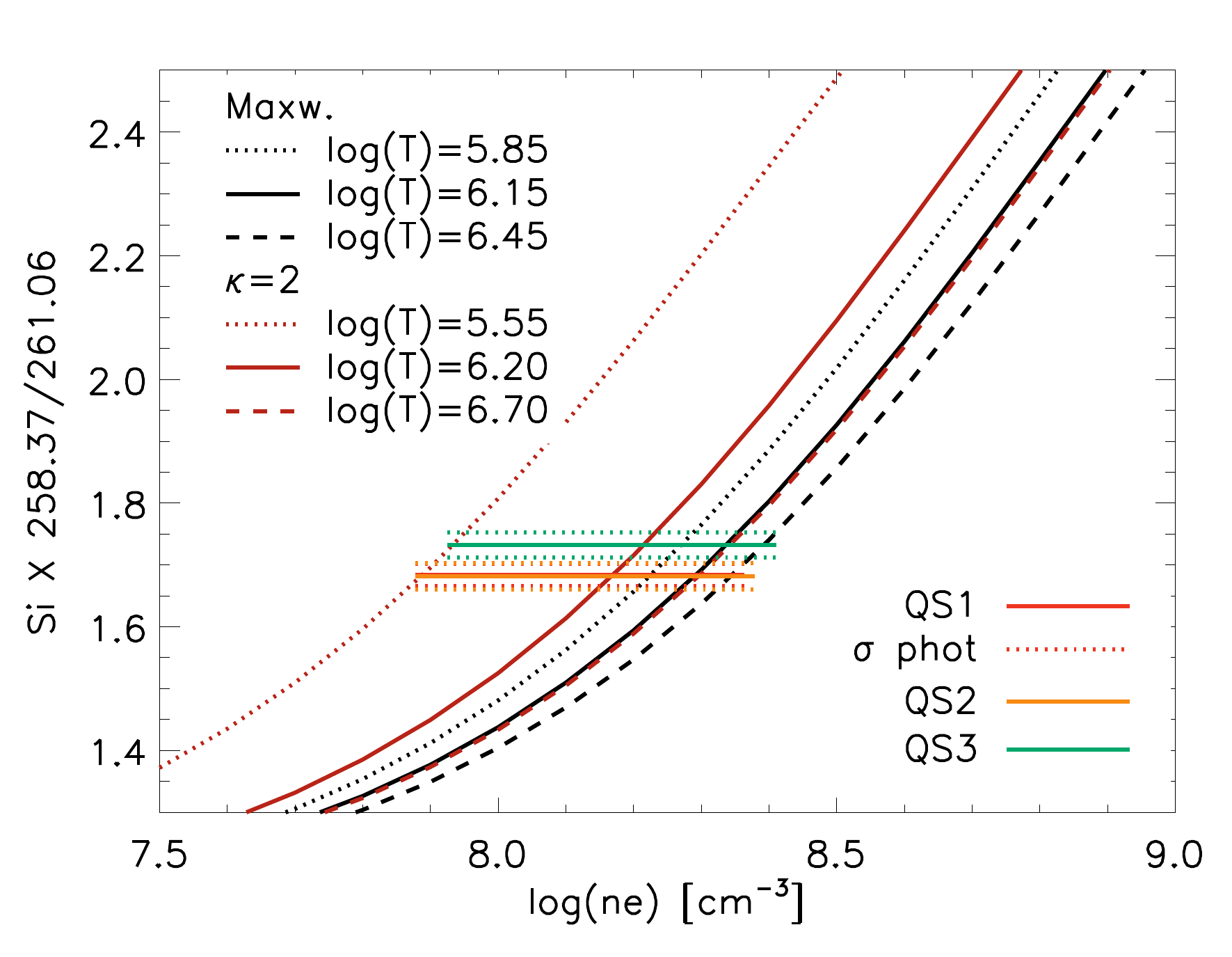}
    \\
    \includegraphics[width=0.41\textwidth, clip,  viewport= 0 0 505 370]{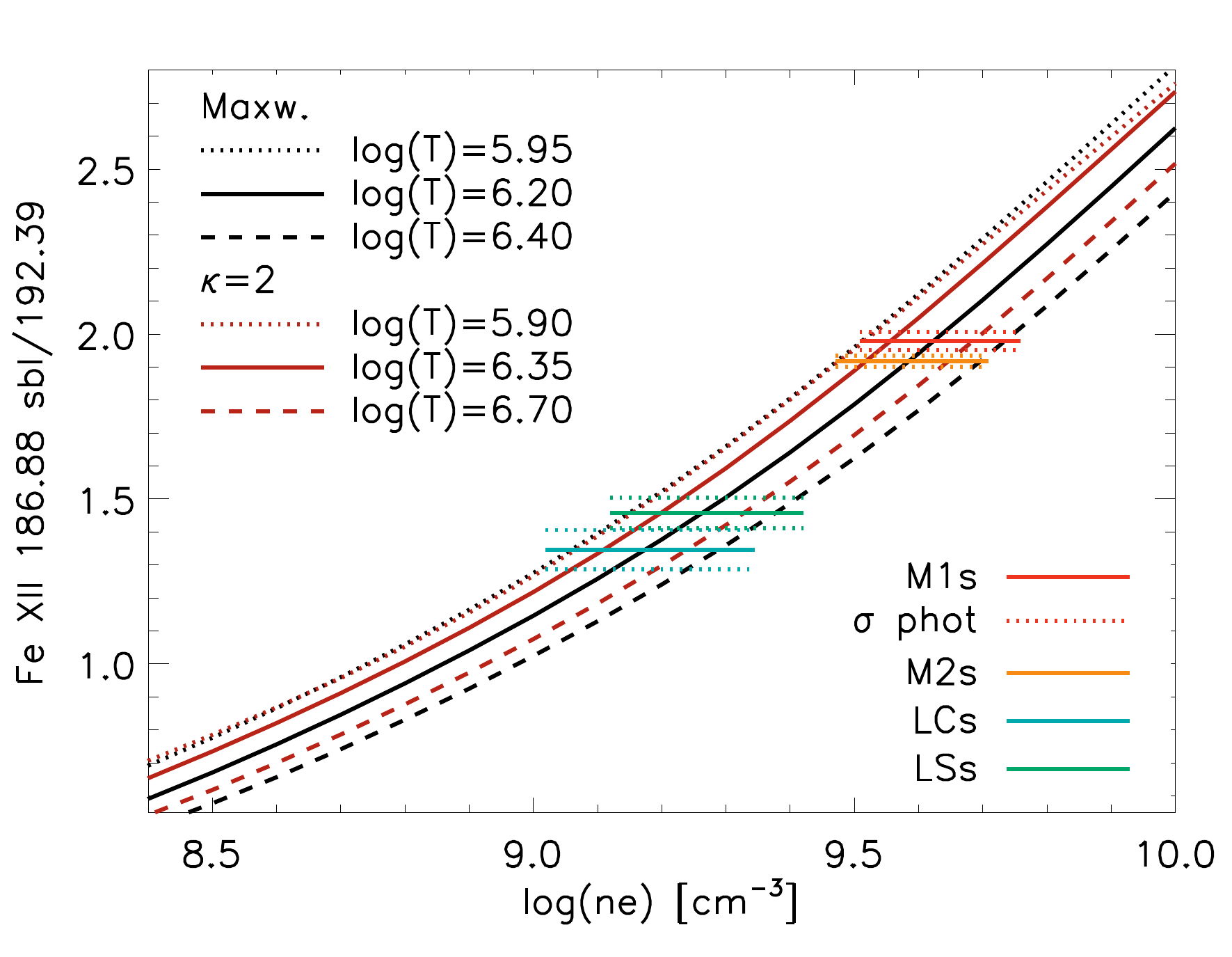}
    \includegraphics[width=0.41\textwidth, clip,  viewport= 0 0 505 370]{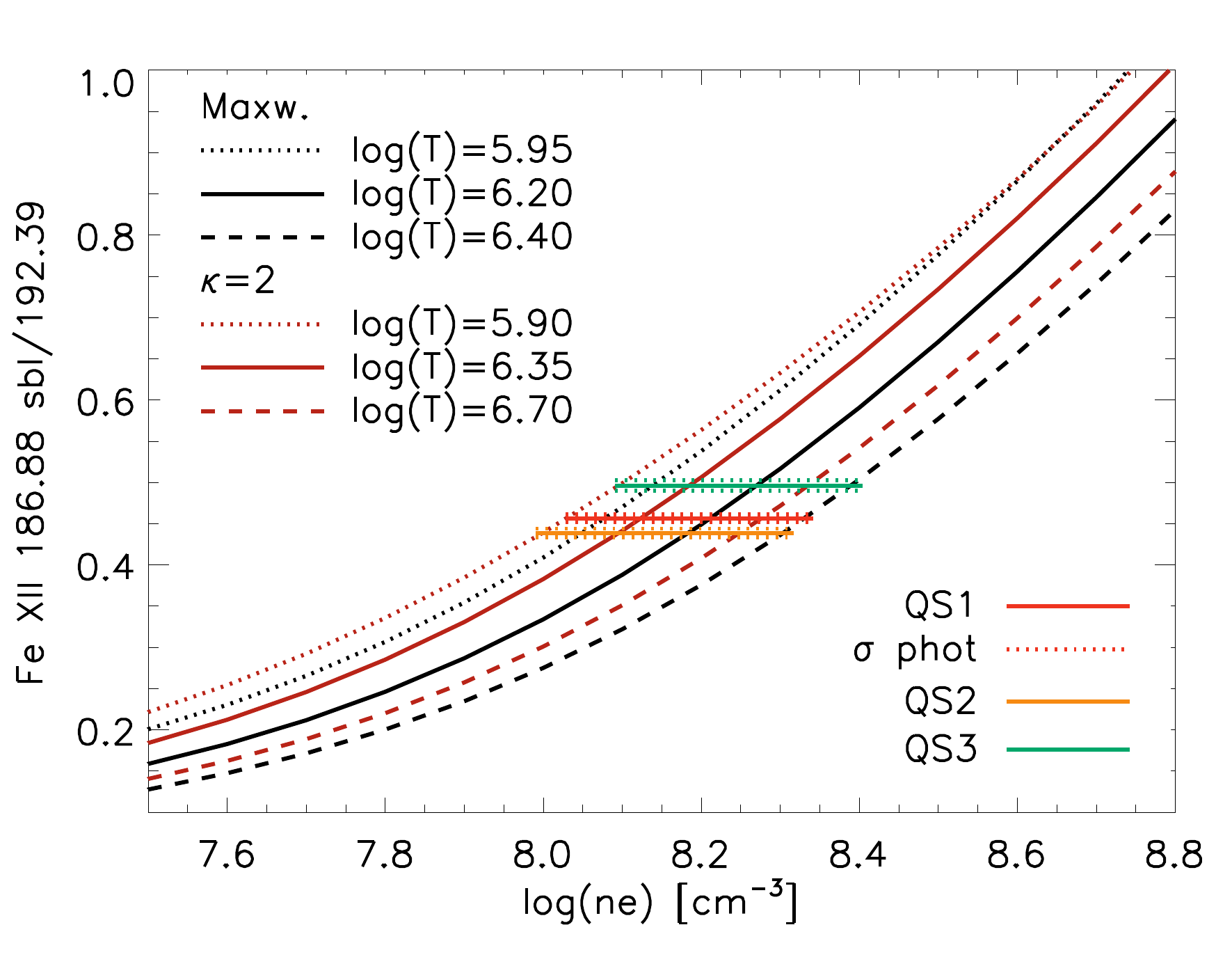}
      \\
    \includegraphics[width=0.41\textwidth, clip,  viewport= 0 0 505 370]{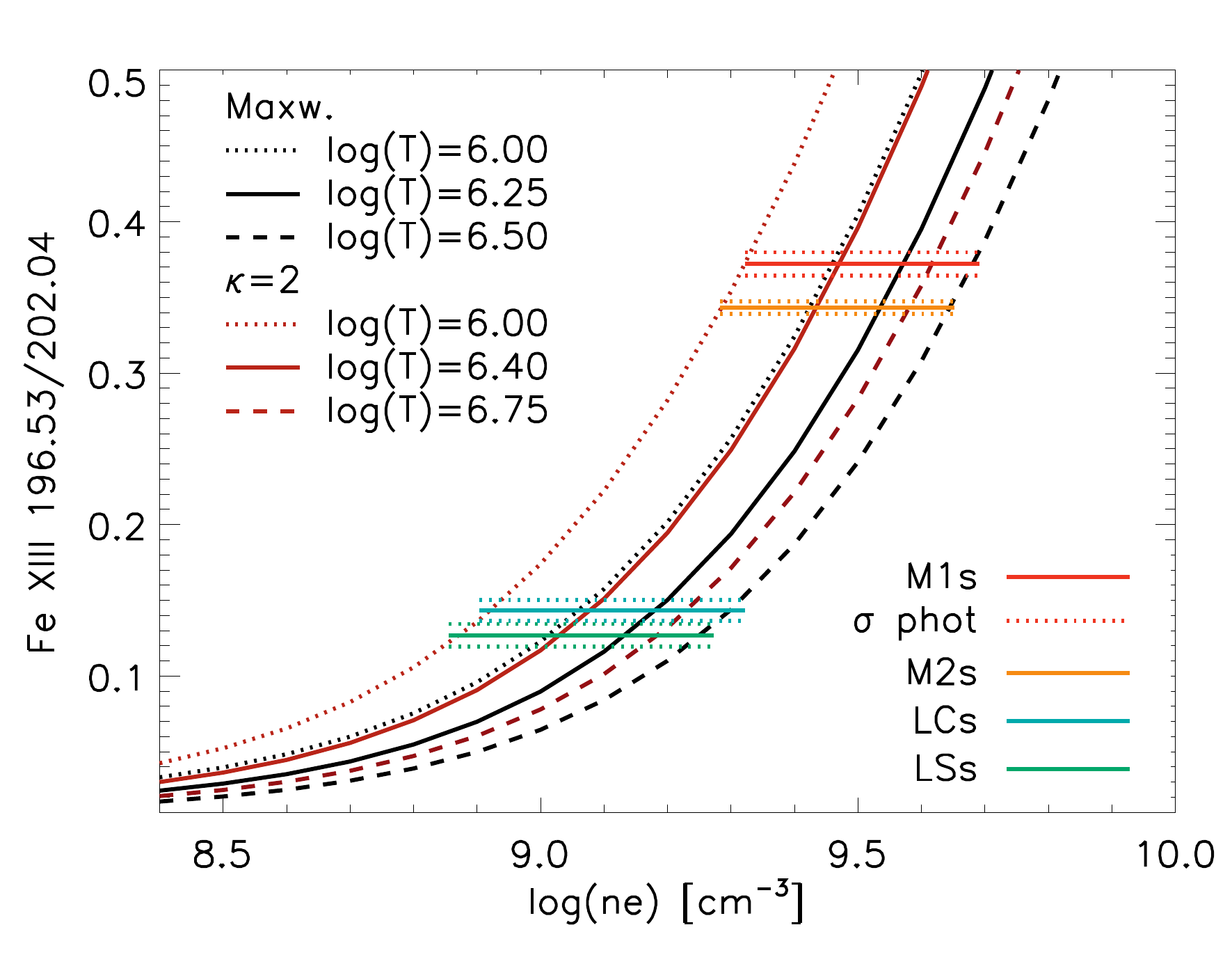}
    \includegraphics[width=0.41\textwidth, clip,  viewport= 0 0 505 370]{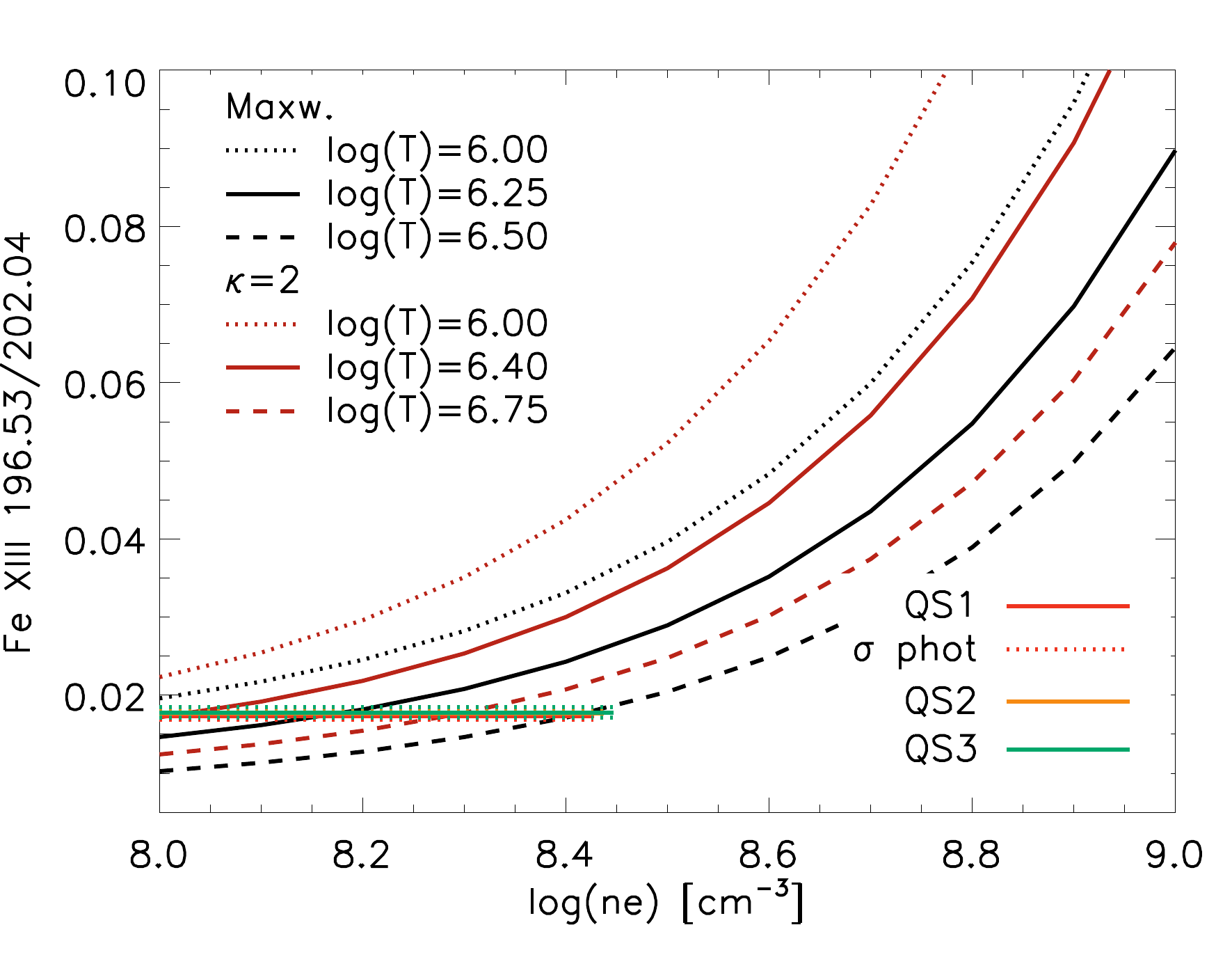}
      \\
    \includegraphics[width=0.41\textwidth, clip,  viewport= 0 0 505 370]{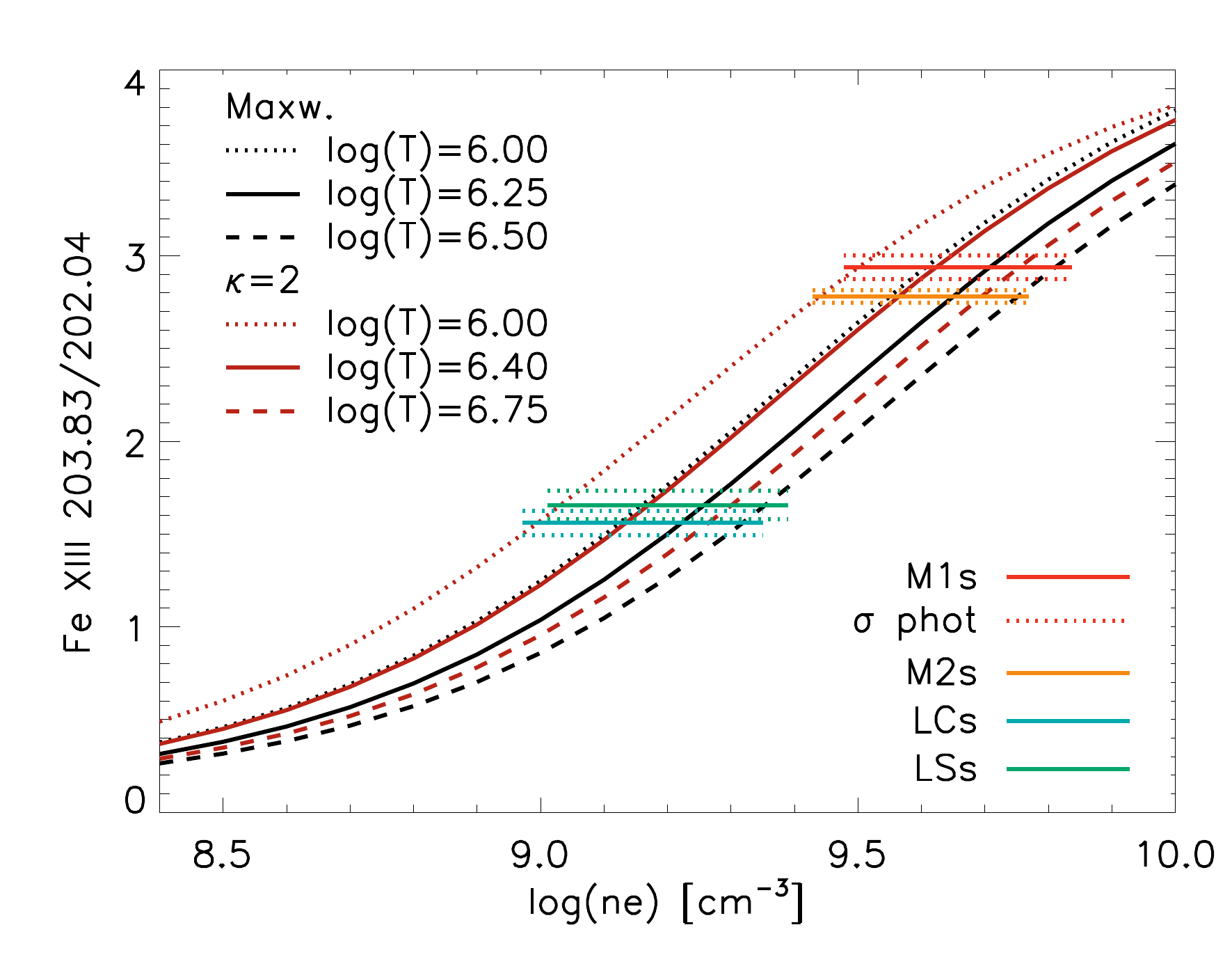}
    \includegraphics[width=0.41\textwidth, clip,  viewport= 0 0 505 370]{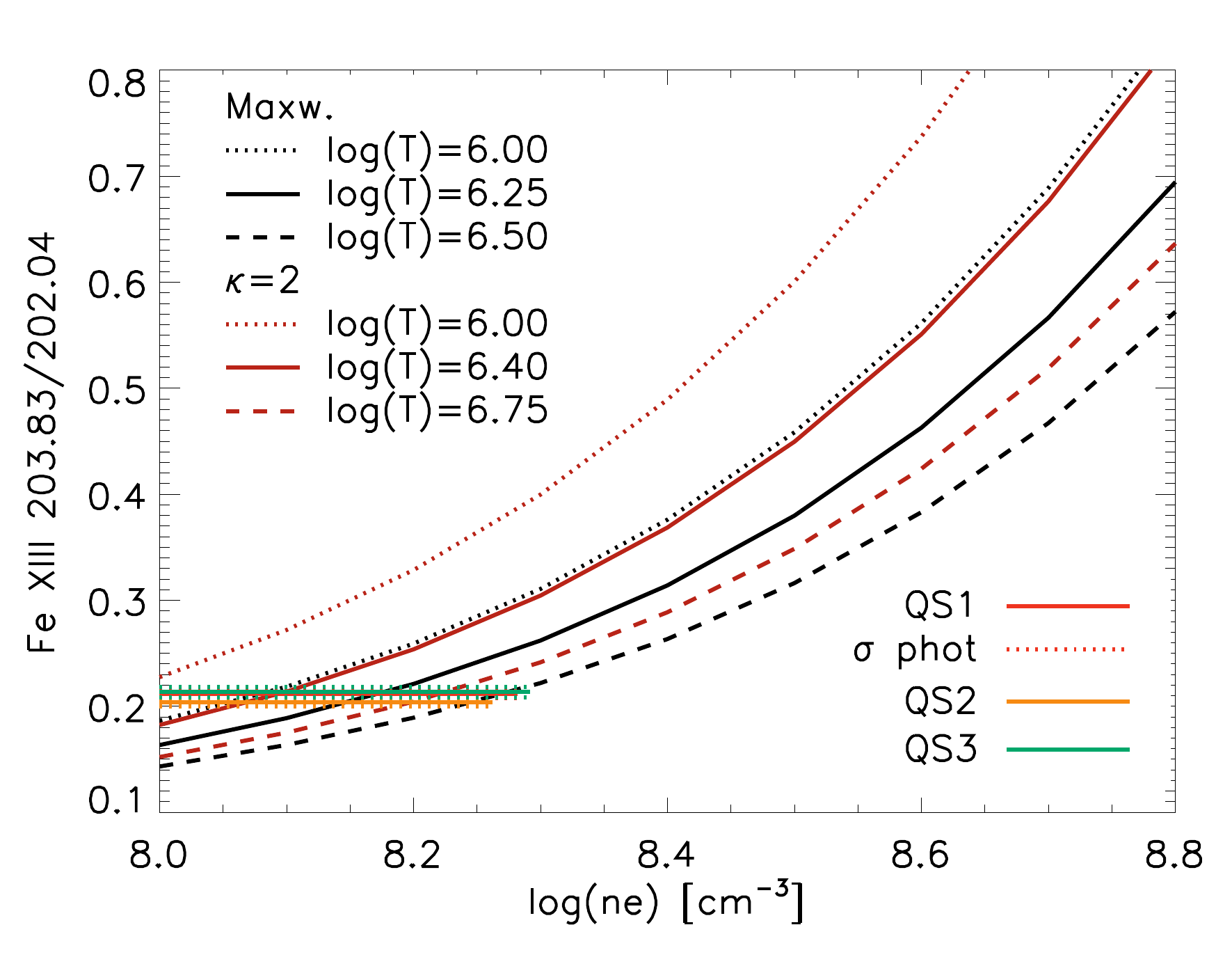}

  \caption{Diagnostics of density in selected boxes. Black and red curves are the theoretical ratios plotted for the Maxwellian and the $\kappa$\,=\,2 distribution. Different line styles code different temperatures for which the ratios were calculated. Colored horizontal solid and dotted lines are the observed ratios and their respective $\sigma_{\text{phot}}$ uncertainties. \label{figure_densdiag}}
\end{figure*}

\begin{figure}[h]
  \centering    
  \includegraphics[height=4.00cm, clip, viewport= 0 55 460 310]{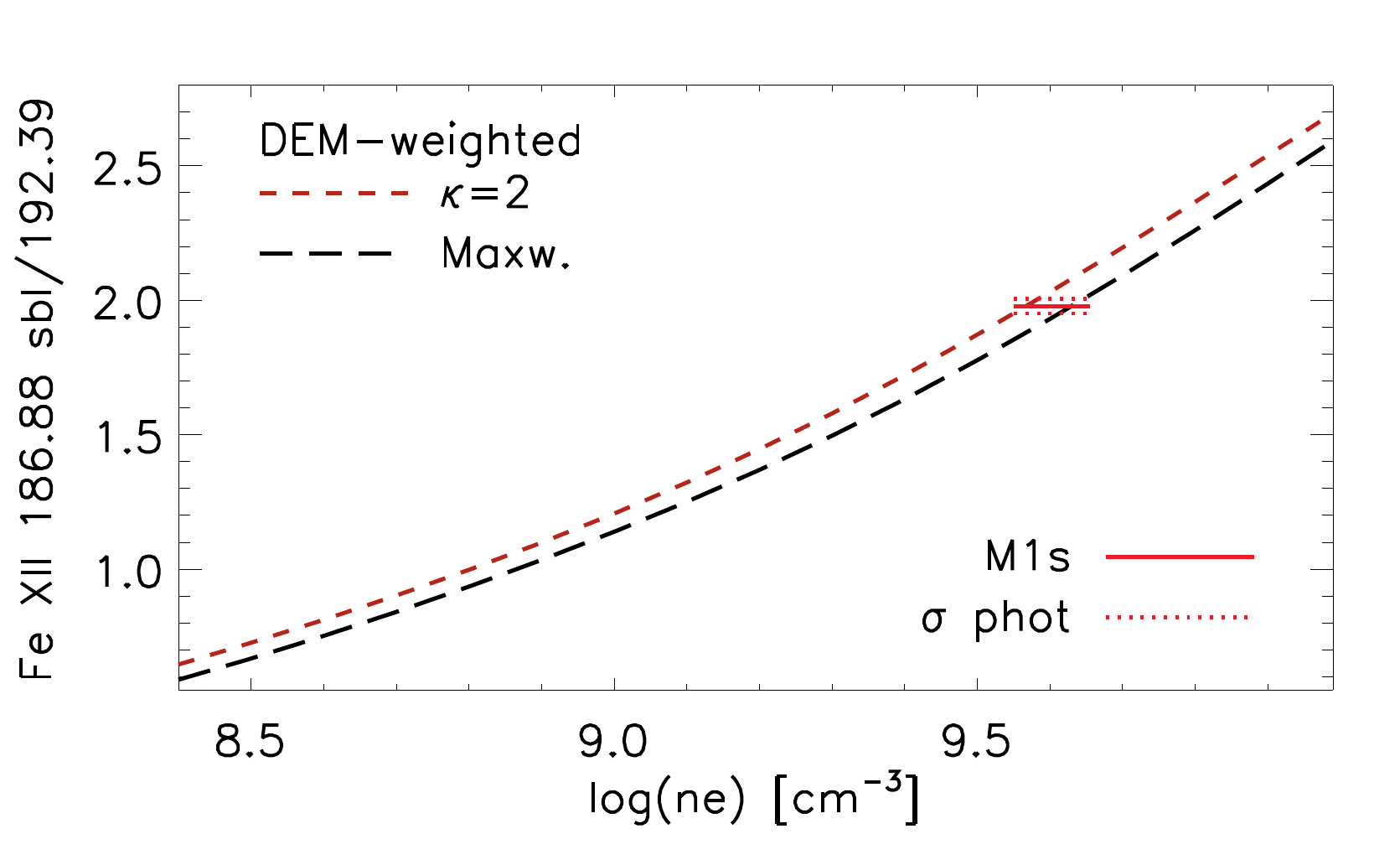}
  \includegraphics[height=3.52cm, clip, viewport= 0 55 460 280]{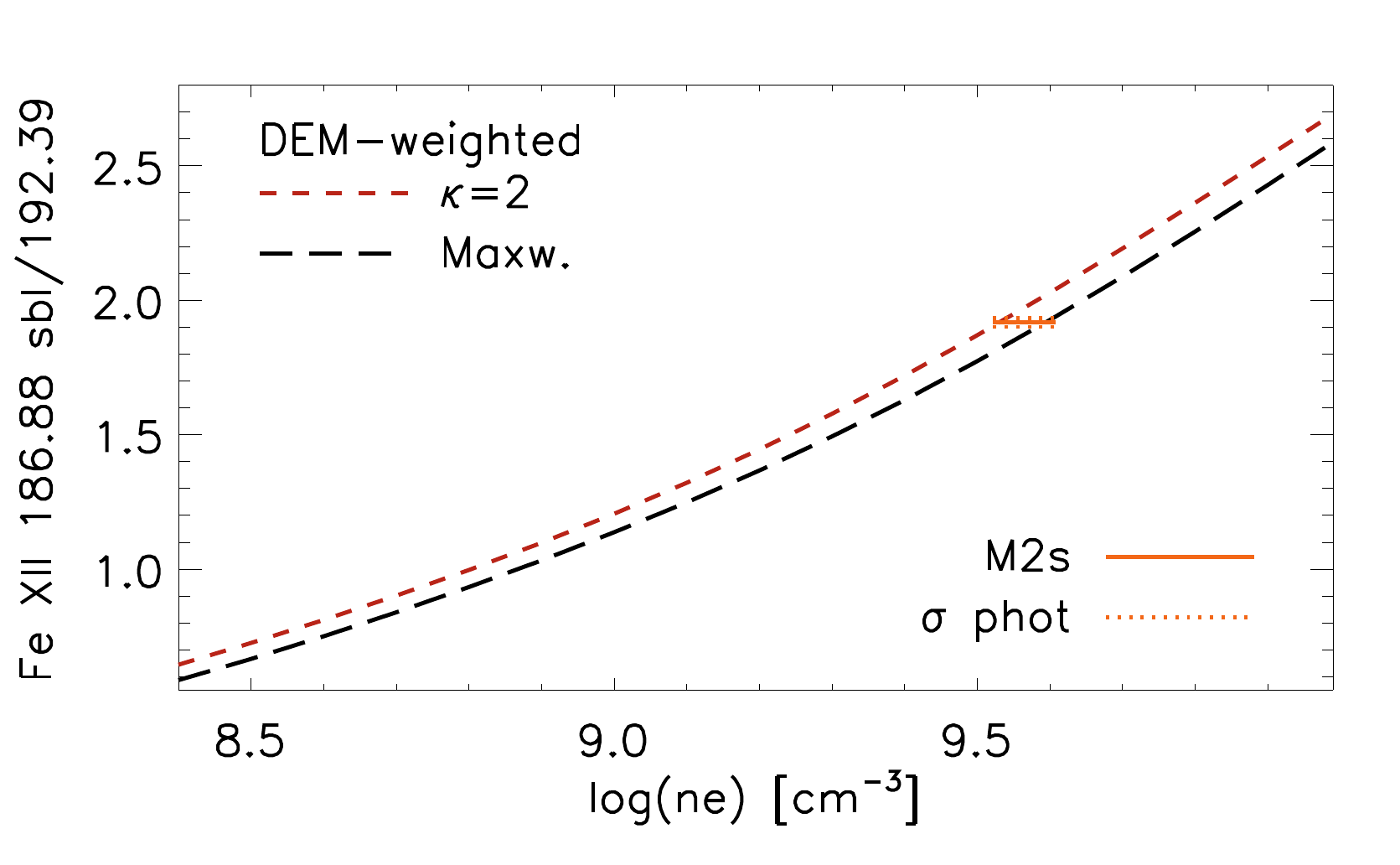}
  \includegraphics[height=3.52cm, clip, viewport= 0 55 460 280]{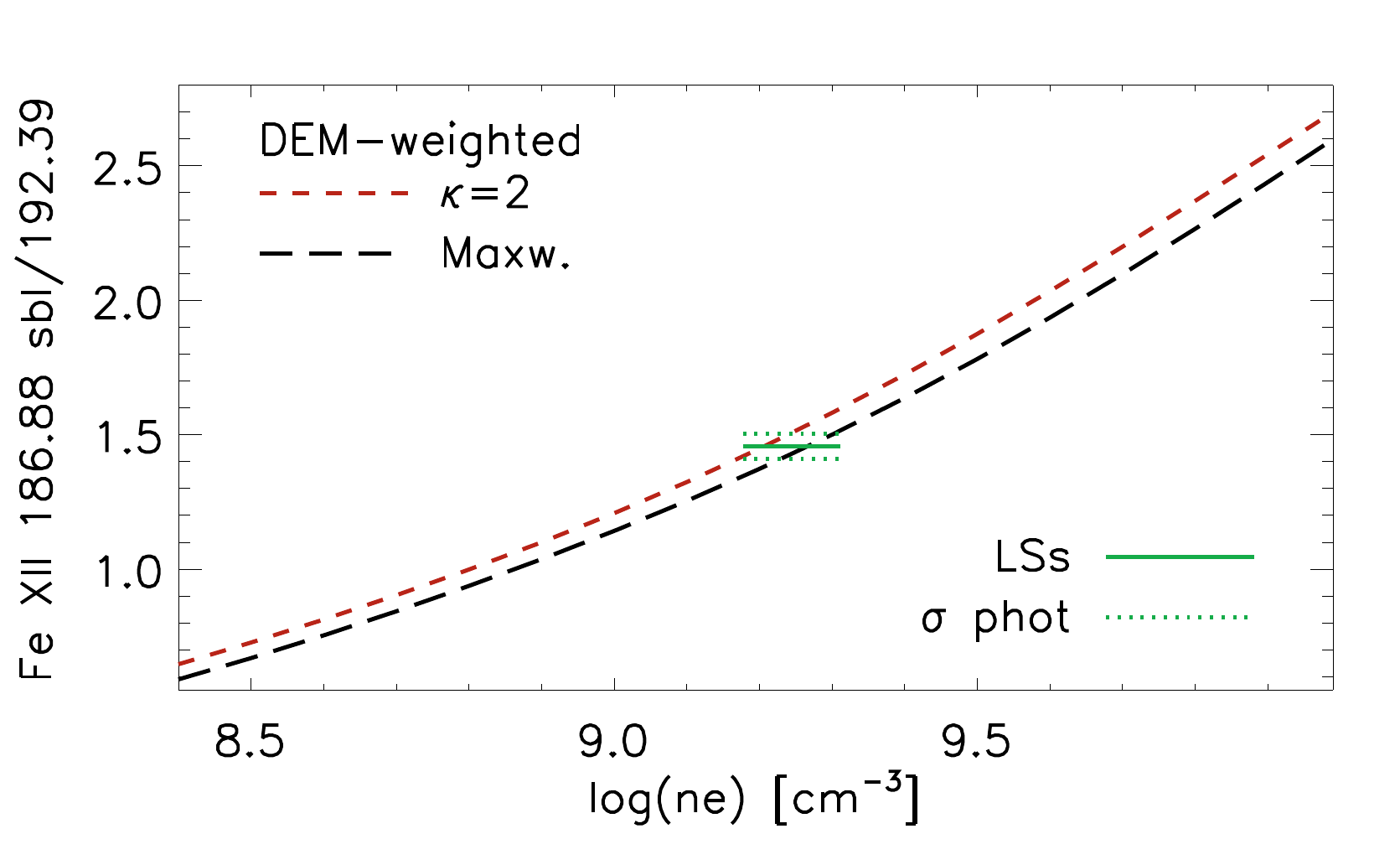}
  \includegraphics[height=4.235cm, clip, viewport= 0 10 460 280]{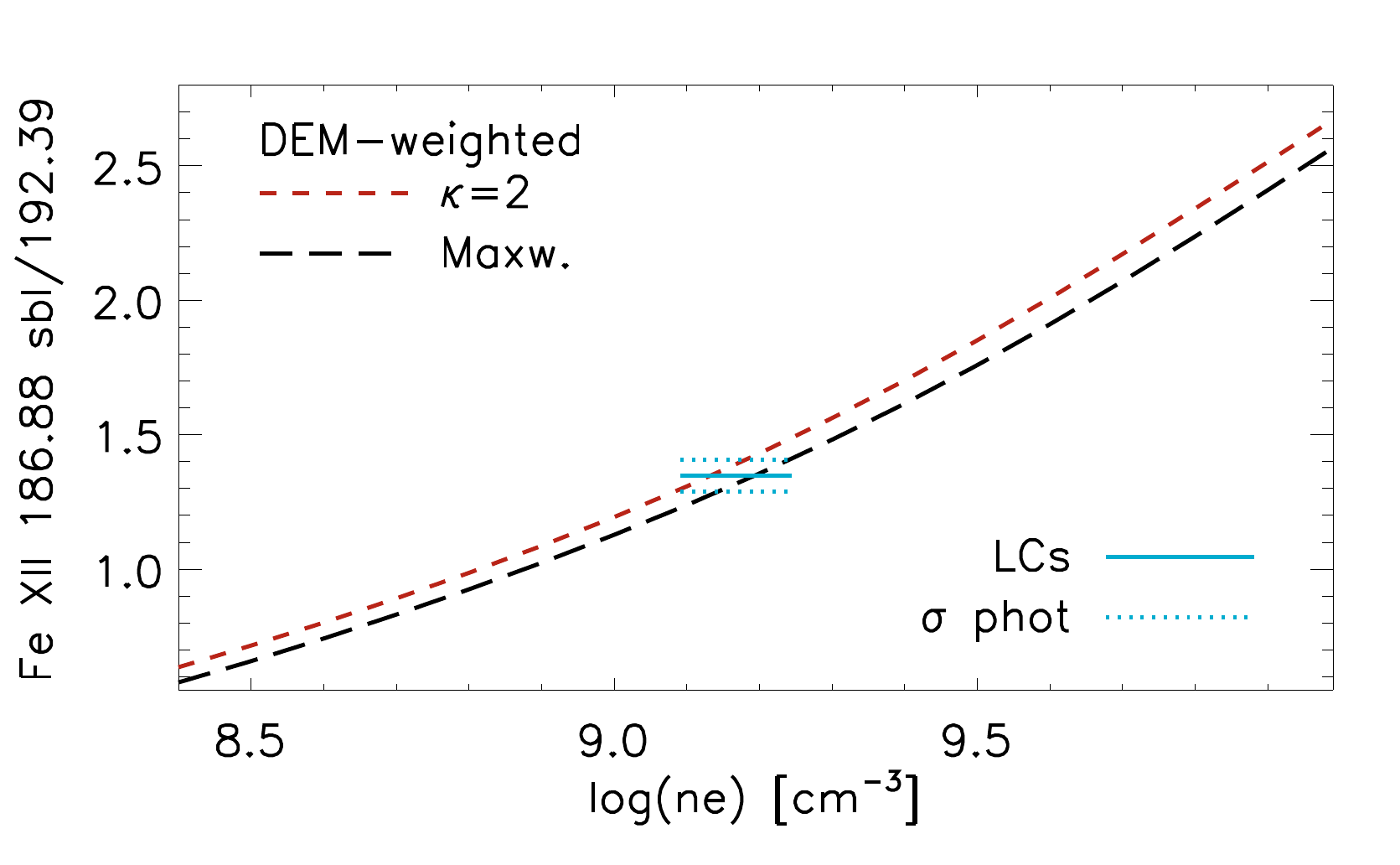}
  \includegraphics[height=4.40cm, clip, viewport= 0  0 460 280]{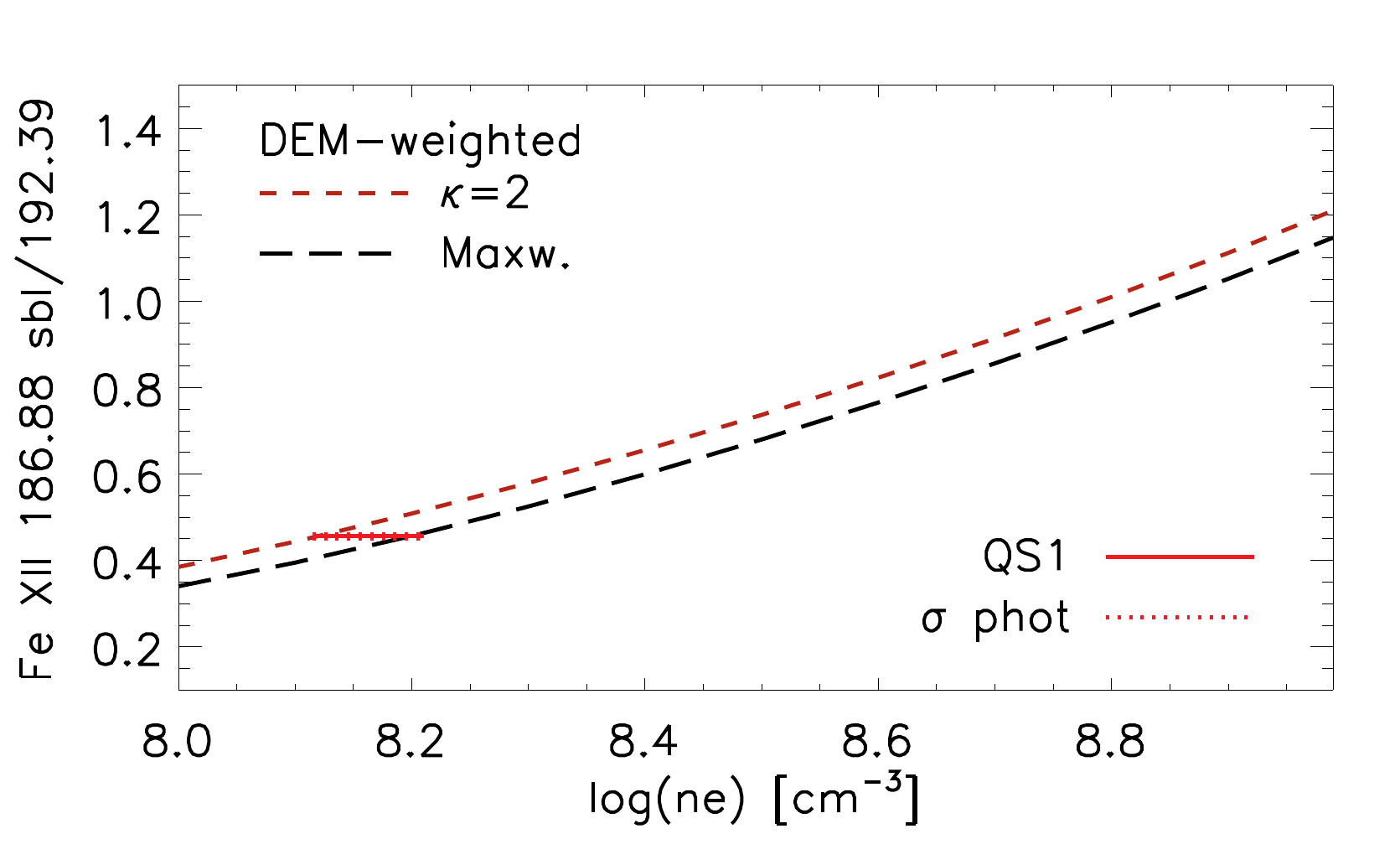}
  \caption{Effects of DEM on diagnostics of density. The black and red curves are DEM-weighted theoretical ratios plotted for the $\kappa$\,=\,2 distribution (red short-dashed) and the Maxwellian distribution (black dashed). Horizontal colored {solid and dotted lines are the ratios and their respective $\sigma_{\text{phot}}$ uncertainties observed} in different structures. Their color-coding is the same as in Figure \ref{figure_densdiag}. \label{figure_dem_densdiag}}
\end{figure}

Figure \ref{figure_densdiag} shows the four density-sensitive ratios used. In this figure, the theoretical calculations are shown by black curves for the Maxwellian distribution, while the red curves stand for $\kappa$\,=\,2. For each distribution, the ratios are shown at three different temperatures, corresponding to the temperatures of the peak of the ionization equilibrium, as well as where the ion abundance reaches 1\,\% of the peak which we take as extreme values \citep[see][]{dzifkuli10,dzifdudik13}. The span of the curves then describes the dependence of these ratios on both $T$ and $\kappa$, and is reduced in following steps of the iteration procedure by the DEM-weighting.
\subsubsection{Differential emission measure}
\label{sec_demdiag}

To fully address the temperature structure of the observed plasma, we examined the differential emission measure DEM and the emission measure distribution EM$_\kappa(T)$ defined as 

\begin{equation}
    \text{EM}(T)=\text{DEM}_\kappa(T) \Delta T=\text{DEM}_\kappa(T) \frac{T\Delta(\text{log }T)}{\text{log }e}.
\end{equation}

Here we employed the regularization inversion method of \citet[][]{hannahkontar12}, used in conjunction with line intensities observed by EIS. The input parameters controlling the regularization were kept at their default setting. We varied the maximum $\chi^2$ permissible to achieve robust solutions in as many temperature bins as possible, while attempting to recover smooth DEMs. In the active region structures the maximum $\chi^2$ of the solutions was set to 3, while the value of 5 was needed for obtaining solutions in the quiet Sun. The uncertainties on the measured intensities included not only the photon noise, but also the 20 $\%$ calibration uncertainty of the instrument, which needs to be taken into account when lines from both the short- and the long-wavelength channels of the instrument are used. 

The DEM$_\kappa(T)$ are calculated as a function of $\kappa$, using $\kappa$\,=\,2, 3, 5, 10, and Maxwellian. The corresponding temperature ranges were chosen to be log($T$\,[K])\,=\,5.6--6.6 for quiet Sun and 5.7--6.8 for active region, with a step of 0.1. The lines of \ion{Fe}{8}--\ion{Fe}{17} are used for producing these DEMs. We note that relying on lines of a single element have the advantage of not introducing additional uncertainties due to elemental abundance variations. The temperature interval covered by these ten Fe ions is also sufficient for quantifying the DEM$_\kappa(T)$ not only at the temperatures of interest for diagnostics of $\kappa$, but also to obtain sufficient constraints at both low and high temperatures for all $\kappa$ values. Additional high-$T$ constraints could in principle be obtained from other lines, such as from \ion{Ca}{14}--\ion{Ca}{17}, \ion{Ni}{17}, or XRT observations, but these would require abundance analysis, which could be coupled to the diagnostics of $\kappa$, and as such is beyond the scope of this work.

An important assumption when analysing DEMs is that the lines used for their construction are independent of density. This assumption is not always satisfied, as numerous lines of \ion{Fe}{11}--\ion{Fe}{13} show at least weak sensitivity to density in the coronal conditions \citep[see e.g.][]{dudik14}. Therefore, we calculated the DEM$_\kappa(T)$ as a function of the electron density in the range of log($N_\mathrm{e}$\,[cm$^{-3}$])\,=\,8--10 for all $\kappa$. Due to the choice of lines (see Appendix \ref{appendix_dem_lines}), the sensitivity of the resulting DEMs to $N_\mathrm{e}$ is on the order of 10\%, lower than uncertainty in the DEMs themselves. This let us use the obtained DEMs to refine the diagnostics of $N_\mathrm{e}$ (see Section \ref{sec_nediag}).

\subsubsection{Diagnostics of $\kappa$ and $T$}
\label{sec_kappadiag}

To diagnose the $\kappa$ parameter, we use the ratio-ratio method \citep{dzifkuli10,dudik14,Dudik15}. The method consists of comparing two observed and theoretical line ratios. Typically, one ratio involves lines from two neighboring ions, which is dominantly sensitive to $T$. The second ratio is sensitive to $\kappa$ due to combining lines formed at different wavelengths. The sensitivity to $\kappa$ arises from these lines having different excitation thresholds. In case of lines observed by EIS, one of the lines is usually observed in the short- and the other in the long-wavelength channel. The sets of theoretical ratios for different $\kappa$ in the ratio-ratio diagram are plotted for a density (or range of densities) diagnosed apriori. Precise measurements of densities are advantageous, as the sets of curves for different densities overlap, which is a source of uncertainty in the resulting $\kappa$ parameter \citep{dudik14,Dudik15}.

Here, we use the combinations of \ion{Fe}{11} and \ion{Fe}{12} lines also used by \citet[][]{Dudik15}. The sensitivity to $\kappa$ is produced by combination of the \ion{Fe}{11} 182.17\,\AA~and 188.22\,\AA~lines observed in the short-wavelength channel with the \ion{Fe}{11} 257.55\,\AA~and 257.77\,\AA~selfblends observed in the long-wavelength channel, identified for the first time in \citet{delzanna10} as a very useful diagnostic to measure $T$. The use of these lines is advantageous since the lines are well-observed, as well as due to the relatively strong sensitivity to $\kappa$ compared to other combinations of EIS lines. In \citet{Dudik15}, these \ion{Fe}{11} lines are coupled with the \ion{Fe}{12} lines such as the 186.89\AA~and 195.12\,\AA~lines in the conjugate ratio to provide strong temperature-dependence.

A multitude of combinations of lines were used by \citet{Dudik15}. However, we opted not to use the \ion{Fe}{11} 188.22\,\AA~line for diagnostics of $\kappa$, because this line is blended with the \ion{Fe}{12} 188.17\,\AA~and \ion{Fe}{11} 188.30\,\AA~lines \citep[see e.g.,][]{delzanna12,dudik14} which are both density dependent and we could not constrain amplitudes of Gaussians fitting these blends. Concerning the temperature-sensitive ratios, we again use the \ion{Fe}{12} 192.39\,\AA~line again instead of the \ion{Fe}{12} 195.12\,\AA~line because of high $\chi^2$ of its fit and possible optical thickness (Section \ref{Sect:3.1.1}).

Finally, note that the sets of ratio-ratio curves for diagnostics of $\kappa$ are plotted as a function of $T$, assuming that the plasma is isothermal. In the case of multithermal plasma, these curves need to be weighted over the respective DEM$_\kappa(T)$ \citep{Dudik15}, producing a single predicted value for each ratio and $\kappa$ (see Section \ref{sec_res_demeffects}).

\begin{deluxetable*}{cccccccc}[t]
\tablecaption{Initial estimates on the density ranges using the line ratio technique \label{tab:density}}
\tablecolumns{9}
\tablenum{1}
\tablewidth{0pt}
\tablehead{
\colhead{ } & \colhead{QS1} & \colhead{QS2} & \colhead{QS3} & \colhead{M1s} & \colhead{M2s} & \colhead{LCs} & \colhead{LSs} }
\startdata
\ion{Si}{10} 258.37/261.06	& 7.9--8.4 	& 7.9--8.4 	& 7.9--8.4 	& 9.1--9.8	& 9.0--9.6	& 8.5--9.2	& 8.5--9.1 	\\ 
\ion{Fe}{12} 186.89/192.39	& 8.0--8.3 	& 8.0--8.3 	& 8.1--8.4 	& 9.5--9.8	& 9.5--9.7 	& 9.0--9.3	& 9.1--9.4	\\ 
\ion{Fe}{13} 196.53/202.04	& 8.0--8.4 	& 8.0--8.4 	& 8.0--8.4 	& 9.3--9.7 	& 9.3--9.7	& 8.9--9.3	& 8.9--9.3	\\ 
\ion{Fe}{13} 203.83/202.04	& 8.0--8.3 	& 8.0--8.3 	& 8.0--8.3 	& 9.5--9.8  & 9.4--9.8	& 9.0--9.4	& 9.0--9.4	\\ \hline
\enddata
\end{deluxetable*}
%
\section{Results} \label{sec_results}

To determine the physical parameters of the emitting plasma in various observed structures in both the active region (Section \ref{sec_obs_ar}) and the quiet Sun (\ref{sec_obs_qs}), we use the iterative technique described in Section \ref{sec_iterations}. Density ranges are obtained first, followed by DEM$_\kappa(T)$ inversions, then we perform the DEM-weighted density-diagnostics and diagnostics of $\kappa$ in two iterations. 

\subsection{Electron densities} \label{sec_res_nediag}

Density-sensitive line ratios as a function of the electron density $N_\mathrm{e}$ are shown in Figure \ref{figure_densdiag}. The theoretical ratios for the Maxwellian (black) and $\kappa$\,=\,2 (red) are intersected by horizontal lines, which denote the observed ratios (solid) and their respective photon noise uncertainties $\sigma_{\text{phot}}$ (dotted lines). The observed ratios intersect the theoretical ratios calculated using the Maxwellian (black curves). The left-hand side of Figure \ref{figure_densdiag} shows the observed ratios in the active region, while right-hand side shows the diagnostics in the quiet Sun.

As apparent from Figure \ref{figure_densdiag} the natural spread of the theoretical ratios due to $T$ and $\kappa$, leads only to ranges of possible densities. These are summarized in Table \ref{tab:density}. The lower limits on the density range for each structure and ratio were obtained from the intersection of the observed ratio minus its $\sigma_{\text{phot}}$ uncertainty with the theoretical ratio calculated assuming the $\kappa=2$ distribution and the lowest temperature (dotted horizontal lines intersecting the dotted red curves in Figure \ref{figure_densdiag}). Conversely, the upper limit on the density range corresponds to the observed ratio plus $\sigma_{\text{phot}}$ intersecting with the rightmost theoretical ratio, typically for the Maxwellian distribution and the highest temperature (dotted upper horizontal lines intersecting the black dashed curves). The initial density ranges can span about $0.3-0.9$ dex in log($N_\mathrm{e}$\,[cm$^{-3}$]) due to the combination of the photon noise uncertainty and the dependence of density-sensitive ratios on $T$ and $\kappa$. Note that since all density-sensitive ratios include lines observed at similar wavelengths in the same channel of EIS, the calibration uncertainty is not considered in density diagnostics.

As a next step in the iterative procedure, we used the DEM$_\kappa(T)$ recovered in all of the observed structures to constrain the diagnostics of density. Examples of the DEM-weighted density-sensitive ratios are shown in Figure \ref{figure_dem_densdiag}. This figure shows the \ion{Fe}{12} ratios, one panel for each of the structures investigated. On each panel, only two lines are shown, the black one for the Maxwellian distribution, and the red for $\kappa$\,=\,2. The horizontal lines again stand for the observed ratios {and their $\sigma_{\text{phot}}$ uncertainties}. For brevity we do not show the analogous {panels} for the other ratios. Instead, the {initial estimates $N_\mathrm{e,0}$ of the densities are summarized} in Table \ref{tab:density_dem}. For each structure and line ratio, two densities are listed, one obtained for the Maxwellian and the other for $\kappa$\,=\,2. These densities differ by about 0.1--0.2 dex, a value typical for density diagnostics for $\kappa$-distributions \citep{dzifkuli10,mackovjak13,dudik14,Dudik15,dzif18}. {In all cases, the effect of the $\sigma_{\text{phot}}$ uncertainties is at most 0.1 dex.}

For the QS1--3, M1s and M2s, the densities diagnosed using all four ratios are consistent, with only minor differences. For LCs and LSs, we obtain consistent densities using the \ion{Fe}{12} and \ion{Fe}{13} ratios. \ion{Si}{10} indicates slightly lower densities, of up to 0.2--0.3 dex compared to the other ratios. The cause of this is small inconsistency is unknown. Nevertheless, we include these results (see below) and note that excluding them would lead to higher densities and ultimately slightly stronger non-Maxwellian diagnostics (see Figure \ref{figure_kappadiag} and Section \ref{sec_res_kappadiag}).

The initial estimate on DEM-weighted density, $N_\mathrm{e0}$ is also listed in Table \ref{tab:density_dem}. It was calculated as the average from the values listed and subsequently used to produce the first estimate on $\kappa$ using the ratio-ratio technique (Section \ref{sec_res_kappadiag}). Then, as described in Section \ref{sec_iterations}, the final DEM-weighted density is obtained as the average of the density-sensitive DEM-weighted ratios using only Maxwellian or $\kappa$\,=\,2 results, depending on the structure investigated. 

The densities obtained are in good agreement with literature. For example, electron densities typically found in the quiet Sun are log($N_\mathrm{e}$\,[cm$^{-3}$]) = 8.2 \citep[e.g.,][]{dere07}, while they are of the order of 9.0 in coronal loops \citep[e.g.,][]{young09,tripathi09,Dudik15}, and 9.5 in the coronal moss \citep[e.g.,][]{depontieu99, warren08, depontieu09}. {Nevertheless, several comments towards the density diagnostics as well as the iterative method itself are in order.}

{First, the uncertainties of the DEM-weighted densities are dominated by the photon noise and the spread of the results from different density-sensitive ratios and are $\pm$\,0.1 dex at most. The uncertainties of the DEM-weighted theoretical ratios were evaluated by considering the DEMs with their respective errors and were found to be lower than the photon noise uncertainty; the DEM-weighted theoretical ratios do not differ from those in which we considered the uncertainties of DEMs by more than 1\%. }

{Second, as already pointed out, the averaged DEM-weighted densities for the Maxwellian and $\kappa$\,=\,2 do not differ for more than 0.2 dex (Table \ref{tab:density_dem}, Section \ref{sec_iterations}). The effect of such small differences in densities on the ratios sensitive to $\kappa$ does not exceed a few percent. Henceforth, averaging of the DEM-weighted densities into $N_\mathrm{e,0}$ does not affect the first estimate on $\kappa$.}

{The differences between the initial estimates $N_\mathrm{e,0}$ and $N_\mathrm{e}$ are 0.1 dex at most. Given the typical $\sigma_{\text{phot}}$ uncertainties of ratios of line intensities (Figure \ref{figure_densdiag}), this indicates that single refinement of the density measurements via DEM and subsequently $\kappa$ yields relatively precise results of density diagnostics, within 0.1 dex.} 

{Finally, for AR, the resulting densities would be lower by about 0.2--0.3 dex if the background were not subtracted.} 


%
\begin{deluxetable*}{ccccccccccccccc}[]
\tablecaption{Density diagnostics using selected line ratios with DEM-weighted theoretical ratios. {The typical uncertainties of these densities \mbox{are $<$\,0.1 dex}. The $N_\mathrm{e,0}$ is} the first estimate on DEM-weighted density, while $N_\mathrm{e}$ is the final density obtained from the iterative procedure. \label{tab:density_dem}}
\tablecolumns{15}
\tablenum{2}
\tablewidth{0pt}
\tablehead{
\colhead{}   		& \multicolumn{2}{c}{QS1} & \multicolumn{2}{c}{QS2} & \multicolumn{2}{c}{QS3} & \multicolumn{2}{c}{M1s} & \multicolumn{2}{c}{M2s} & \multicolumn{2}{c}{LCs} & \multicolumn{2}{c}{LSs} \\
\colhead{Line ratio} & $\kappa=2$ & Mxw & $\kappa=2$ & Mxw & $\kappa=2$ & Mxw & $\kappa=2$ & Mxw & $\kappa=2$ & Mxw & $\kappa=2$ & Mxw & $\kappa=2$ & Mxw}
\startdata
\ion{Si}{10} 258.37/261.06   & 	8.2 & 8.3 	& 	 8.2 & 8.3	& 	8.2 & 8.3	& 	9.5 & 9.6	& 9.3 & 9.5		& 8.9 & 9.0		& 8.9 & 9.0  \\ 
\ion{Fe}{12} 186.89/192.39   & 	8.1 & 8.2 	& 	 8.1 & 8.2	& 	8.2 & 8.3	& 	9.6 & 9.7	& 9.5 & 9.6		& 9.1 & 9.2		& 9.2 & 9.3  \\ 
\ion{Fe}{13} 196.53/202.04   & 	8.0 & 8.1 	& 	 8.0 & 8.2	& 	8.0 & 8.1	& 	9.5 & 9.6	& 9.4 & 9.6		& 9.1 & 9.2		& 9.0 & 9.2  \\ 
\ion{Fe}{13} 203.83/202.04   & 	8.1 & 8.2 	& 	 8.0 & 8.1	& 	8.1 & 8.2	& 	9.6 & 9.7	& 9.6 & 9.7		& 9.1 & 9.3		& 9.2 & 9.3  \\
\hline
$N_\mathrm{e,0}$ & \multicolumn{2}{c}{8.2} & \multicolumn{2}{c}{8.1} & \multicolumn{2}{c}{8.2} & \multicolumn{2}{c}{9.6} & \multicolumn{2}{c}{9.5} & \multicolumn{2}{c}{9.1} & \multicolumn{2}{c}{9.1} \\ 
$N_\mathrm{e}$ 		& \multicolumn{2}{c}{8.2} & \multicolumn{2}{c}{8.2} & \multicolumn{2}{c}{8.3} & \multicolumn{2}{c}{9.5} & \multicolumn{2}{c}{9.4} & \multicolumn{2}{c}{9.0} & \multicolumn{2}{c}{9.0} \\ 
\enddata
\end{deluxetable*}
\begin{figure*}[!t]
  \centering
    
    \includegraphics[width=6.3cm, clip,   viewport= 0 0 311 280]{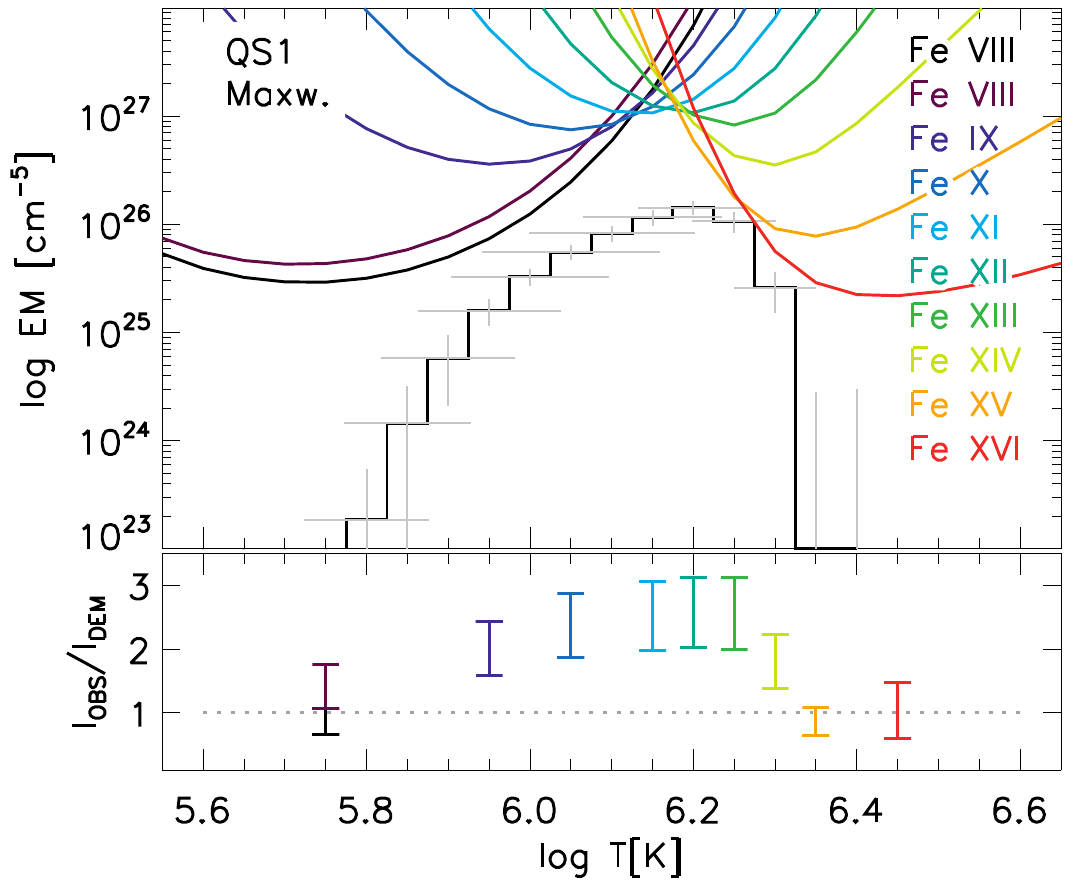}
    \includegraphics[width=5.83cm, clip,  viewport= 23 0 311 280]{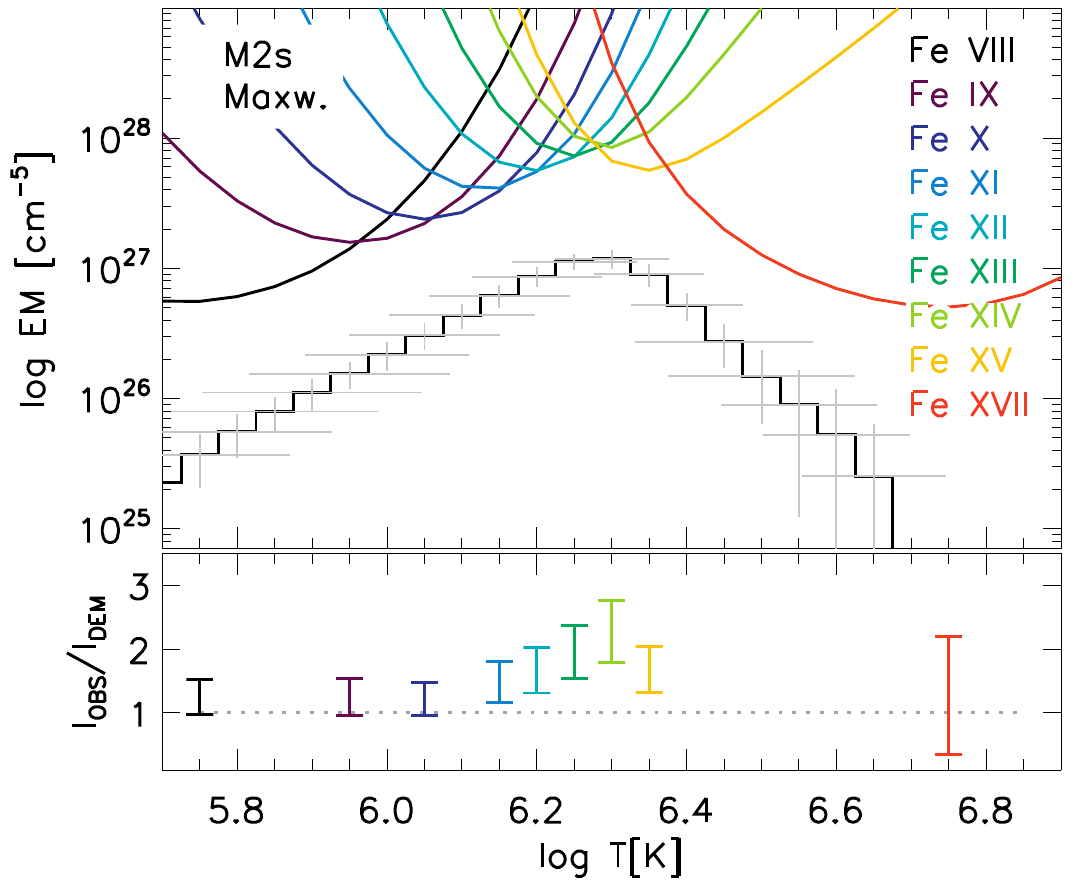}
    \includegraphics[width=5.39cm, clip,  viewport= 45 0 311 280]{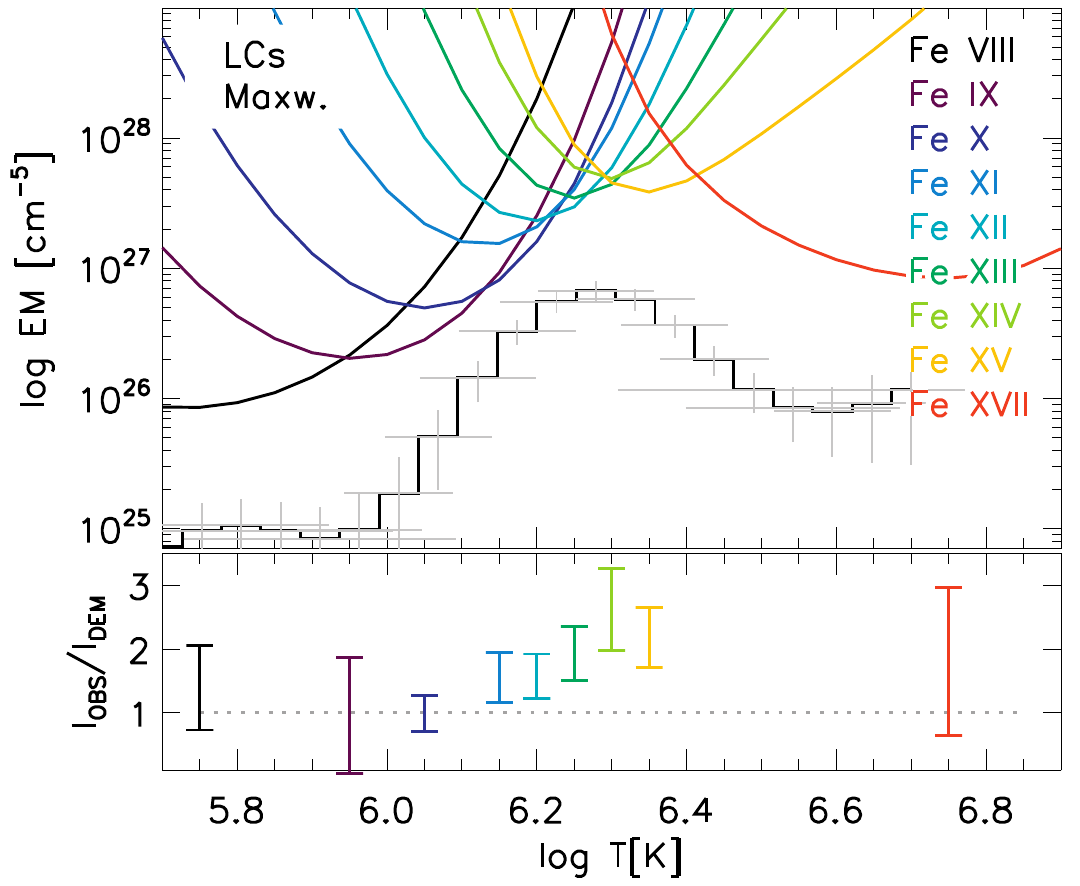}    
    \\
    \includegraphics[width=6.3cm, clip,   viewport= 0 0 311 255]{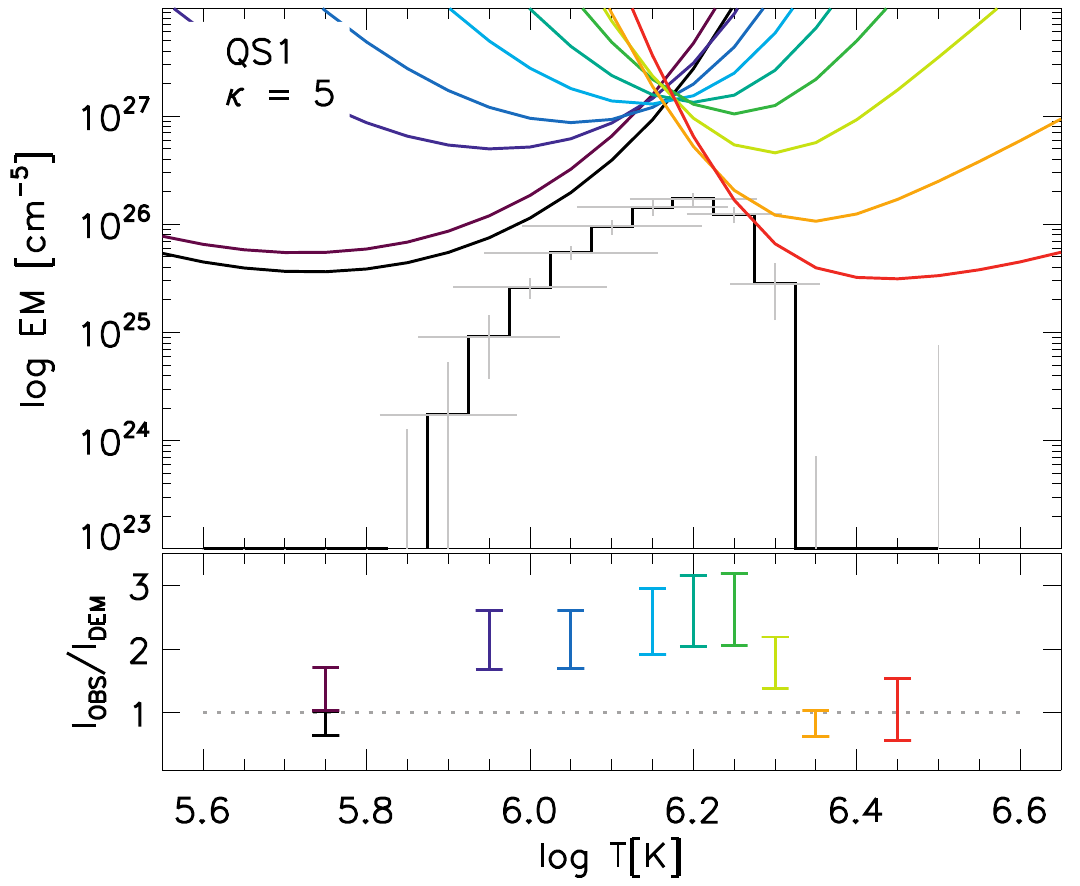}
    \includegraphics[width=5.83cm, clip,  viewport= 23 0 311 255]{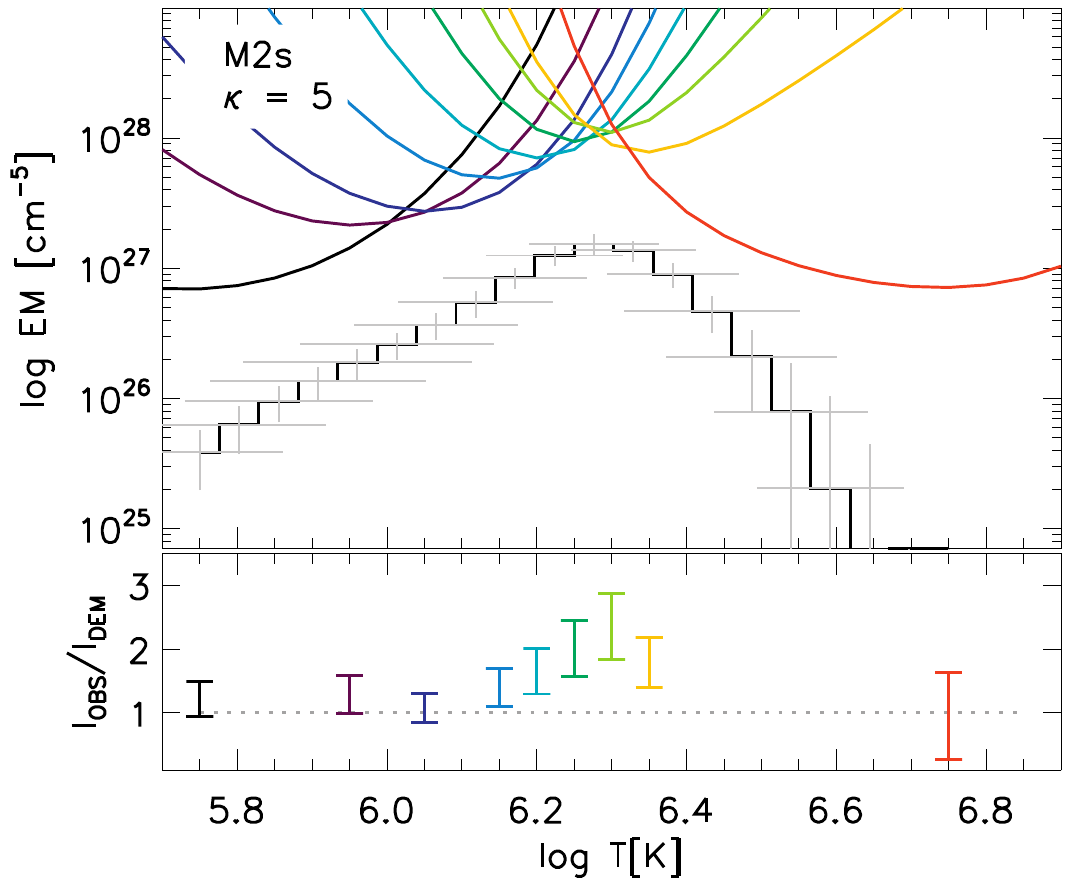}
    \includegraphics[width=5.39cm, clip,  viewport= 45 0 311 255]{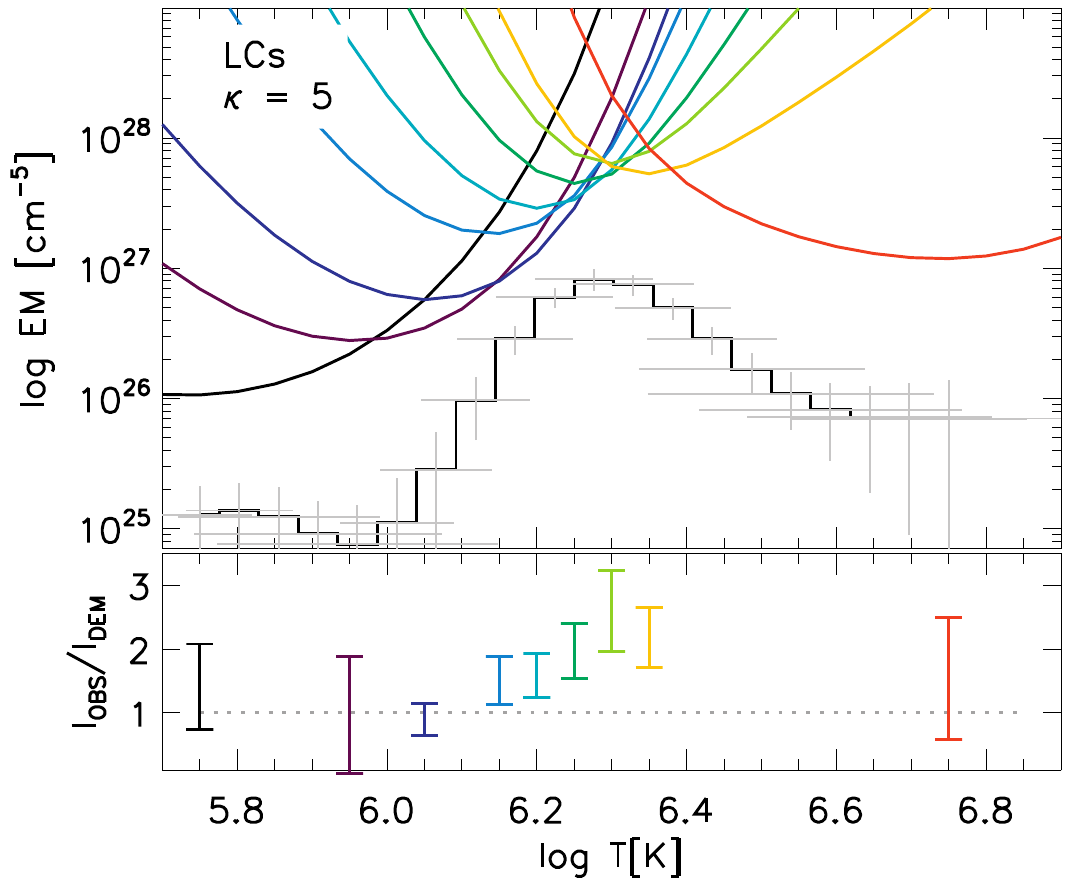}
    \\
    \includegraphics[width=6.3cm, clip,   viewport= 0 0 311 255]{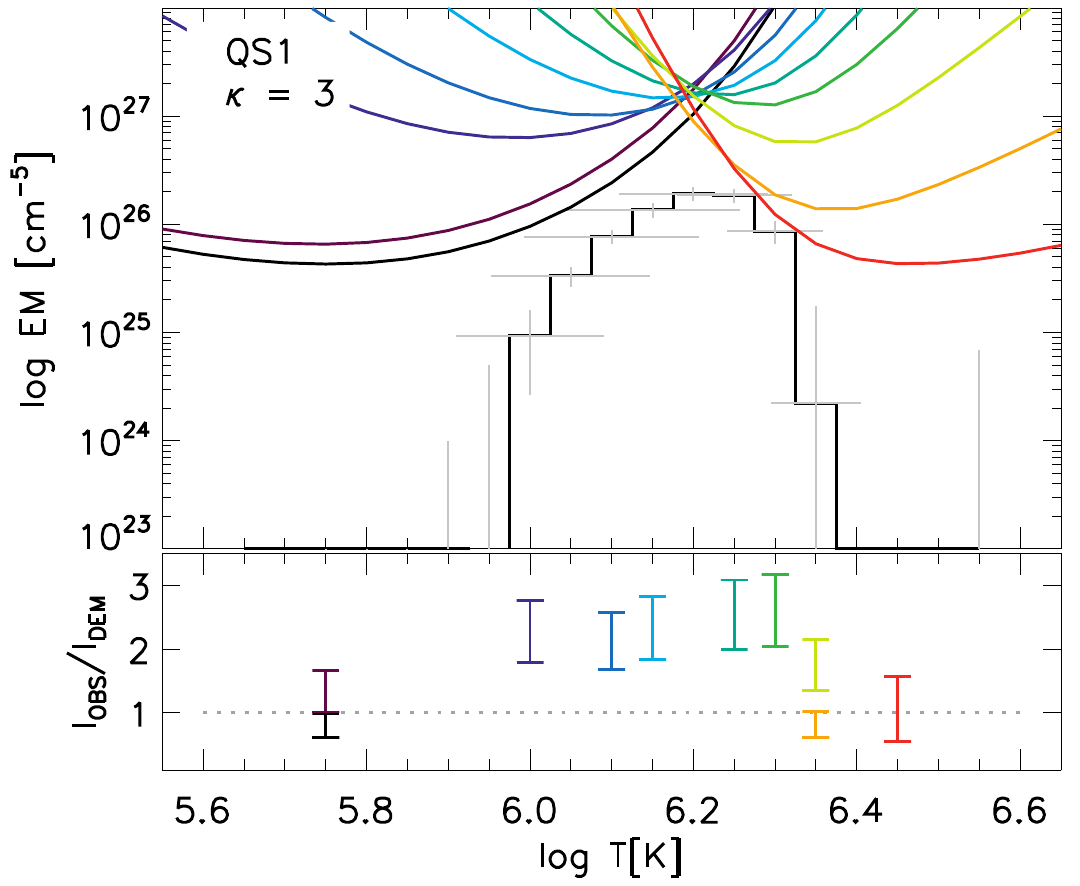}
    \includegraphics[width=5.83cm, clip,  viewport= 23 0 311 255]{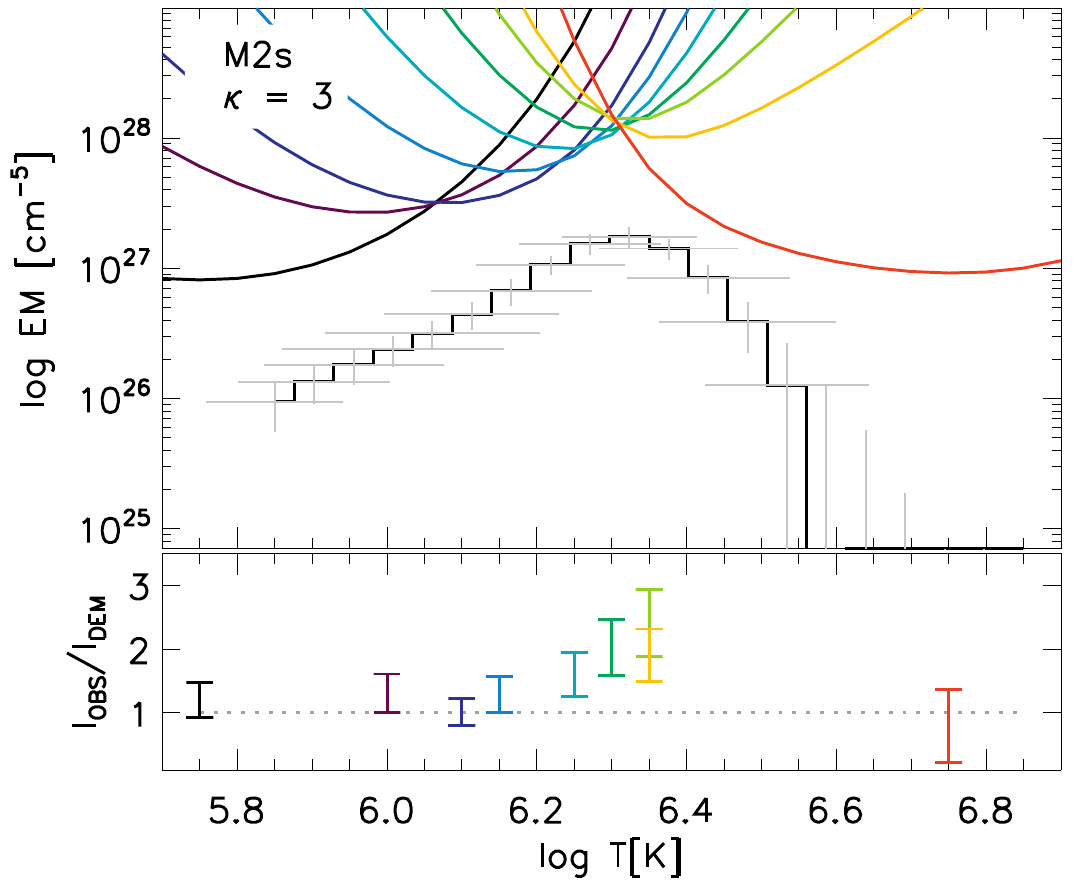}
    \includegraphics[width=5.39cm, clip,  viewport= 45 0 311 255]{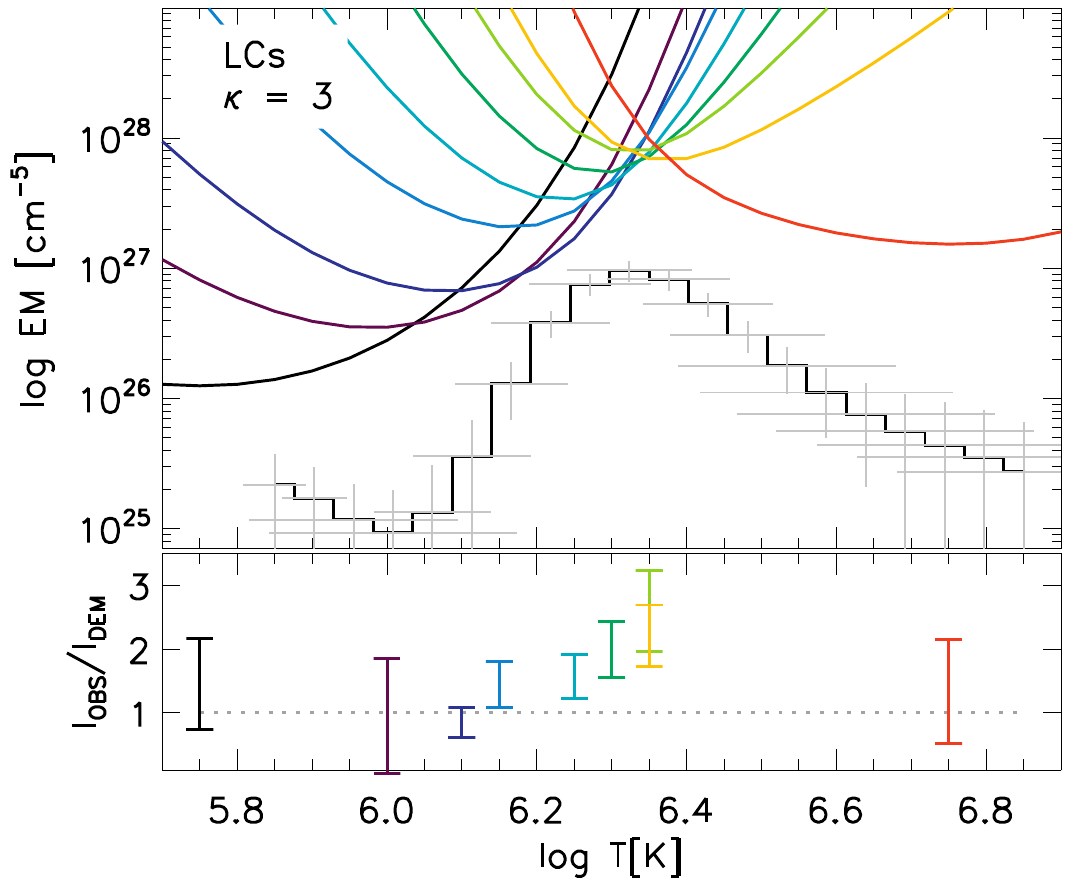}
    \\
    \includegraphics[width=6.3cm, clip,   viewport= 0 0 311 255]{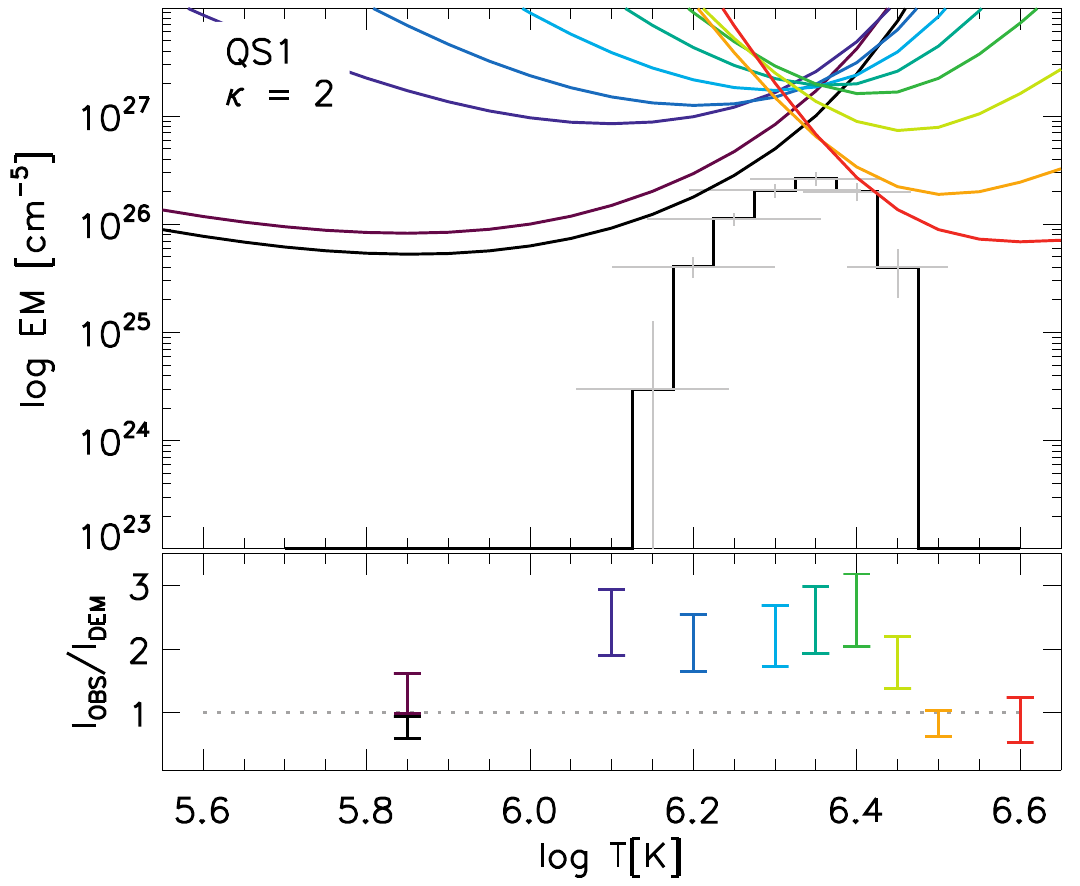}
    \includegraphics[width=5.83cm, clip,  viewport= 23 0 311 255]{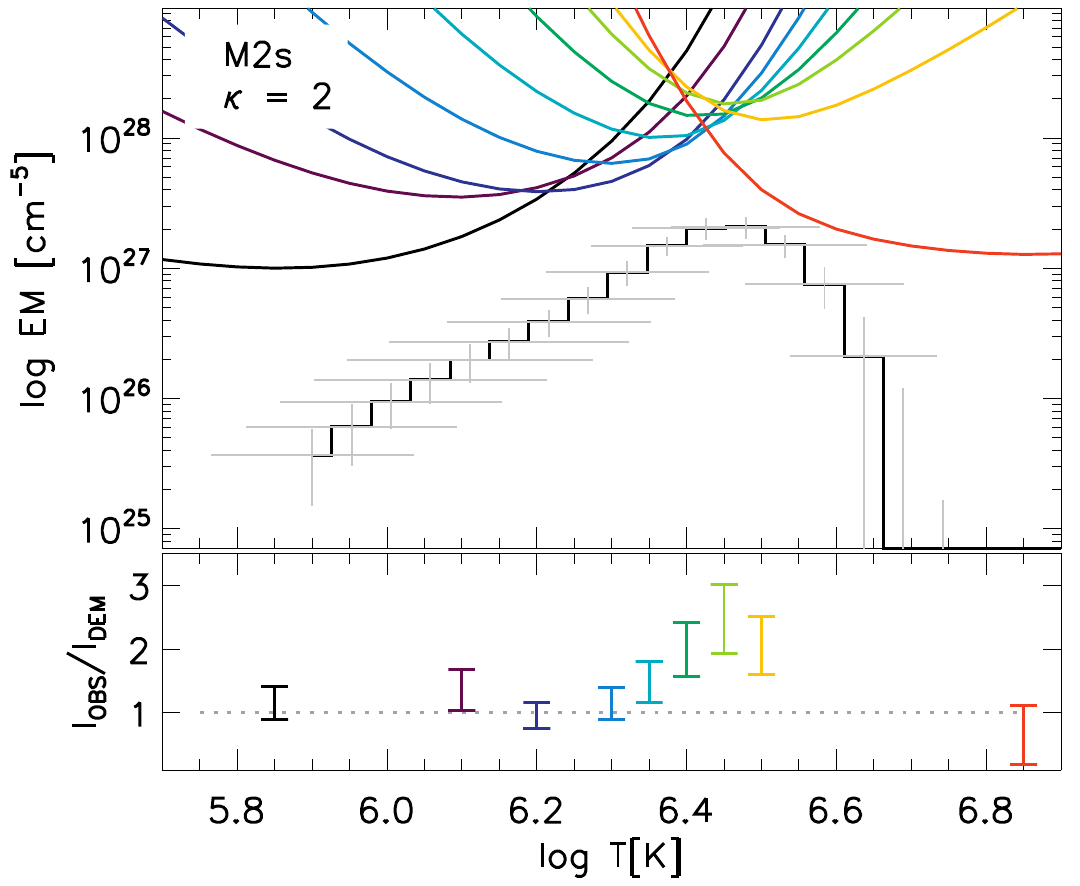}
    \includegraphics[width=5.39cm, clip,  viewport= 45 0 311 255]{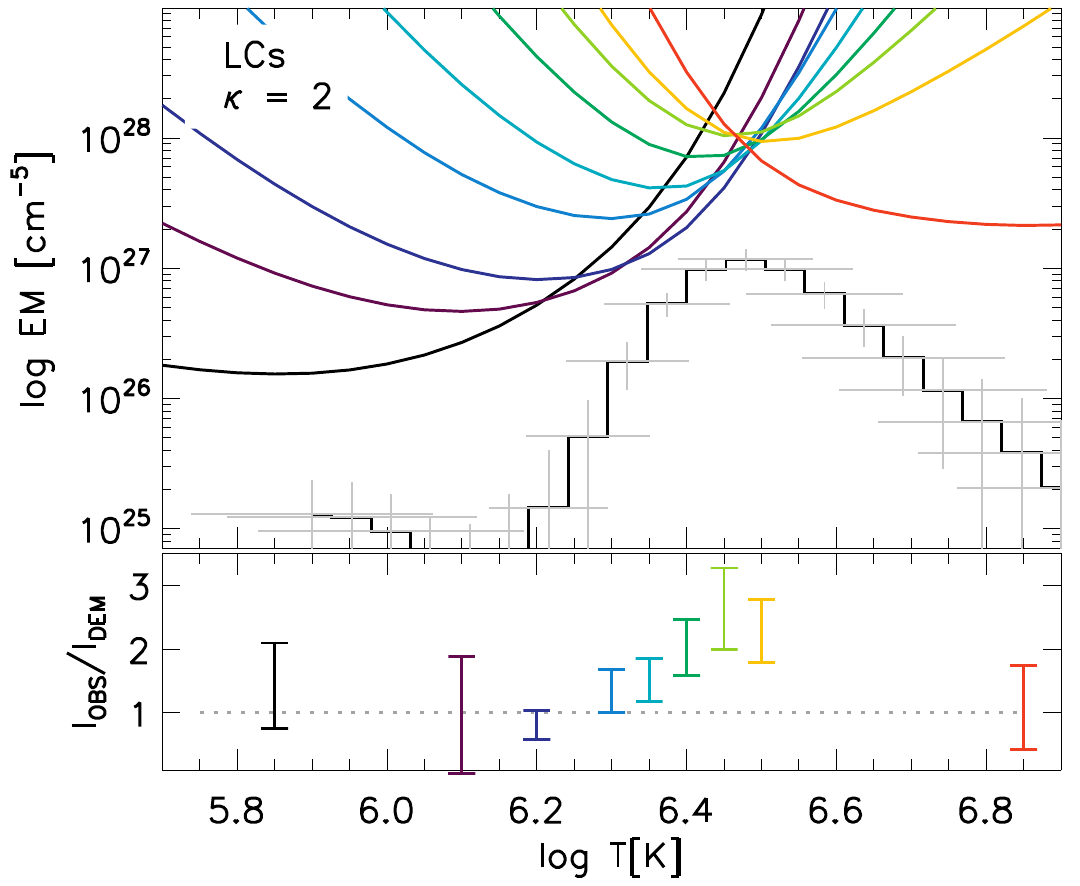}
  \caption{Emission measure distribution EM($T$) for different values of $\kappa$. The EM loci curves have been color-coded in order to distinguish between the different lines used. \textit{Left} column shows the solutions recovered for the quiet Sun QS1, \textit{middle} column for the background-subtracted moss M2s, and the \textit{right} column shows the solutions for the background-subtracted loop LCs. Below each solution, the ratios $I_\mathrm{OBS}$ / $I_\mathrm{DEM}$ for each line used for construction of DEM are shown. \label{figure_demdiag}}
\end{figure*}
\subsection{Emission measure distribution in the observed structures}
\label{sec_res_demdiag}

Recovered emission measure EM$_\kappa(T)$ curves are shown in Figure \ref{figure_demdiag}. These are shown for the {appropriate densities diagnosed} (Table \ref{tab:density_dem}), although we note that in practice, these curves are almost independent of $N_\mathrm{e}$. {The curves are plotted together with their respective errors in temperature (grey horizontal lines) and EM$_\kappa(T)$ (grey vertical lines).} Each EM$_\kappa(T)$ panel is accompanied by the ratios $I_\mathrm{OBS}$ / $I_\mathrm{DEM}$ of the DEM-predicted to the observed intensities. Due to similarities between DEMs of individual structures, we only show illustrative examples of results recovered for QS1, M2s, and LCs. The EM$_\kappa(T)$ are shown for four different values of $\kappa$\,=\,2, 3, 5, as well as the Maxwellian distribution. 

\subsubsection{Quiet Sun}

In the quiet Sun QS1, DEM for the Maxwellian distribution converged in a temperature range of log($T$\,[K])\,=\,5.7--6.4. The corresponding EM$(T)$ curve (top-left panel in Figure \ref{figure_demdiag}) peaks at log($T$\,[K])\,$\approx$\,6.15. For the $\kappa$-distributions, we found that the temperature at which the DEM$_\kappa(T)$ curves peak shift to higher temperatures, an effect that is well known \citep{mackovjak14,Dudik15}. For the case of $\kappa$\,=\,2, the EM$_\kappa(T)$ curve peaks at log($T$\,[K])\,$\approx$\,6.35, with the solution credibly converging in 7 temperature bins only (bottom left panel of Figure \ref{figure_demdiag}). Many of the EM loci curves plotted for different values of $\kappa$ intersect in one point, indicating a possible near-isothermal nature of the QS plasma if $\kappa$-distributions are taken into account. 

There are two reasons why the QS DEMs are only recovered in a narrow temperature range. First, signal observed in the quiet Sun boxes is typically lower than signal in the active region structures, which affects the credibility of DEMs \citep{hannahkontar12}. Second, convergence of QS DEMs in a narrow temperature range is expected as most of the strong lines observable there are formed in a narrow temperature range \citep[][]{landiyoung10, mackovjak14}. Indeed, almost isothermal DEMs peaking at about 1 MK are typical for the quiet Sun \citep[see, e.g.,][]{Landi03,Warren09,delzanna12,delzanna18}. Even though that the temperature range in which we recovered DEMs in the quiet Sun boxes is narrow, it is sufficient enough for our purposes of predicting intensities of lines used in Section \ref{sec_res_demeffects}.

Finally, we note that the solutions in QS2 and QS3 are very similar to QS1.

\subsubsection{Active Region}

For the moss M2s, the EM$(T)$ recovered for the Maxwellian distribution is broad and peaks at about log($T$\,[K])\,=\,6.25 (Figure \ref{figure_demdiag}). At high temperatures, it is difficult to be constrained. The \ion{Fe}{17} line at 254.9\,\AA~is used for this purpose; however, this line is weak, which results in large uncertainties of its intensities. The DEM solutions converged with high confidence in the temperature range of log($T$\,[K])\,=\,5.7--6.7. For the $\kappa$-distributions, a shift of the EM$_\kappa(T)$ peaks to higher temperatures is again evident. For $\kappa$\,=\,2, the EM$_\kappa(T)$ peaks at log($T$\,[K])\,$\approx$\,6.45, and then rapidly decreases for log($T$\,[K])$>$ 6.6. 

The EM$_\kappa(T)$ curves for the loop LCs are shown in the right column of Figure \ref{figure_demdiag}. The maxima of EM$_\kappa(T)$ are similar as for the moss M2s, but their shape differs for both high- and low-temperatures. First, the EM$_\kappa(T)$ curves for LCs contain a dip at the temperature roughly corresponding to the $T_{\text{max}}$ of \ion{Fe}{9}, in agreement with this loop being faint at these temperatures. Second, with the low intensity of the \ion{Fe}{17} 254.9\,\AA~line, it is still difficult to properly constrain the Maxwellian DEM at high temperatures using Fe lines only. The Maxwellian solution can, in some cases, even rise again at log($T$\,[K])\,$\approx$\,6.7. For this reason, in LSs (not shown), where the 254.9\AA~line was not observed at all, we exceptionally had to use the \ion{Ca}{15} 200.97\,\AA~line. Despite this, the recovered EM$_\kappa(T)$ are sufficient for predicting the line intensities (see the $I_\mathrm{OBS}$\,/\,$I_\mathrm{DEM}$ ratios). In particular, note that the high- and low-temperature limits of the EM$_\kappa(T)$ do not play a role for recovering the intensities of \ion{Fe}{11}--\ion{Fe}{12} lines, which are critical for diagnostics of $\kappa$ (see Section \ref{sec_res_demeffects}). 

In summary, our EM$_\kappa(T)$ curves are broad, indicating that both M2s and LCs can be multi-thermal \citep[cf.,][]{fletcher99,tripathi09,tripathi10,odwyer11,Dudik15}, at least if the Maxwellian distribution is considered.  We note that the behavior of EM$_\kappa(T)$ with $\kappa$ for M2s is similar to \citet{mackovjak14}: The curves for low $\kappa$ are similar, only shifted to higher temperatures. This is not true for LCs, where the low-$T$ shoulder of the EM$_\kappa(T)$ becomes progressively less steep for low $\kappa$, as many of the EM-loci curves (\ion{Fe}{10}--\ion{Fe}{17}) nearly cross at the same point for $\kappa$\,=
\,2--3, possibly indicating plasma close to isothermality for such low values of $\kappa$.

\begin{figure*}[h]
  \centering
    
    \includegraphics[width=8cm, clip,   viewport= 0 0 481 460]{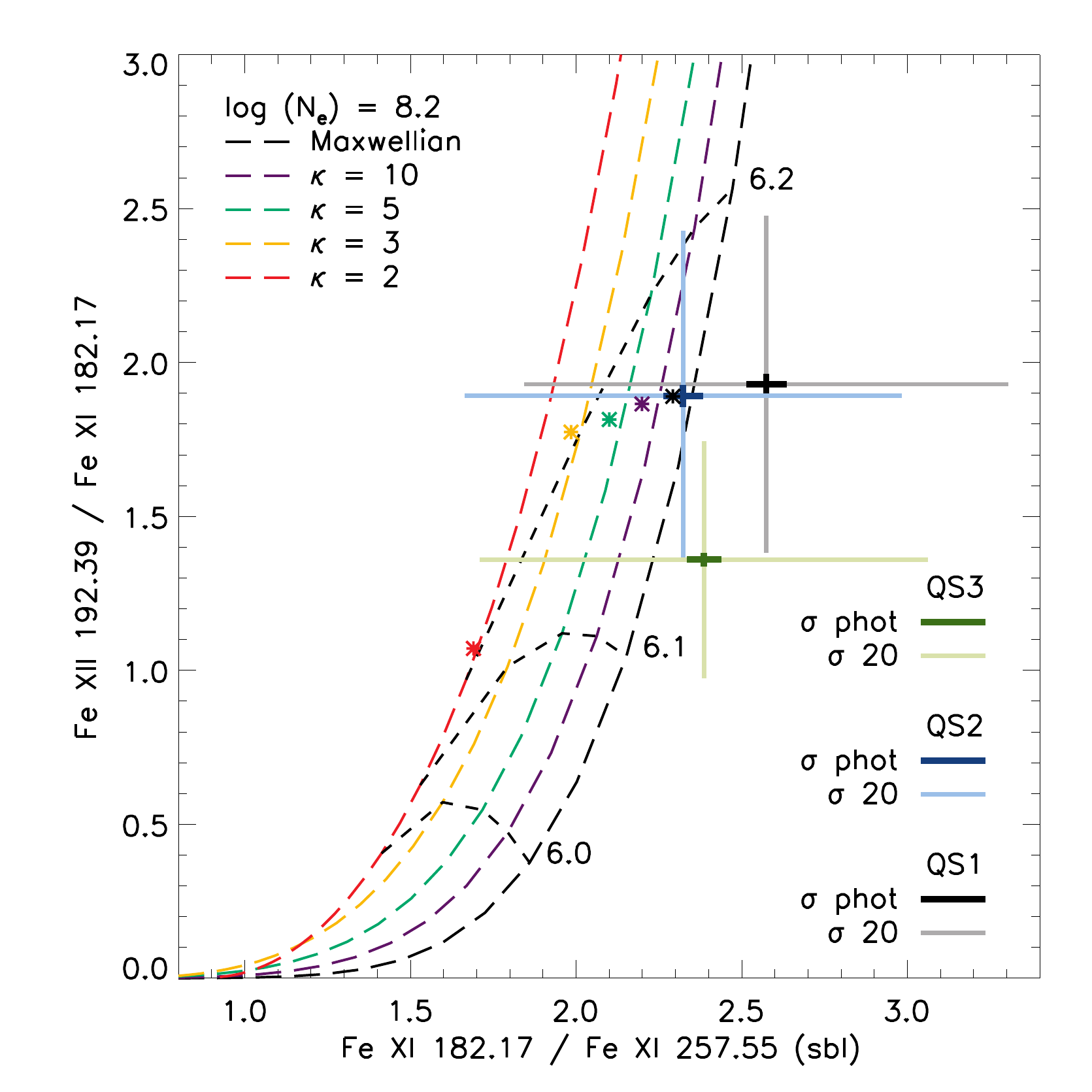}
    \includegraphics[width=8cm, clip,   viewport= 0 0 481 460]{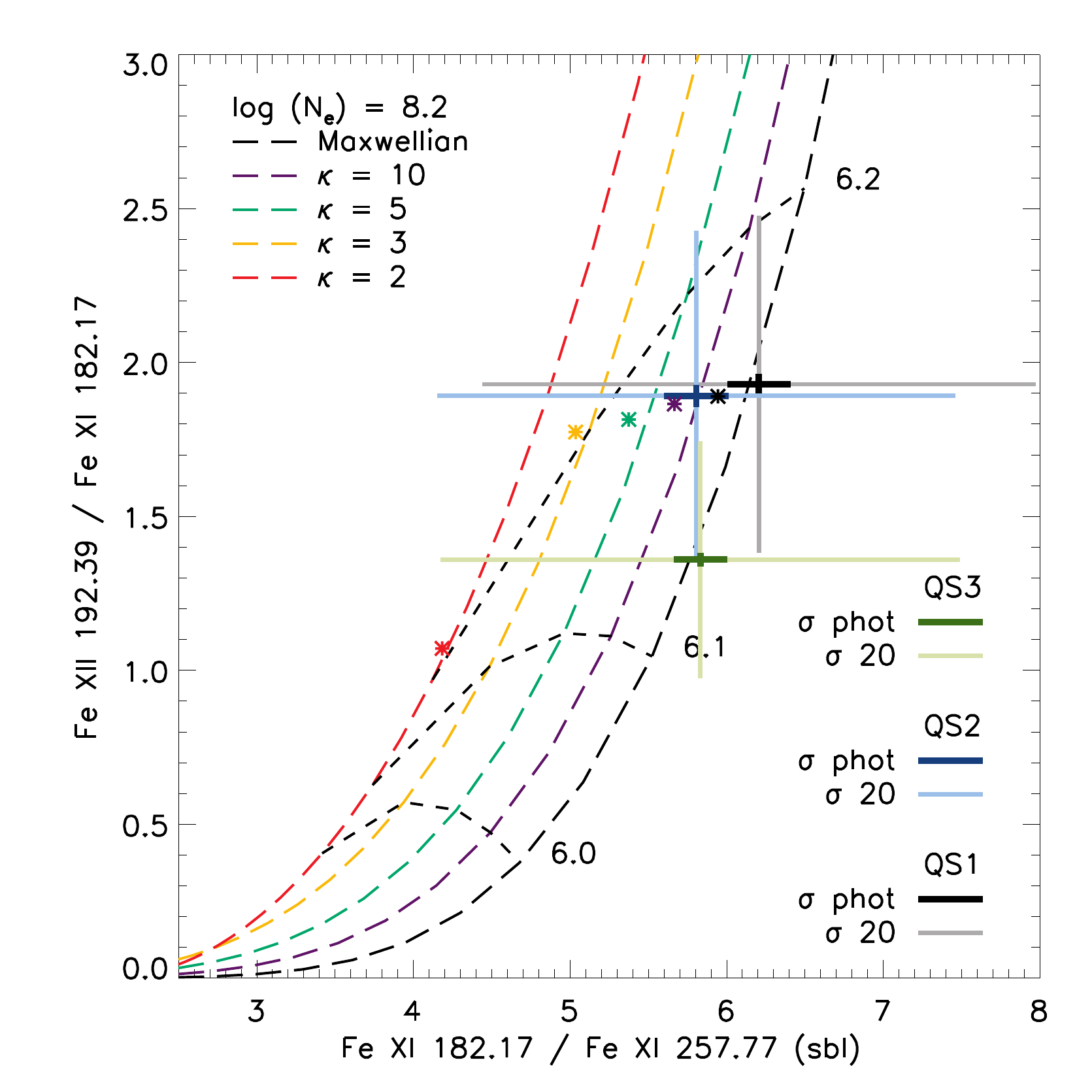}
    \\
    \includegraphics[width=8cm, clip,   viewport= 0 0 481 460]{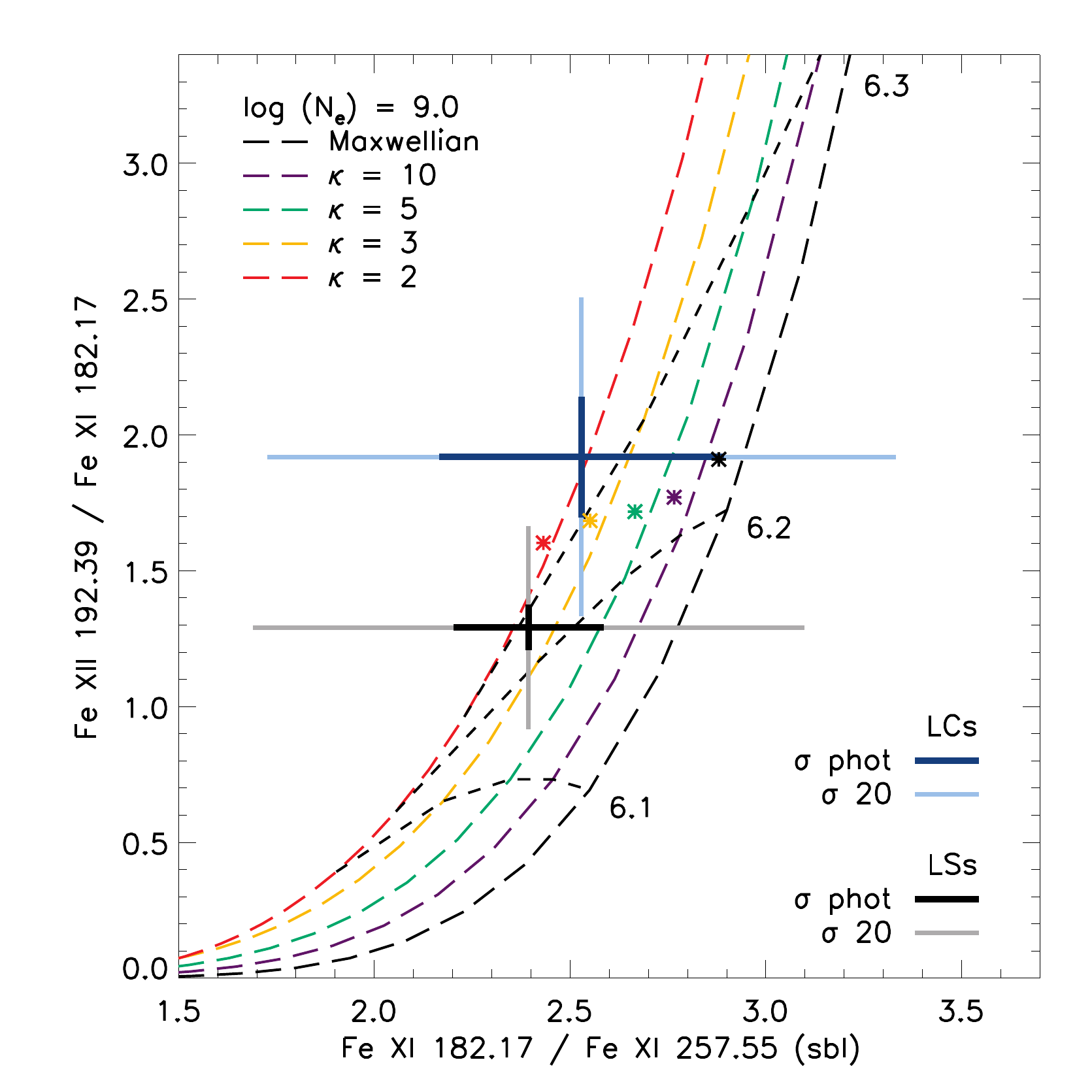}
    \includegraphics[width=8cm, clip,   viewport= 0 0 481 460]{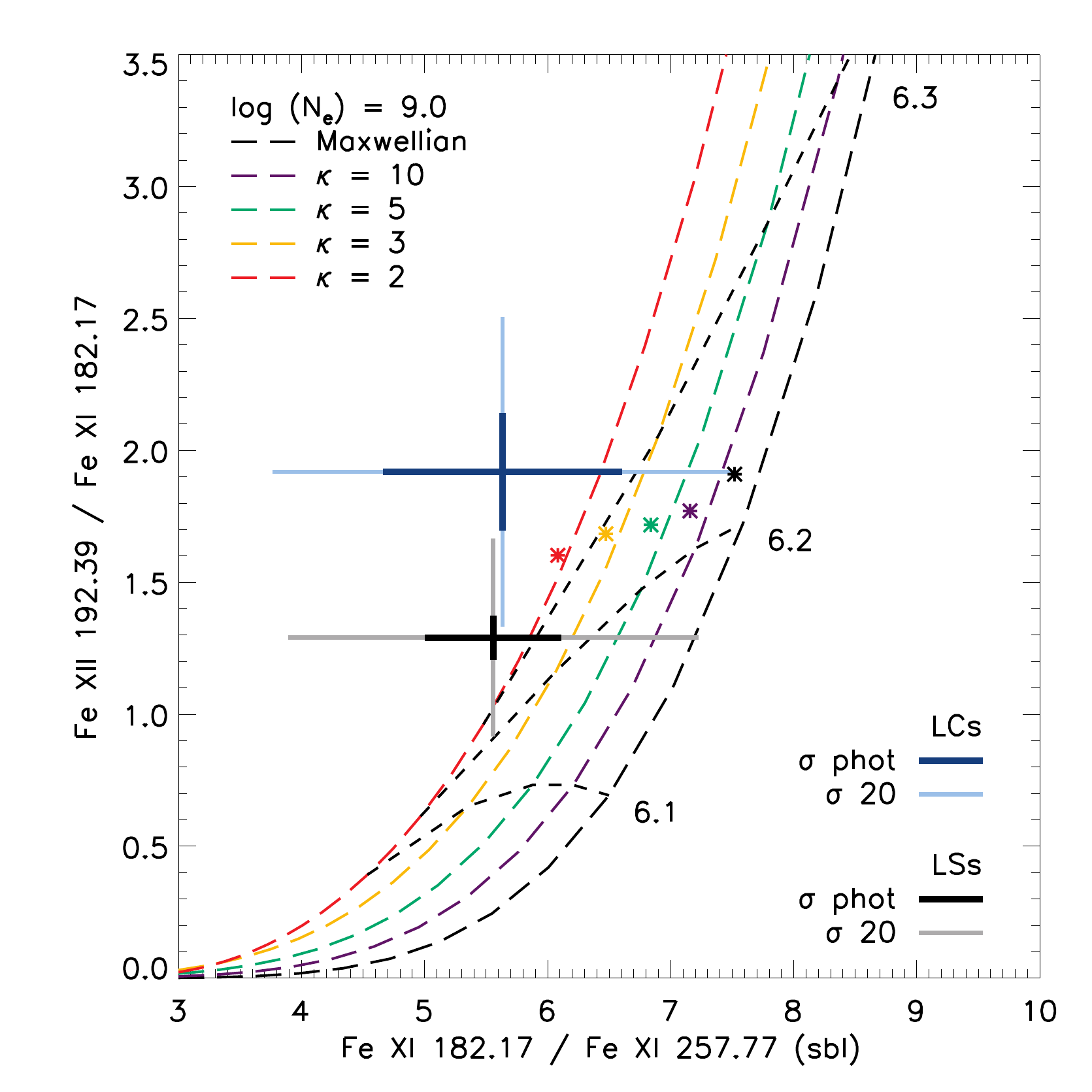}
      \\
    \includegraphics[width=8cm, clip,   viewport= 0 0 481 460]{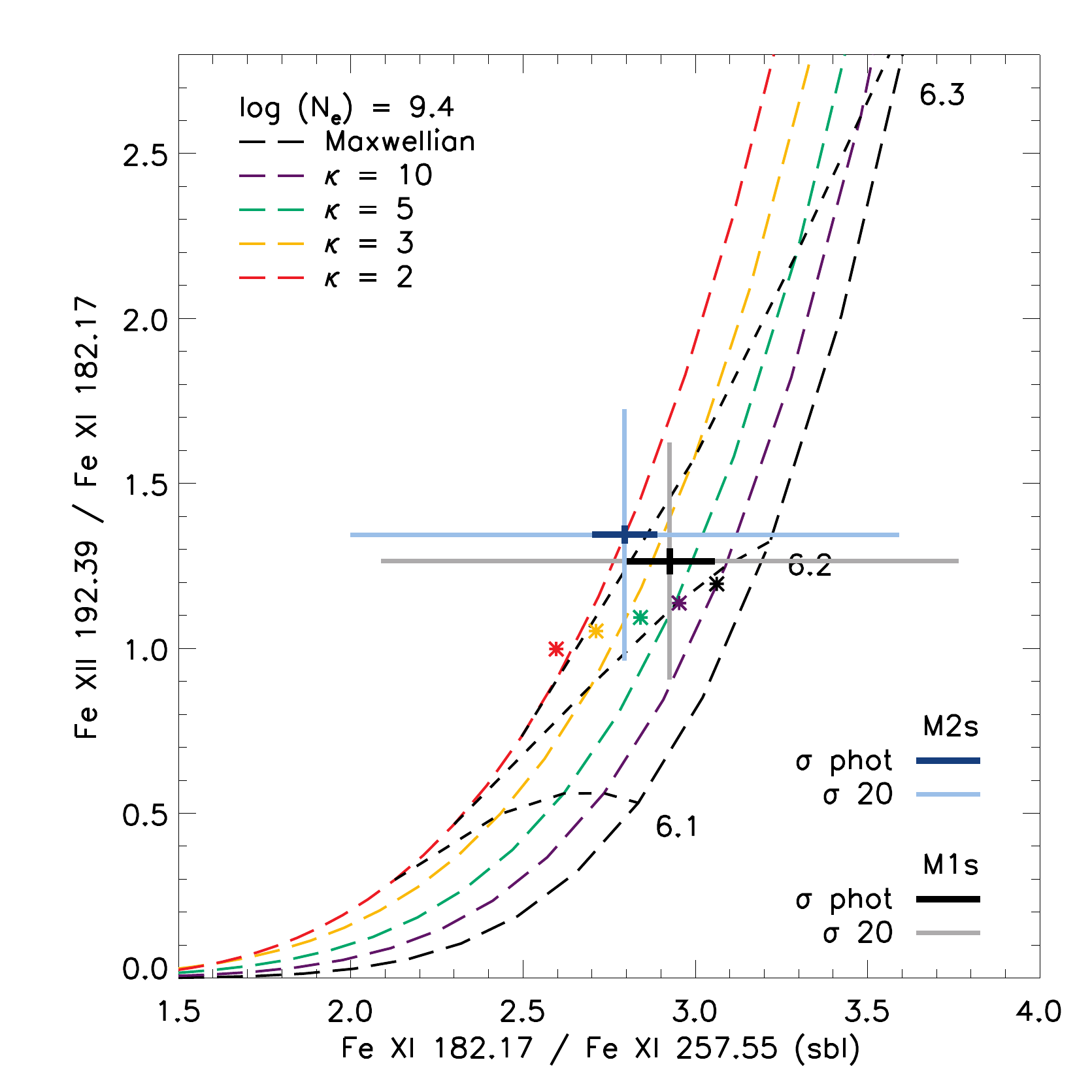}
    \includegraphics[width=8cm, clip,   viewport= 0 0 481 460]{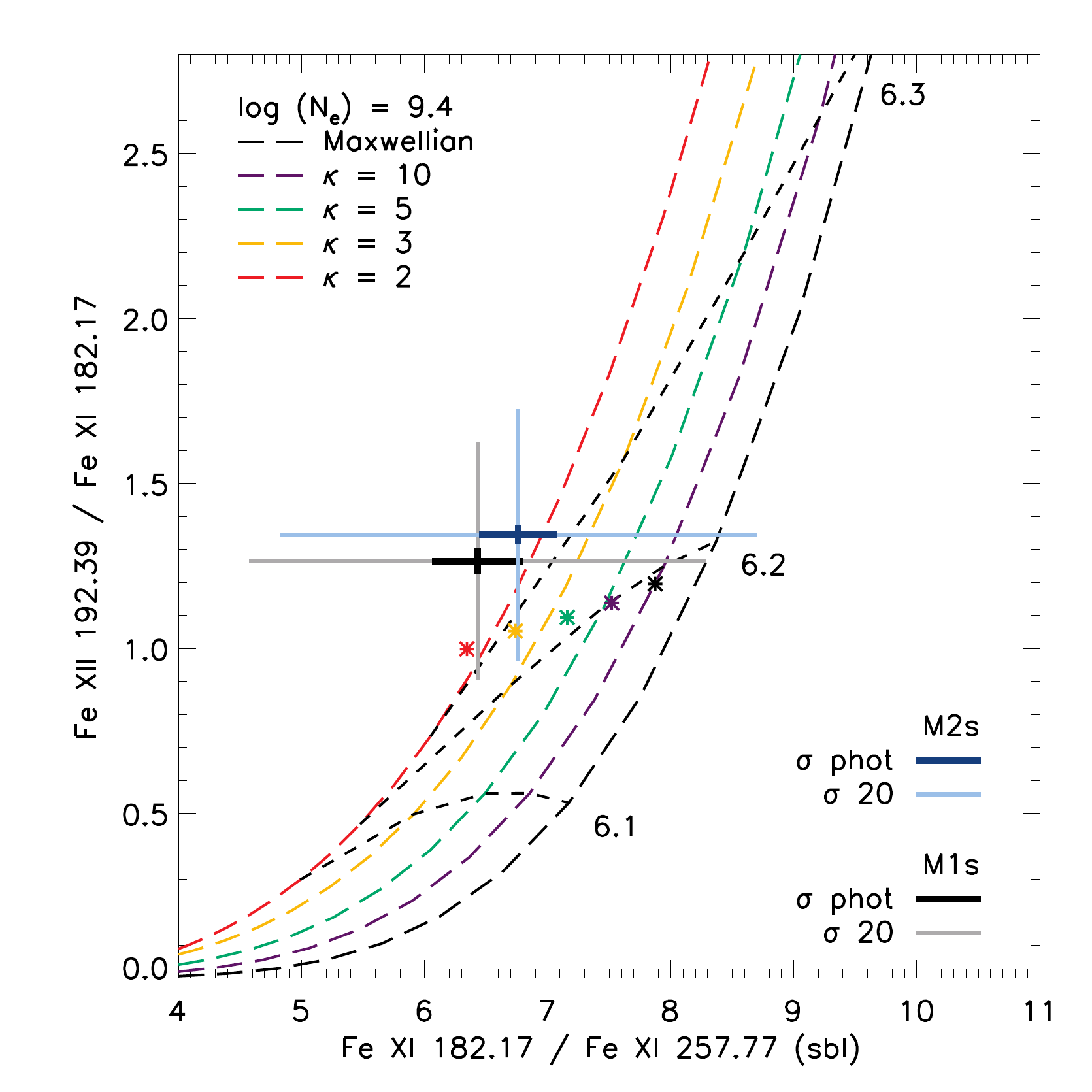}
      \\

  \caption{Ratio-ratio diagrams used for simultaneous diagnostics of $\kappa$ and $T$ plotted for densities measured apriori. Dashed colored lines are the theoretical ratios, which are intersected by the isotherms (black short-dashed lines). The blue, black, and green crosses represent the observed ratios with their $\sigma_{\text{phot}}$ and  $\sigma_{\text{20}}$ uncertainties. Colored asterisks are the ratios predicted from DEMs produced for different $\kappa$.}
  \label{figure_kappadiag}
\end{figure*}
\subsection{Diagnostics of $\kappa$ and $T$  \label{sec_res_kappadiag}}

The ratio-ratio diagrams for diagnostics of $\kappa$ are shown in Figure \ref{figure_kappadiag}. The dashed curves are the theoretical ratios, with colors denoting the values of $\kappa$. Black curves stand for the Maxwellian distribution, while the violet, green, orange, and red are for $\kappa$\,=\,10, 5, 3, and 2, respectively. Pairs of ratio-ratio diagrams are shown. The left panels use the \ion{Fe}{11} 182.17\,\AA\,/\,257.55\,\AA~ratio, while the right panels use the \ion{Fe}{11} 182.17\,\AA\,/\,257.77\,\AA. On each panel, these curves are shown for the appropriate density $N_\mathrm{e}$ (Table \ref{tab:density_dem}, Section \ref{sec_nediag}). For simplicity, we assumed that the densities between the individual areas of the quiet Sun, moss, and loops do not differ. This is justified, as these do not differ for more than 0.1 dex in log $N_\mathrm{e}$. Black dashed curves intersecting different values of $\kappa$ are the isotherms, with the corresponding temperatures shown in units of log($T$\,[K]).

The observed ratios are for each structure shown by crosses, whose size depends on the uncertainties. The thick inner crosses denote the photon noise uncertainties $\sigma_{\text{phot}}$. For completeness, the thin large crosses correspond to adding the 20\% calibration uncertainty $\sigma_{\text{20}}$.

The diagnostic diagrams for the quiet Sun (top row of Figure \ref{figure_kappadiag}) are shown for log($N_\mathrm{e}$\,[cm$^{-3}$])\,=\,8.2. The observed ratios for QS1--3 are clustered near the Maxwellian curve, within one or several times the photon noise uncertainty. In the diagnostic diagram using the \ion{Fe}{11} 257.55\AA~line (left), the observed ratios are located either on the right-hand side of the theoretical ratios, or intersect the theoretical ratios calculated for the Maxwellian distribution. The ratios involving the \ion{Fe}{11} 257.77\AA~line (right) are slightly shifted toward the left, still indicating either the Maxwellian or the $\kappa$\,=\,10 distribution. The observed ratios also indicate temperatures log($T$\,[K])\,=\,6.1--6.2. These temperatures correspond well to the peak temperatures of DEMs recovered using the contribution functions calculated assuming the Maxwellian distribution. 

The ratio-ratio diagrams for the loops LCs and LSs are shown in the middle row of Figure \ref{figure_kappadiag}, plotted for the density of log($N_\mathrm{e}$\,[cm$^{-3}$])\,=\,9.0. Both ratios indicate $\kappa$\,$\leq$\,2 distribution, but the $\sigma_{\text{phot}}$ uncertainties are large. {The $\sigma_{\text{phot}}$ uncertainties result in different constraints on $\kappa$ from different ratios. These are summarized in Table \ref{tab_kappa}, where we list the results of diagnostics from multiple line ratios, as indicated within the $\sigma_{\text{phot}}$ uncertainty.} Note that the diagram involving the \ion{Fe}{11} 257.77\,\AA~line (Figure \ref{figure_kappadiag} middle right) indicates strongly non-Maxwellian plasma with $\kappa \leq$ 2, which is in accordance with results of \citet{Dudik15}. 

The results obtained in the coronal moss boxes are similar. The ratio-ratio diagrams plotted for the density of log($N_\mathrm{e}$ [cm$^{-3}$])\,=\,9.4 are shown in the bottom row of Figure \ref{figure_kappadiag}. Again, non-Maxwellian plasma is indicated, with {$\kappa$\,$\lessapprox$\,3} determined using the 257.55\AA~line, while the 257.77\AA~line suggests $\kappa\,\lessapprox$\,2. Note that for M1s, the densities diagnosed were 9.5 (Table \ref{tab:density_dem}). For such densities, the theoretical ratios are shifted {toward the right-hand side, leading to slightly} higher departures from the Maxwellian distribution. Within the $\sigma_{\text{phot}}$ uncertainty, the observed ratios typically intersect the isotherms corresponding to the temperatures of log ($T$ [K]) = 6.2, 6.3, or more (not shown). These again correspond to the peak temperatures of the EM($T$) curves of both LCs and M2s in DEMs obtained for the $\kappa = 3, 5$, and the Maxwellian distributions. Note that as the emission in the active region structures is distributed over a wide range of temperatures, this {temperature diagnostic is indicative only and the effects of DEM will be discussed shortly}.

{In} order to supplement our diagnostics of $\kappa$, we also constructed the ratio-ratio diagrams in which we used the \ion{Fe}{12} 186.89\AA~line instead of the \ion{Fe}{12} 192.39\AA~line. Since the ratio-ratio diagrams are comparable to those presented in Figure \ref{figure_kappadiag}, we do not show them here. To supplement our diagnostics, we have however included them {in Table \ref{tab_kappa}.}

{Finally, as is apparent in Figure \ref{figure_kappadiag}, the thin crosses, standing for the $\sigma_{\text{20}}$ uncertainty, cross all of the theoretical ratios sensitive to $\kappa$ as well as multiple isotherms. Strictly speaking, no constraints on $\kappa$ can be obtained from error analysis alone. The only indication that the distribution in AR is different to that of the QS would then be that the measured \ion{Fe}{11} ratios are observed to be nearly same in both AR and QS. This result holds despite \textit{an order of magnitude} difference in electron density. Therefore, if the distribution would be Maxwellian in both the AR and QS, the observed ratios should be different. We however note that the data we use here were acquired shortly after the start of the mission, indicating that the in-flight calibration and the associated calibration uncertainty should not play a large role in diagnostics.}

\begin{deluxetable*}{cccccc}[t]
\tablecaption{Results of {diagnostics of $\kappa$ as} indicated by the ratio-ratio diagrams within $\sigma_{\text{phot}}$. \label{tab_kappa}}
\tablecolumns{6}
\tablenum{3}
\tablewidth{0pt}
\tablehead{
\colhead{Lines [\AA~]}  & QS & M1s & M2s & LCs & LSs} 
\startdata
182.17, 257.55, 192.39  & Maxwellian	& $\kappa \leq 3$ & $\kappa = 2$ 	& $\kappa \leq 10$ &	$\kappa \leq 5 $ \\ 
182.17, 257.77, 192.39  & $\kappa \geq 10$ & $\kappa < 2$    & $\kappa \leq 2$ & $\kappa \leq 2 $   &	$\kappa \leq 2$	 \\ 
182.17, 257.55, 186.89  & Maxwellian	& $\kappa \leq 3$ & $\kappa \leq 2$	& $\kappa \leq 5$  	 &	$\kappa \leq 3$	 \\ 
182.17, 257.77, 186.89  & $\kappa \geq 10$	& $\kappa < 2$    & $\kappa < 2$ 	& $\kappa < 2$ 	 	 &	$\kappa < 2$ 	 \\ 
\enddata
\end{deluxetable*}

\subsection{Effects of DEM$_\kappa(T)$ on diagnostics of $\kappa$ \label{sec_res_demeffects}}

The ratio-ratio diagrams provide diagnosics of $T$ and $\kappa$ if analyzed plasma is isothermal. As we however reported in Section \ref{sec_res_demdiag}, in many cases the observed plasma is multithermal, with the emission measures EM$_\kappa(T)$ shown in Figure \ref{figure_demdiag}. Therefore, we investigated the influence of the multi-thermality of plasma on the diagnostics of $\kappa$. To do that, we used the EM$_\kappa(T)$ obtained to predict the intensity ratios as a function of $\kappa$. These are shown as colored asterisks on each panel of the Figure \ref{figure_kappadiag} to facilitate comparisons with the observed ratios.

For the QS1, the predicted ratios converge the observed one as the parameter $\kappa$ increases. This conforms to the result that the quiet Sun is nearly Maxwellian, while the EM-predicted ratios for $\kappa$\,=\,2 (red asterisks) are farthest from the observed ones.

In case of LCs, the EM-predicted ratios confirm the results of diagnostics of $\kappa$. If the 257.55\,\AA~line is used, the observed ratios are closest to the EM-predicted ratios for $\kappa$\,$\leq$\,5. Conversely, if the 257.77\,\AA~line is used, the closest match is found for $\kappa$\,$\leq$\,2.

Contrary to these, the results obtained for M2s are ambiguous. For both the 257.55\,\AA~and the 257.77\,\AA~lines, none of the predicted ratios converges on the observed one. Moreover, in the ratio-ratio diagram involving the 257.77\,\AA~line, the observed ratio is far from the EM-predicted ones. The origin of this inconsistency is not known. Perhaps the electron distribution in the moss is not a $\kappa$-distribution. Alternatively, opacity effects due to unresolved chromospheric absorbing structures \citep{depontieu09} could explain this departure.
\begin{figure}[h]
  \centering    
    \includegraphics[width=8.0cm, clip,   viewport= 0 0 481 481]{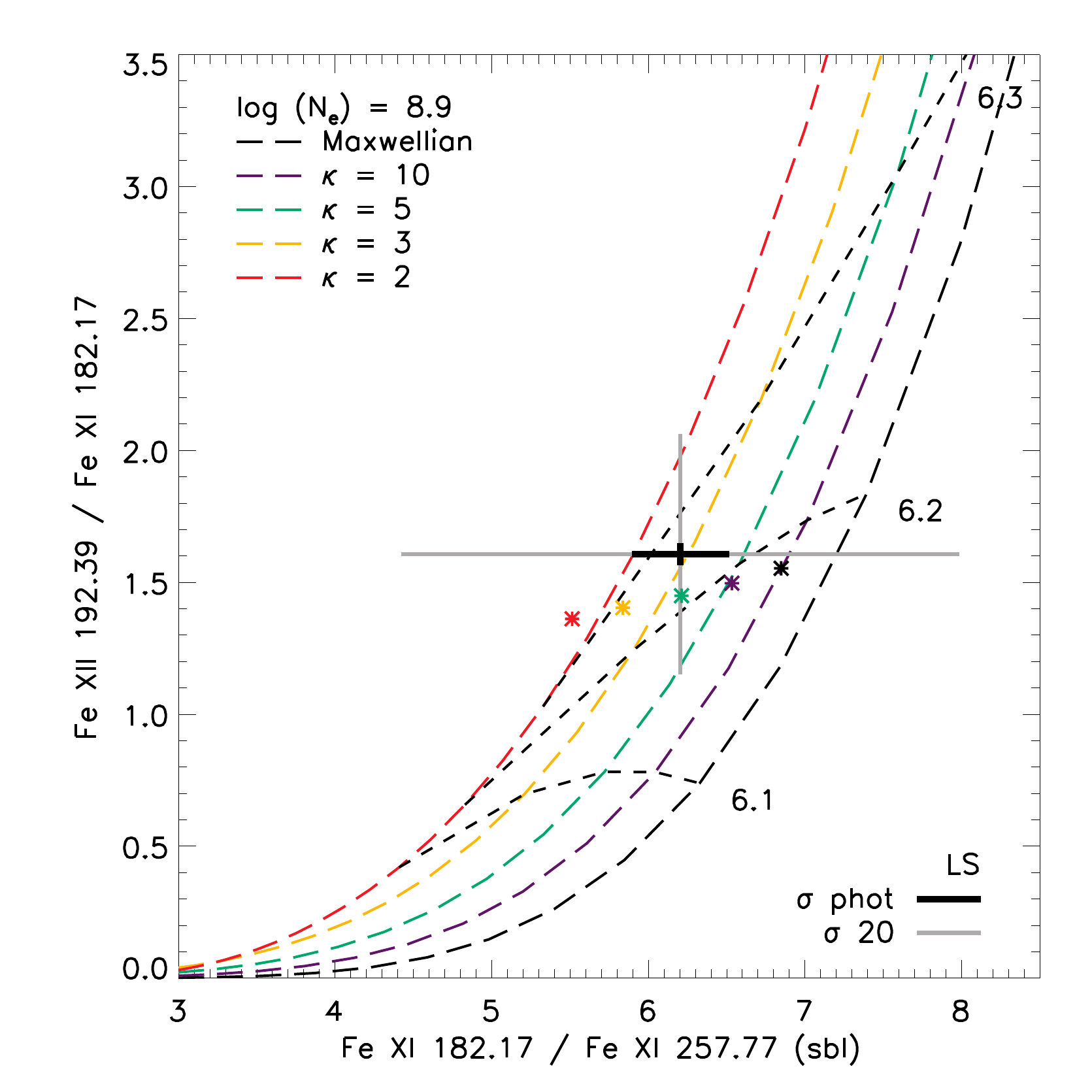}
  \caption{Ratio-ratio diagram used for simultaneous diagnostics of
    $\kappa$ and $T$ in the non background-subtracted loop LS plotted for densities which were adopted using DEM-weighted theoretical ratios for the $\kappa$\,=\,2 distribution. Dashed colored lines are the theoretical ratios, which are intersected by the isotherms (black short-dashed lines). The black cross represents the observed ratios with their $\sigma_{\text{phot}}$ and  $\sigma_{\text{20}}$ uncertainties. Colored asterisks are the ratios predicted from DEMs produced for different $\kappa$. \label{figure_nonsub}}
\end{figure}
\section{Discussion}
\label{sec_discussion}

\subsection{Effects of the Background Subtraction on Diagnostics}
\label{sec_discussion_background}

Since the results of the diagnostics for the active region can be dependent on the choice of background, we investigated the effect of background subtraction on the results of diagnostics. As an example, we used the intensities observed in the loop LS, but did not perform the background subtraction. Plasma diagnostics were performed again using the same iterative scheme (Section \ref{sec_iterations}). The DEM$_\kappa(T)$ curves obtained with $\chi^2$ = 3 for all values of $\kappa$ are smooth and have broad peaks. The DEM-weighted electron densities obtained are log($N_\mathrm{e,0}$ [cm$^{-3}$]) $\approx$ 9.0 for the Maxwellian and 8.9 for the $\kappa$\,=\,2 distribution. An initial estimate of $\kappa$ using these densities led to finding non-Maxwellian plasma with $\kappa \approx$ 3, which in turn led us to adopt the lower of the two densities diagnosed therein. An example ratio-ratio diagram involving the \ion{Fe}{11} 257.77\,\AA~line is shown in Figure \ref{figure_nonsub}. As the difference between the estimated and final measured densities is only 0.1 dex in log($N_\mathrm{e,0}$ [cm$^{-3}$]), we still obtain $\kappa \approx$\,3 in LS. 

Recalling the results of diagnostics obtained using the background-subtracted intensities, we measured the density log($N_\mathrm{e,0}$ [cm$^{-3}$]) = 9.0 and $\kappa \leq$ 2. Note that the \ion{Fe}{11} ratio plotted on the \textit{X}-axis, which is dominantly sensitive to  $\kappa$, is 5.6$\pm$0.6 for the
background-subtracted (5.6$\pm$0.6) case, while if the background is not subtracted, it is 6.2$\pm$0.3. The difference between the measured values of $\kappa$ is therefore due to a combination of the background
subtraction and difference in the measured density, which both serve to decrease the value of $\kappa$ and thus lead to the diagnosis of higher departures from the Maxwellian. This result shows the importance of the background subtraction.

\subsection{Atomic Data Uncertainties}
\label{sec_discussion_atomic_data}

The atomic data are not in their state-of-the-art form, since they lack at least some of the high-energy levels. We now briefly summarize the effect of atomic data uncertainties on diagnostics of $\kappa$. A
full discussion can be found in \citet{Dudik15}, who repeated the diagnostics of $\kappa$ using older atomic datasets, corresponding to CHIANTI v7.1 \citep{Dere97,Landi13}, which have a lower number of energy levels compared to present version 9 of CHIANTI. These authors found that when these older atomic data are used, the curves in the ratio-ratio diagrams are shifted rightward, increasing the departure from the Maxwellian distribution. Furthermore, the older atomic data would lead to difficulties in density diagnostics, with the \ion{Fe}{12} density-sensitive ratio indicating inconsistent densities with respect to \ion{Fe}{13}, typically higher by about 0.5 dex \citep[see][for further details]{DelZanna12a}.

The present atomic data, corresponding to CHIANTI v9, still lack $n$\,$\geq$\,5 energy levels. As discussed by \citet{Dudik15} however, including these higher energy levels and the cascading from them is unlikely to change the results of diagnostics, as the synthetic line intensities would not increase by more than about 10\%, with details depending on the line.

\subsection{Interpretation of the Results}
\label{sec_discussion_interpretation}

The result that the quiet Sun is Maxwellian with nearly the same temperatures as seen with previous instruments \citep[e.g.,][]{Landi03} is an important one, because it indicates that both our iterative diagnostic procedure and the latest atomic data work well. Moreover, since we used data observed shortly after the start of the \textit{Hinode} mission, the degradation of the instrument could have been neglected. 

The fact that the quiet Sun is consistently Maxwellian regardless of location, while the active region structures tend to be non-Maxwellian, can have implications for the mechanism heating the corona. Since the active region spectra show departures from the Maxwellian, even after accounting for multithermal plasma, this could indicate that the coronal heating mechanism operating there accelerates particles more efficiently than in the quiet Sun. Presumably, the heating frequency in active regions is higher than in the quiet Sun, and sufficient for the plasma to remain energized and non-Maxwellian. In both these cases, the non-Maxwellian effects need to be taken into account in modelling the coronal heating and coronal loop evolution, since at $\kappa$\,=\,2, about 80\% of the kinetic energy of particles is carried by the high-energy tail \citep{oka13}. In addition, for $\kappa$\,=\,2, the coronal ions are formed at higher temperatures than for the Maxwellian \citep{dzifdudik13}, which also has consequences for loop energetics.

An alternative explanation for both the non-Maxwellianity and the multithermality of plasma is that the plasma is out of the ionization equilibrium. Such a situation can arise for example due to effects of a periodic electron beam as investigated by \citet{dzif16}. To an initially undisturbed bulk of
coronal plasma, high-energetic electrons in the periodic beam are injected. This drives the plasma out of the ionization equilibrium at all times, irrespectively of the frequency of the beam. As the plasma is out of equilibrium, it appears multithermal, and since there are energetic electrons the spectra are also non-Maxwellian. The degree of these effects is manifested in the shapes of DEMs, which authors recovered using the same method and ions as we did here, but with synthesized line intensities. DEMs presented in the bottom two rows of Figure 6 and 7 therein are very similar to those we obtained in the structures selected in the active region (Figure \ref{figure_demdiag}, middle and right columns). 

Finally, an additional effect might contribute to the observed difference in terms of $\kappa$ between the quiet Sun and active region. The total cooling time $\tau_\mathrm{cool}$ is inversely proportional to the plasma pressure as $\tau_\mathrm{cool}$\,$\sim$\,$P^{-1/6}$ \citep[Equation (A2) in][]{cargill14}. Since $P$\,$\simeq$\,$N_\mathrm{e}T$, given the low densities observed in the quiet Sun, this implies that the cooling time of the quiet Sun plasma should be longer compared to the active regions. As a consequence, if the frequencies of heating events in active regions and quiet Sun were the same, the quiet Sun plasma would also be observed as non-Maxwellian, which does not conform to our observations. Note that this comparison is only speculative, since the active region appears to be hotter than the quiet Sun, at least judging by the peaks of the DEM$_\kappa(T)$ -- an effect more pronounced if the $\kappa$\,=\,2 in the active region spectra is considered compared to the Maxwellian quiet Sun (Figure \ref{figure_demdiag}).


\section{Summary} \label{sec_conclusions}

In this manuscript we present diagnostics of the non-Maxwellian $\kappa$-distributions in an active region and quiet Sun observed by \textit{Hinode}/EIS. Our results indicate that the plasma in the quiet Sun is Maxwellian, while active region loops and moss show strong departures from the Maxwellian distribution with $\kappa$ $\lessapprox$ 3.

The method we used for diagnostics of $\kappa$ involves emission lines observed in both wavelength channels of the EIS instrument. To avoid problems with the decay of sensitivity and changes in the in-flight calibration, we used spectral atlases taken near in time and soon after the start of the mission. We chose three quiet Sun areas QS1--3 for analysis, along with two coronal moss areas M1s and M2s, as well as a closed loop LCs and a fan loop LSs. From the intensities observed in the moss and loops, we subtracted their respective backgrounds.

Since the diagnostics of $\kappa$ is contingent on the diagnostics of electron density, the density had to  be diagnosed first. However, the density-sensitive ratios of coronal lines are themselves dependent on both temperature and $\kappa$. In addition, the observed structures can be multithermal, with the DEM$_\kappa(T)$ being dependent on $\kappa$. Therefore, we developed a simple iterative procedure that progressively constrains the parameters to be diagnosed - electron density, DEM$_\kappa(T)$, and finally the non-Maxwellian parameter $\kappa$. Since the DEMs are largely insensitive of density and the density diagnostics is not strongly dependent on $\kappa$, the iterative procedure converges in two iterations. 

In addition to being Maxwellian, the quiet Sun is also nearly isothermal, with EM$(T)$ peaking at log($T$\,[K]) $\approx$ 6.2. The densities there were found to be log($N_\mathrm{e}$\,[cm$^{-3}$])\,=\,8.2--8.3. 

In active region, both LCs and LSs were found to be strongly non-Maxwellian, and both have densities of about 9.0. The corresponding EM$_{\kappa=2}$ peaks at log($T$\,[K])\,=\,6.3, with strong increase at lower temperatures and a more gradual drop at higher temperatures. The EM-loci plot for $\kappa$\,=\,2 is also closer to isothermal than the corresponding Maxwellian one. In the moss, we find densities of about 9.4--9.5. The EM$_\kappa(T)$ curves are again more steep for lower $\kappa$ than for Maxwellian, with the EM-loci curve for $\kappa$\,=\,3 indicating a near-isothermal plasma for \ion{Fe}{10}--\ion{Fe}{17}. Although the moss appears to be strongly non-Maxwellian, after accounting for multithermal effects, our results are not conclusive, as the DEM-predicted ratios are far from the observed ones. This could indicate either that the moss can not be described by a $\kappa$-distribution, or a presence of unresolved, low-lying absorbing structures.

Our results that the quiet Sun is Maxwellian could be taken as an indication that both the atomic data, the ground calibration of the instrument, as well as the iterative diagnostic procedure work well. Furthermore, these results strengthen the presumption that the active region loops can be strongly non-Maxwellian; all the more so since the \ion{Fe}{11} ratios used for diagnostics of $\kappa$ are observed to be nearly the same in the quiet Sun and the active region, independently of the electron density. In particular, the \ion{Fe}{11} 182.17,\AA\,/\,257.55\,\AA~ratio is observed to be about 2.5, while the \ion{Fe}{11} 182.17\,\AA\,/\,257.77\,\AA~ratio is about 6.0 in both quiet Sun and active region loops. Even though we performed averaging of intensities over many pixels in structures selected within both observations, and the signal-to-noise ratio of the relatively-weak \ion{Fe}{11} 257.55\,\AA~and 257.77\,\AA~lines is high, the diagnostics of $\kappa$ presented here is severely limited by the 20\% calibration uncertainty of the instrument, which leads to the observed ratios having larger uncertainties than the spread of the ratio-ratio curves for diagnostics of $\kappa$. Nevertheless, since both the quiet Sun and the active region were observed close in time, {we are convinced that our results are not influenced by the calibration issues (and their uncertainties)}.


\acknowledgments
This work was supported by the Charles University, project GA UK 1130218. J.L., J.D., and E. Dz. acknowledge support from Grant No. 18-09072S of the Grant Agency of the Czech Republic, as well as institutional support RVO:67985815 from the Czech Academy of Sciences. J.D., J.L., G.D.Z., and H.E.M. acknowledge support from STFC
(UK) and the Royal Society via the Newton Fellowships Alumni Programme. \textit{Hinode} is a Japanese mission developed and launched by ISAS/JAXA, collaborating with NAOJ as a domestic partner, NASA and STFC (UK) as international partners. Scientific operation of the \textit{Hinode} mission is conducted by the \textit{Hinode} science team organized at ISAS/JAXA. This team mainly consists of scientists from institutes in the partner countries. Support for the post-launch operation is provided by JAXA and NAOJ (Japan), STFC (U.K.), NASA (U.S.A.), ESA, and NSC (Norway). 
 

\bibliographystyle{aasjournal}
\bibliography{kappa_arqs}

\appendix
\section{Line Fitting and Intensities}
\label{sec_appendix}

The intensities of the observed spectral lines used for diagnostic purposes were obtained via fitting of the observed spectra with Gaussian fits. The line profiles were averaged in the structures selected in Sections \ref{sec_obs_ar} and \ref{sec_obs_qs}. We fitted the spectra manually using the fitting routine \texttt{xcfit} to be able to control and constrain the parameters of fits of line blends, such as the number of Gaussians, their widths, positions, maxima of amplitudes, but also the level of the continuum. Fitted intensities of spectral lines further used in this work are listed in Table \ref{tab_fit_int_ar}. We fitted the averaged line profiles within each of the box of interest including coronal background. Background subtraction was performed by subtracting the fitted background intensities.


\subsection{Lines used for diagnostics of $N_{\text{e}}$ and $\kappa$}

\subsubsection{\ion{Si}{10}}

In coronal conditions, the \ion{Si}{10} 258.37\,\AA~and 261.06\,\AA~lines are neither blended or self-blended. The 258.37\,\AA~line was fitted using a single Gaussian with reduced {chi}-squared (hereafter, $\chi_\mathrm{red}^2$) of $\approx$ 4. Major contribution to the residuals originated in wings of the line, which were not well fitted by the Gaussian. We attempted to improve this fit by adding an additional low and broad Gaussian to fit the wings separately. The highest difference between the intensities obtained using one- and two-Gaussian fits were about 4\%, which is by a factor of 5 lower than the calibration uncertainty of the instrument. We note that we were unable to fit the wings of some of the other lines which fitting is described in the following sections. As the fitting of their wings using additional Gaussian led to similar results as in the case of the \ion{Si}{10} 258.37\,\AA~ line, we for simplicity used single-Gaussian fits only, were applicable. In spectra averaged in the active region structures we also found an unknown weak line with centroid at $\approx$ 258.2\,\AA, which was fitted using one Gaussian. The weaker 261.06\,\AA~line was fitted using a single Gaussian with $\chi_\mathrm{red}^2$ of the fit $\approx$ 2. 

\subsubsection{\ion{Fe}{11}}

The \ion{Fe}{11} 182.17\,\AA, 257.55\,\AA, and 257.77\,\AA~lines were in this work used for diagnostics of $\kappa$. The 182.17\,\AA~is not blended and was fitted with a single Gaussian. In the long-wavelength channel of EIS, both the 257.55\,\AA~and 257.77\,\AA~lines are located close one to another in neighborhood of multiple spectral lines, such as \ion{Si}{10} 257.2\,\AA, \ion{Fe}{10} doublet at 256.26\,\AA, \ion{Fe}{14} 257.4\,\AA, and \ion{Fe}{11} 257.9\,\AA. Both lines were therefore fitted within one spectral window, broad enough to apply multiple single-Gaussian fits of the neighboring lines and constrain the value of the continuum. Single-Gaussian fits were also sufficient for the \ion{Fe}{11} 257.55\,\AA~line, which is self-blended with three additional transitions at 257.54\,\AA, 257.55\,\AA, and 257.56\,\AA~\citep{dudik14}, i.e., below EIS wavelength resolution. Single-Gaussian fits were also applied for the 257.77\,\AA~line. The weak self-blend at 257.73 was not discerned in the spectra.

\subsubsection{\ion{Fe}{12}}

The strong lines of \ion{Fe}{12} are found close to the peak of the effective area of the short-wavelength channel of EIS \citep[see e.g.,][]{delzanna13} and are commonly used for diagnostics of density. The 186.89\,\AA~line consists of three self-blending transitions at 186.86\,\AA, 186.89\,\AA, and 186.93\,\AA. The line is also blended in both wings with the \ion{Si}{11} 186.84\,\AA~line \citep{young09} and with an unknown line at $\approx$ 186.98\,\AA. The latter resulted in pronounced red wing in all of the investigated areas and required fitting with an additional Gaussian. The strong 195.12\,\AA~line is self-blended with 195.18\,\AA~line, which intensity is $<$ 10 \% of the stronger line for log($N_{\text{e}}$ [cm$^{-3}$]) $<$ 10 \citep{young09}. Unfortunately, even in structures in which this blend was taken into account, we were not able to fit this line with $\chi_\mathrm{red}^2$ lower than 20, reaching up to $\approx$ 100 in some cases. Again, significant contribution to the residuals originated in wings of the line, for which the Gaussian fit is not adequate. The situation has repeated itself in case of another strong \ion{Fe}{12} 193.51\,\AA~line. 
We finally opted to use the 192.39\,\AA~line, which is neither blended, nor self-blended, and we managed to fit it with $\chi_\mathrm{red}^2$\,=\,2--17, depending on the analyzed structure. 

\subsubsection{\ion{Fe}{13}}

We used the well-known lines at 196.53\,\AA, 202.04\,\AA, and 203.83\,\AA. The 196.53\,\AA~line was in both the AR and QS found to be weaker than its self-blended companion, the \ion{Fe}{12} 196.6\,\AA~line. However, the lines are well-separated and can easily be fitted each with a single Gaussian. In all of the analyzed structures, we obtained fits of this spectral window with $\chi_\mathrm{red}^2$ close to 1. The 202.04\,\AA~line is self-blended with the 202.0\,\AA~line, but can be fitted using a single Gaussian. $\chi_\mathrm{red}^2$s of its fits were relatively-high, reaching up to $\approx$ 20 based on the analysed structure. High residuals were again found in the wings of the line. The 203.83\,\AA~line is a complex self-blend composed of five transitions at 203.77\,\AA, 203.8\,\AA, 203.81\,\AA, 203.83\,\AA, and 203.84\,\AA~\citep{young09}. Moreover, the blue wing of the line is blended by the \ion{Fe}{12} 203.73\,\AA~line which intensity reaches 15--20 \% of its stronger companion. Two Gaussians were needed to fit this multiplet and $\chi_\mathrm{red}^2$ of the recovered fits were found to be around 15 for the coronal moss and close to 1 in the quiet Sun, loops, and background.

\subsection{Lines used in DEM analysis } \label{appendix_dem_lines}

\subsubsection{Low $T$ range}

Both in the quiet Sun and the active region structures, the DEMs were at log($T$\,[K]) $\approx$ 5.75 constrained using \ion{Fe}{8} 185.21\,\AA~and 194.66\,\AA~lines. The 185.21\,\AA~line is blended by the \ion{Ni}{16} 185.23\,\AA, which intensity in synthetic spectra reaches up to 18\% intensity of the 185.21\,\AA~line. As this contribution is within the calibration uncertainty of EIS, we did not remove it. The weaker 194.66\,\AA~line is weakly blended with the \ion{Fe}{12} 194.61\,\AA~line. This line is also close to the \ion{Ni}{12} 194.82\,\AA~and \ion{Fe}{12} 194.90\,\AA~lines, which are within the blue wing of the strong \ion{Fe}{12} 195.12\,\AA~line. After adding a Gaussian to all the lines in this spectral window, we obtained good fits with $\chi_\mathrm{red}^2 \rightarrow $ 2 in some cases. In both loops LS and LC, where the \ion{Fe}{8} lines were weak, the fitting resulted in high $\sigma_{\text{phot}}$ uncertainties comparable to the 20\% calibration uncertainty of EIS. Concerning the weak blends of both \ion{Fe}{8} lines, note that the positions of the minima of their EM loci curves (see Q1 DEMs in Figure \ref{figure_demdiag}) are similar. Therefore, even though we did not exclude the contributions of the blending lines, both lines served equally good as the lower-temperature constraints of DEMs. At temperatures of log($T$\,[K])\,$\approx$\,6.0 we used the \ion{Fe}{9} 197.86\,\AA~and \ion{Fe}{10} 184.54\,\AA~lines. In the active region spectra, the \ion{Fe}{9} 197.86\,\AA~was found to be surrounded with multiple weaker lines, such as \ion{Fe}{8} 197.36\,\AA, \ion{Fe}{13} 197.43\,\AA, and \ion{Ni}{11} 198.39\,\AA, all of which required adding an additional Gaussian. We also found a weak unknown blend at $\approx$ 197.7\,\AA. Despite the complexity of the fit of the \ion{Fe}{9} 197.86\,\AA~line, we obtained $\chi_\mathrm{red}^2$ $<$ 10 in all areas within the active region. In the quiet Sun spectra, this line was only accompanied with an unknown weak line at $\approx$ 198.09\,\AA. The \ion{Fe}{10} 184.54\,\AA~line is in the active regions blended with the \ion{Ar}{11} 185.52\,\AA~line, but the contribution of this blend reaches only about 1\% for log($N_{\text{e}}$ [cm$^{-3}$]) $\approx$ 9.5 and vanishes for lower densities. This line was in all active region structures fitted using a single Gaussian with $\chi_\mathrm{red}^2$ $\approx$ 1. The $\chi_\mathrm{red}^2$ was higher in the quiet Sun ($\approx$10) due to high residuals in the red wing of the line, due to an unknown blend.

We note that for constraining DEMs in this range of temperatures, the \ion{Si}{7} 257.37\,\AA~line can also be used. The line is not blended and can be without problems fitted with $\chi_\mathrm{red}^2$ $\approx$ 1. However, to avoid possible inconsistencies linked to element abundances, we in all structures but the background-subtracted loop LS used lines of iron ions only. 

\subsubsection{Mid $T$ range}

Constraints for DEM$_\kappa(T)$ near their peaks were obtained by using the \ion{Fe}{11} 182.17, \ion{Fe}{12} 192.39, and \ion{Fe}{13} 202.04\,\AA~lines whose fitting was already described. At higher temperatures near the peak, EIS observes multiple strong lines of \ion{Fe}{14}, most of which are either density-sensitive or blended \citep[e.g.,][]{delzanna13calib}. In accordance with \citet{delzanna13} we used the 211.32\,\AA~line, which we found to be the least sensitive to density. We note that in the quiet Sun spectra we observed this line to be accompanied with the \ion{Ni}{11} 211.43\,\AA~line, which we fitted with a single Gaussian. Fitting of the \ion{Fe}{15} 284.16\,\AA~line in active region resulted in obtaining high $\chi_\mathrm{red}^2$, reaching up to $\approx$ 100 in M1. Just as in the case of the \ion{Si}{10} 258.37\,\AA~or \ion{Fe}{12} 195.12\,\AA~lines, the Gaussian curve is likely not suitable for reproducing the wings of the line. This line is known to be blended with the \ion{Al}{9} 284.03\,\AA~line, but based on the symmetry of the residuals observed in wings of this line, this blend was either weak or not present at all in active region spectra. On the other hand, in the quiet Sun, the intensity of this blend was found to reach $\approx$ 30\% of the intensity of the 284.16\,\AA~line. We note that this observation is not consistent with the synthetic spectra, because the spectral synthesis performed with Maxwellian quiet Sun and active region DEMs suggest much lower contribution of this blend, about 1--2\% only. 

\subsubsection{High $T$ range}

In the quiet Sun we at log($T$\,[K]) $>$ 6.3 observed only very little emission. The \ion{Fe}{16} 262.98\,\AA~line which we used for constraining the DEMs at corresponding temperatures was very weak, reaching typically only about 1--2 erg cm$^{-2}$ s$^{-1}$ sr$^{-1}$ in fits with $\chi_\mathrm{red}^2$\,=\,1--11. In the quiet Sun, around 15\% of the intensity of this line is due to the \ion{Fe}{13} 262.99\,\AA~blend, which contribution we did not exclude because it is lower than the $\sigma_{\text{phot}}$ of the \ion{Fe}{16} 262.98\,\AA~line intensities and the calibration uncertainty of the instrument. Moreover, the ratios of this blend with other strong lines of \ion{Fe}{13} observable with EIS are density-sensitive in the range log($N_{\text{e}}$ [cm$^{-3}$]) 8.0--10.0, which corresponds to the densities of the observed structures. Therefore, the deblending of this line would be, if needed, difficult to perform. Even though the intensities of this line measured in the quiet Sun should be taken with a grain of salt, we used them as high-temperature constraints ensuring the convergence of the DEMs. To constrain DEMs at high temperatures in active regions, lines of \ion{Ca}{14}--\ion{Ca}{17} or \ion{Ni}{17} are often used \citep[see e.g.][]{warren12,delzanna13,mackovjak14}. As we were trying to avoid using lines of ions other than iron, we used the \ion{Fe}{17} 254.9\,\AA~line instead. This line is weak and we were only able to distinguish it from the continuum in spectra of M1, M2, and LC. Based on the shape of the line, it could be blended in its blue wing, even though no blend is suggested by the synthetic spectra. The blend is however  well-separated from the peak of \ion{Fe}{17} and we could fit the profile using two Gaussians. The intensities obtained were typically only a few ergs cm$^{-2}$ s$^{-1}$ sr$^{-1}$ in fits with $\chi_\mathrm{red}^2$ of the order of 10. Nevertheless, intensities of this line of this order of magnitude are comparable to those we obtained using forward modelling and were sufficient as a high-temperature constrain for our DEMs.

\begin{deluxetable}{cc|cc|ccccccc}[h!]
\tablecaption{Intensities of emission lines observed in the analyzed structures. \label{tab_fit_int_ar}}
\tablecolumns{111}
\tablenum{4}
\tablewidth{0pt}
\tablehead{
\multicolumn{2}{c}{Line} & \multicolumn{2}{c}{log ($T_\mathrm{max}$ [K])}  & \multicolumn{7}{c}{Structure} \\
\colhead{$\lambda$ [\AA{}]} & \colhead{Ion} & \colhead{Maxw.} & \colhead{$\kappa$ = 2} & \colhead{M1s} & \colhead{M2s} & \colhead{LCs} & \colhead{LSs} & \colhead{QS1} & \colhead{QS2} & \colhead{QS3}}
\startdata
182.17 & \ion{Fe}{11} & 6.15 & 6.30 & 1480 $\pm$ 44 & 1271 $\pm$ 30 &  412 $\pm$ 47 &  788 $\pm$ 49 & 	227 $\pm$ 4 & 208 $\pm$ 4 & 259 $\pm$ 4   \\
184.54 & \ion{Fe}{10} & 6.05 & 6.15 & 1262 $\pm$ 30 &  974 $\pm$ 21 &  211 $\pm$ 32 &  829 $\pm$ 37 & 	338 $\pm$ 3 & 327 $\pm$ 4 & 518 $\pm$ 4  \\
185.21 & \ion{Fe}{8} & 5.65 & 5.60 &  528 $\pm$ 28 &  512 $\pm$ 17 &   79 $\pm$ 30  &  295 $\pm$ 35 & 	27 $\pm$ 1 &  25 $\pm$ 1 &  68 $\pm$ 2   \\
186.89 & \ion{Fe}{12} & 6.20 & 6.35 & 3703 $\pm$ 36 & 3278 $\pm$ 24 & 1064 $\pm$ 39 & 1482 $\pm$ 37 & 	200 $\pm$ 2 & 173 $\pm$ 2 & 175 $\pm$ 2  \\
192.39 & \ion{Fe}{12} & 6.20 & 6.35 & 1872 $\pm$ 18 & 1709 $\pm$ 12 &  790 $\pm$ 20 & 1017 $\pm$ 19  & 438 $\pm$ 2 & 393 $\pm$ 2 & 352 $\pm$ 2  \\
194.66 & \ion{Fe}{8} & 5.65 & 5.60 &   92 $\pm$ 12 &  102 $\pm$  4 &   unobs.   &   88 $\pm$ 13   &  9  $\pm$ 1 &   8 $\pm$ 1 &  18 $\pm$ 1    \\
196.53 & \ion{Fe}{13} & 6.25 & 6.40 & 1059 $\pm$ 11 &  929 $\pm$  7 &  298 $\pm$ 12 &  217 $\pm$ 11   &  17 $\pm$ 0.4 &  15 $\pm$ 0.4 &  12 $\pm$ 0.5  \\
197.86 & \ion{Fe}{9} & 5.90 & 6.00 &  155 $\pm$ 16 &  129 $\pm$ 15 &   19 $\pm$ 17  &  156 $\pm$ 15   &  37 $\pm$ 1 &  35 $\pm$ 1 &  76 $\pm$ 1   \\
202.04 & \ion{Fe}{13} & 6.25 & 6.40 & 2847 $\pm$ 50 & 2706 $\pm$ 31 & 2078 $\pm$ 53 & 1710 $\pm$ 51  & 1001 $\pm$ 4 & 860 $\pm$ 4 & 670 $\pm$ 3   \\
203.83 & \ion{Fe}{13} & 6.25 & 6.40 & 8367 $\pm$ 102 & 7523 $\pm$ 63 & 3241 $\pm$ 107 & 2833 $\pm$ 98   & 212 $\pm$ 4 & 175 $\pm$ 3 & 143 $\pm$ 3  \\
211.32 & \ion{Fe}{14} & 6.30 & 6.45 & 5813 $\pm$ 354 & 5642 $\pm$ 178 & 3693 $\pm$ 407 & 2308 $\pm$ 332  & 294 $\pm$ 7 & 248 $\pm$ 8 & 174 $\pm$ 8  \\
249.17 & \ion{Ni}{17} & 6.50 & 6.70 &  446 $\pm$ 31 &  287 $\pm$ 25 &  742 $\pm$ 34 & unobs. & unobs. & unobs. & unobs. \\
254.90 & \ion{Fe}{17} & 6.60 & 6.65 &   13 $\pm$ 13 &    3 $\pm$  3 &    5 $\pm$  3 & unobs. & unobs. & unobs. & unobs. \\
257.55 & \ion{Fe}{11} & 6.15 & 6.30 &  506 $\pm$ 17 &  455 $\pm$ 16 &  163 $\pm$ 14 &  329 $\pm$ 17   &  88 $\pm$ 1 &  89 $\pm$ 2 & 108 $\pm$ 2   \\
257.77 & \ion{Fe}{11} & 6.15 & 6.30 &  230 $\pm$ 11 &  188 $\pm$ 10 &   73 $\pm$ 10 &  142 $\pm$ 11   &  37 $\pm$ 1 &  36 $\pm$ 1 &  44 $\pm$ 1    \\
258.37 & \ion{Si}{10} & 6.15 & 6.15 & 1975 $\pm$ 29 & 1602 $\pm$ 27 &  601 $\pm$ 24 &  972 $\pm$ 27   & 318 $\pm$ 2 & 288 $\pm$ 2 & 330 $\pm$ 2   \\
261.06 & \ion{Si}{10} & 6.15 & 6.15 &  598 $\pm$ 16 &  508 $\pm$ 15 &  230 $\pm$ 15 &  369 $\pm$ 17   & 189 $\pm$ 2 & 171 $\pm$ 2 & 190 $\pm$ 2   \\
262.98 & \ion{Fe}{16} & 6.45 & 6.60 &  768 $\pm$ 19 &  695 $\pm$ 19 & 1036 $\pm$ 20 &  130 $\pm$ 18   &   2 $\pm$ 1 &   1 $\pm$ 4 &   1 $\pm$ 0.5   \\
275.37 & \ion{Si}{7} & 6.15 & 6.15 &  126 $\pm$ 11 &  186 $\pm$  9 &   unobs.  &  230 $\pm$ 12   &  15 $\pm$ 1 &  14 $\pm$ 2 &  44 $\pm$ 1   \\
284.16 & \ion{Fe}{15} & 6.35 & 6.50 & 9496 $\pm$ 93 &11616 $\pm$ 103 & 8054 $\pm$ 93 & 3767 $\pm$ 95   & 164 $\pm$ 2 & 140 $\pm$ 2 & 113 $\pm$ 2   \\
\enddata
\end{deluxetable}
\end{document}